\documentclass[fleqn,usenatbib]{mnras}

\usepackage{xcolor}

\usepackage[T1]{fontenc}
\usepackage{booktabs}

\DeclareRobustCommand{\VAN}[3]{#2}
\let\VANthebibliography\thebibliography
\def\thebibliography{\DeclareRobustCommand{\VAN}[3]{##3}\VANthebibliography}

\usepackage{graphicx}	
\usepackage{amsmath}	
\usepackage{amssymb}	

\usepackage{color,ulem}  
\definecolor{deepblue}{rgb}{0,0,0.5}  
\definecolor{deepred}{rgb}{0.6,0,0}   
\definecolor{deepgreen}{rgb}{0,0.5,0} 
\definecolor{darkgreen}{rgb}{0,0.6,0} 

\def\fcont{F$_{\rm C}$}

\def\fbr{F$_{\rm B,R}$}
\def\fb{F$_{\rm B}$}
\def\fr{F$_{\rm R}$}

\def\ha{{\sc{H}}$\alpha$\/}
\def\habc{{\sc{H}}$\alpha_{\rm BC}$}
\def\hanc{{\sc{H}}$\alpha_{\rm NC}$}

\def\hb{{\sc{H}}$\beta$\/}
\def\hbbc{{\sc{H}}$\beta_{\rm BC}$}
\def\hbnc{{\sc{H}}$\beta_{\rm NC}$}

\def\he{{\sc{H}}$\epsilon$\/}
\def\kms{km s$^{-1}$\/}
\def\l{$\lambda$}

\def\lbol{L$_{\rm bol}$\/}
\def\ledd{L$_{\rm Edd}$}
\def\loiii{L$_{\rm [OIII]}$}
\def\lhabc{L$_{\rm H\alpha BC}$}
\def\l5100{L$_{\rm 5100}$}
\def\oi{{[O\sc{i}]}\/}

\def\oib{{[O\sc{i}]}$\lambda$6364\/}
\def\oiii{{[O\sc{iii}]}\/}

\def\oiiib{{[O\sc{iii}]}$\lambda$5007\/}
\def\oiiill{{[O\sc{iii}]}$\lambda\lambda$4959,5007\/}

\def\nii{{[N\sc{ii}]}\/}
\def\niill{{[N\sc{ii}]}$\lambda\lambda$6548,6584\/}

\def\niib{{[N\sc{ii}]}$\lambda$6584\/}
\def\neiii{{[Ne\sc{ii}]}$\lambda$3967\/}
\def\mbh{M$_{\rm BH}$}
\def\msun{M$_\odot$}
\def\rk{R$_{\rm K}$}
\def\rx{R$_{\rm X}$}
\def\sii{{[S\sc{ii}]}\/}
\def\siill{{[S\sc{ii}]}$\lambda\lambda$6716,6731\/}
\def\siia{{[S\sc{ii}]}$\lambda$6716\/}

\title[Identification and Multiwavelength properties of 47 type 1 AGN]{SDSS-IV MaNGA: Identification and Multiwavelength Properties of Type-1 AGN in the DR15 sample}

\author[E. Cortes-Suárez et al.]{
Edgar Cortes-Suárez$^{1}$\thanks{E-mail: ecortes@astro.unam.mx}, C. A. Negrete$^{2}$,
H. M. Hernández-Toledo$^{1}$, H. Ibarra-Medel$^{3}$, and I. Lacerna$^{5,6}$
\\
$^{1}$Instituto de Astronomía, Universidad Nacional Autónoma de México, A.P. 70-264, 04510 CDMX, Mexico\\
$^{2}$CONACyT Research Fellow -- Instituto de Astronomía, Universidad Nacional Autónoma de México, A.P. 70-264, 04510 CDMX, Mexico. \\
$^{3}$University of Illinois Urbana-Champaign, Department of Astronomy, 1002 W Green St, Urbana, Illinois, 61801, United States\\
$^{5}$Instituto de Astronom\'ia y Ciencias Planetarias, Universidad de Atacama, Copayapu 485, Copiap\'o, Chile\\
$^{6}$Millennium Institute of Astrophysics, Nuncio Monsenor Sotero Sanz 100, Of. 104, Providencia, Santiago, Chile \\
}

\date{Accepted 2022 May 26. Received 2022 May 13; in original form 2021 December 8}

\pubyear{2015}

\begin{document}
\label{firstpage}
\pagerange{\pageref{firstpage}--\pageref{lastpage}}
\maketitle

\begin{abstract}
We present a method to identify type-1 active galactic nuclei (AGN) in the central 3 arcsec integrated spectra of galaxies in the MaNGA DR15 sample. It is based on flux ratios estimates in spectral bands flanking the expected \ha\ broad component \habc. The high signal-to-noise ratio obtained (mean S/N = 84) permits the identification of \habc\ without prior subtraction of the host galaxy (HG) stellar component. A final sample of 47 type-1 AGN is reported out of 4700 galaxies at $z$ < 0.15. The results were compared with those from other methods based on the SDSS DR7 and MaNGA data. Detection of type-1 AGN in those works compared to our method goes from 26\% to 81\%. Spectral indexes were used to classify the type-1 AGN spectra according to different levels of AGN-HG contribution, finding 9 AGN-dominated, 14 intermediate, and 24 HG-dominated objects. Complementary data in NIR-MIR allowed us to identify type I AGN-dominated objects as blue and HG-dominated as red in the WISE colors. From NVSS and FIRST radio continuum data, we identify 5 HERGs (high-excitation radio galaxies) and 4 LERGs (low-excitation radio galaxies), three showing evidence of radio-jets in the FIRST maps. Additional X-ray data from ROSAT allowed us to build \oiii\ and \habc\ versus X-ray, NIR-MIR, and radio continuum diagrams, showing that L(\habc) and L(\oiii) provide good correlations. The range in \habc\ luminosity is wide 38 < logL(\habc) < 44, with log FWHM(\habc)$\sim$ 3–4, covering a range of Eddington ratios of -5.15 < log \lbol/\ledd\ < 0.70. Finally, we also identify and report ten possible changing-look AGN candidates.

\end{abstract}




\section{Introduction}
\label{sec:intro}
The identification of active galactic nuclei (AGN), mostly type-2 AGN (which only show narrow emission lines) but also AGN in general, frequently uses narrow emission lines that emerge from gas that is photoionized by the nuclear ionizing continuum, estimating their flux ratios and placing them into the BPT diagnostic diagrams \citep{Baldwin1981, Kewley2001, Kauffmann2003, Kewley2006}. 
Other studies have shown however, the importance of taking into account the contribution from post-asymptotic giant branch (AGB) emission and shocks that may lead to misinterpretations in these diagrams. To this purpose, the introduction of the WHAN diagram \citep[EW(\ha) Vs. log \nii/\ha;][]{CidFernandes2011} has been useful to disentangle these contributions. 

On the other hand, the identification of type-1 or broad line AGN, based on the presence of broad permitted emission lines (mainly \ha\ and \hb\ in the optical range) 
is also fraught with difficulties. Broad emission lines (BELs) show complex profiles that are difficult to identify correctly. The shape and the width of the BEL profiles depend, among other parameters, on the geometry of the line emitting region, 
on obscuring effects, on the superposition of line emission from different regions, and on the anisotropy of the line emission clouds.  
Furthermore, the observed velocity field might be a superposition of different components, such as Doppler motions, turbulence, shocks, inflows/outflows, and rotation, such that different velocity components result in different profiles, and thus, the final profile is a convolution of all these. \citep{Sulentic2000,Zamfir2010,Marziani2018}.


Among previous important attempts to identify type-1 AGN in the nearby universe, we mention \citet[][e.g. the OSSY catalogue]{Ho1995,Hao2005,Greene2007,Onori2017,Oh2011}. 
More recently, attempts such as \cite{Stern2012}, \cite{Oh2015} and \cite{Liu2019}  have increased significantly the number of these objects, 
based on the analysis of available spectra from the Sloan Digital Sky Survey Data Release 7 (SDSS DR7) database. \cite{Stern2012} implemented a flux ratio method around the \ha\ region, interpreting the excess flux over an interpolated continuum as due to the presence of a broad line component. \cite{Oh2015} proposed the estimate of a flux ratio by considering a spectral band and an adjacent continuum, further refining their criteria to recover low luminosity type-1 AGN. Notice also that both \cite{Stern2012} and \cite{Oh2015}, applied a host galaxy subtraction to the spectra as part of their methodologies.

Other projects like SPIDERS \citep[SPectroscopic IDentification of eROSITA Sources;][]{Dwelly2017} identify 
type-1 AGN as optical counterparts of X-ray surveys like ROSAT, XMM-Newton and more recently from eFEDS (the eROSITA Final Equatorial-Depth Survey). Notice however, that the identification of SPIDERS type-1 AGN uses the spectroscopic coverage around \hb\ and/or MgII emission lines, instead of the \ha\ region.

With the advent of surveys using the integral field spectroscopy (IFS) technique, more detailed methods for identifying and analyzing AGN (in particular of type-1 AGN), and 
their relation to the host galaxies are possible. Among the attempts to identify AGN using data from the MaNGA survey, we mention \cite{Rembold2017} and \cite{Sanchez2018}. They identified AGN in the MaNGA Product Launch 5 sample (MPL-5, which contains 2792 galaxies) by using the BPT and WHAN 
diagrams. While \cite{Rembold2017} consider only candidates above the Kewley line \citep[defined in][]{Kewley2001} in the \nii\ BPT diagram and with EW(\ha)$>$3, \cite{Sanchez2018} consider all the candidates above the Kewley lines in the three BPT diagrams (\nii, \oi\ and \sii) but with a more relaxed threshold in EW(\ha) ($>$1.5). They identified 62 and 98 AGN, respectively, regardless of whether they were type-1 or 2 AGN. 

In contrast, \cite{Wylezalek2018,Wylezalek2020} followed a different methodology to find AGN in the MaNGA survey. They took advantage of the IFS data by building spatially-resolved BPT diagrams, identifying 308 AGN candidates from MPL-5. More recently, \cite{Comerford2020} made a cross-matching of the MaNGA MPL-8 data (6261 galaxies) with other catalogs at different wavelengths, finding 406 AGN, most of them identified as radio sources. 

In this work, we present a method to identify type-1 AGN in the MaNGA survey by using the integrated spectra of the central 3 arcsec of the IFS data. It is based on the estimate of two flux ratios placed at the red and blue positions aside from the expected \ha\ broad emission line. 
This is a ''non-invasive'' method 
that avoids the host galaxy subtraction of the optical spectra and considers only the prominence of the \ha\ emission line broad component for a given signal-to-noise ratio (S/N). The method was applied to the MaNGA DR15 (MPL-7) data sample and further tested by using data from the SDSS DR7, showing comparable results to other methods that identify type-1 AGN in the DR7 catalogs.  


This paper is organized as follows. Section (\ref{sec:data}) summarizes the description of MaNGA survey, the spectra used in this work and its multiwavelength information. Section (\ref{sec:method}) describes the selection method used to identify AGN with broad emission lines, as well as comparisons considering the host galaxy subtraction and the BPT and WHAN diagrams selection. Section (\ref{sec:comparison}) reports a comparison of our final sample with previous SDSS and MaNGA AGN catalogues, applying also our method in those catalogues. Section (\ref{sec:multiwavelength}) presents a characterization of our type-1 AGN sample in terms of properties in the infrared, radio, X-Ray and optical properties. Finally, in Section (\ref{sec:conclusions}) we summarize the results and present our conclusions.
We assume a Hubble constant H$_0$ = 70 \kms\ Mpc$^{-1}$, $\Omega_m$ = 0.3, and $\Omega_\Lambda$ = 0.7 throughout.

\section{Data Sample }
\label{sec:data}

The MaNGA survey is part of the SDSS-IV project that was dedicated to observe around 10 000 nearby galaxies, in a redshift range z < 0.15, 
with integral field spectroscopy. 
The survey covers a wide interval from 3 600 \AA\ to 10 000 \AA\ in the optical wavelength, simultaneously using 17 integral field units (IFUs) each one composed of arrays of 2 arcsec diameter fibers to map in detail the core and the galaxy within the field-of-view. Each IFU feeds a dual channel spectrograph for the red and blue arms respectively \citep{Smee2013}. The average spectral resolution ($\lambda/\delta\lambda$) in the blue channel is 1915, and 2250 in the red one. Detailed description of the observations, as well as the selection criteria can be found in \cite{Bundy2015}. \citet{Yan2016b} describe the survey spectrophotometric calibrations, while \citet{Law2015} explains the observational strategy as well as the pipeline used for the data reduction. 

Three main sub-samples were  generated in the MaNGA sample. A primary sample (initially about 50\% of the targets), with coverage up to R = 1.5 Re (effective radius) having a flat distribution in K-corrected i-band absolute magnitude (M-i). A secondary sample, (about 33\% of the initial targets), having a flat distribution in M-i with coverage up to R = 2.5 Re, and a third Color-Enhanced supplement sub-sample designed to add galaxies in regions of the (NUV - i) vs M-i color-magnitude diagram that are under-represented in the primary sample, such as high-mass blue galaxies and low-mass red galaxies \citep[about 17\% of the initial targets;][]{Wake2017}.

In this paper we carry out an analysis of the spectroscopic data in the MaNGA DR15 \citep[][]{Aguado2019} IFS survey, for 4636 galaxies, about half of the total MaNGA sample. 
The integrated spectra in the central 3 arcsec circular aperture was synthesized for each galaxy. 
To test the results, we used the raw and stellar subtracted spectra generated with the Starlight code \citep{CidFernandes2005}. We also use the data from the MPL-10 version of the Pipe3D Valued Added Catalog\footnote{ Table SDSS17Pipe3D\_v3\_1\_1.fits downloaded from
\url{http://ifs.astroscu.unam.mx/MaNGA/Pipe3D_v3_1_1/tables/}
} 
\citep[VAC, ][]{Sanchez2016a,Sanchez2021,Lacerda2022}, a fitting tool for the analysis of the stellar populations and the ionized gas derived from moderate resolution IFS spectra of galaxies.  
Multiwavelength information was gathered for our final type-1 AGN sample from different databases; the WISE catalog \citep[Wide-field Infrared Survey Explorer, ][]{Wright2010} at IR wavelengths, radio continuum from FIRST \citep[Faint Images of the Radio Sky at Twenty centimeters, ][]{Becker1995} and NVSS \citep[NRAO Very Large Array Sky Survey, ][]{Condon1998}, as well as X-ray catalogs mainly ROSAT \citep{Boller2016} in order to carry out a first analysis of their multiwavelength properties.

\section{Type-1 AGN Selection Method}
\label{sec:method}
 
Among the reasons for studying broad-line (or type-1) AGN are that they permit us to retrieve kinematic information of the region closest to the central supermassive black hole (SMBH).
Another advantage is that under certain circumstances we can use a Power Law, (PL) to estimate the flux of the ionizing continuum of an AGN, a valid approximation if the continuum is synchrotron and  non-thermal in nature. However, \cite{Malkan1982} showed that the optical-UV continuum is thermal, due to the accretion disk. 
In general, the SED of AGN is more complex than a single power law specially for sources accreting above the accretion limit.  This information in combination with the multi-dimentional space of spectroscopic, photometric and kinematic parameters derived from the IFU analysis, allows for a more detailed analysis of the properties of AGN and their relation to their host galaxies.

A frequent method to identify active galaxies in the nearby Universe, uses diagnostic diagrams such as the BPT diagrams \citep{Baldwin1981,Kewley2006,Sanchez2018}. They are useful to determine the origin of the photons which ionize the gas producing emission lines. The source of these ionizing photons could be an AGN radiation field, the formation of massive stars, or a mix of both, for which BPT diagrams successfully isolate them in well-confined regions. However, for luminous type-1 AGN (with bolometric luminosities L$_{bol}$\ $>$ 10$^{45}$ ergs), BPT diagrams fail to diagnose the AGN origin because the line ratios do not consider the presence of strong, broad emission components. In this case, the peak flux of \ha\ and \hb\ narrow components (NCs; \hanc, \hbnc) show a larger value due to the contribution of the broad components. In many cases, the surrounding forbidden lines close to \ha\ and \hb\ broad components (BCs; \habc, \hbbc), namely \oiiill, \niill, and \siill, have a flux increment too due to the presence of the broad component of the Balmer lines. For the highest luminosity AGN, where the non-thermal nuclear emission dominates over that from the host galaxy (HG), the narrow components could be completely buried into the broad ones \citep[e.g., the narrow line Seyfert 1 galaxies - NLSy1;][]{VandenBerketal2001,Marziani2010}. Hence, to isolate the NCs properly, it is necessary to make a good spectral line decomposition.

In terms of the broad line emitting region, the AGN unified model by \citet{UrryPadovani1995} proposes that the Broad Line Region (BLR) is hidden from our line-of-sight by a molecular torus surrounding the accretion disk (AD). For small angles (towards an AD face-on), we can see the BLR directly and thus a type-1 AGN spectra, while for larger angles (towards an AD edge-on), the torus obscures the BLR showing only narrow lines, which is characteristic of type-2 AGN. In the latter case, the broad lines could be detected in polarized light \citep{Tran2010}, although this is not the case for all type-2 AGN \citep{Tran2011}.
    
Aside from the physical (presence of gas and dust) and geometrical reasons that may prevent the detection of the BLR, in practice, the spectra coming from the 3-arcsec SDSS observations or other observations using similar apertures contain a fraction of the host galaxy light. A most common procedure to remove that contribution is by using stellar population synthesis methods \citep[e.g.,][]{Stern2012}. However, these procedures are model-dependent, yielding non-unique results with different contributions. The host galaxy fitting could be translated to degenerations on the pure AGN spectra. The residuals of the HG+PL subtraction could modify the spectral region close to the broad components. Given these difficulties, it would be desirable to have a method that allows for identifying broad emission lines in the observed spectra without making any a-priori assumption on both the Host Galaxy stellar and AGN contributions. In line with this idea, 
we present a method for identifying the broad \ha\ emission line in the nuclear 
spectra of MaNGA galaxies that led us to the identification of a type-1 AGN sample.

\subsection{Flux ratio method}
\label{sec:fluxratio}

We next outline a straightforward method that considers the estimate of flux ratios in two 
spectral regions around the expected position of the \ha\ broad emission line. While one of the selected regions avoids any emission or absorption lines, the other two regions are selected close to the expected broad emission line. If a broad component is present, the evaluation of those flux ratios, should yield larger values than expected if 
\habc\ is absent. 
This method which is a variant of methods presented in \cite{Stern2012}; \cite{Oh2015}; \cite{Liu2019} is optimized around the \ha\ region line. 
It is applied to the integrated spectra from the central 3-arcsec aperture without any host stellar subtraction and exploits the significantly higher S/N achieved from the MaNGA data.



The first step of our method is to extract the central 3 arcsec spectra of the 4636 MaNGA MPL-7 data cubes, with an aperture size similar to that of SDSS DR7 single aperture fiber. Since the size of each spaxel is 0.5x0.5 arcsec, the full aperture spans over 29 spaxels, allowing us to achieve significantly higher S/N values.

\begin{figure}
  \centering
   \includegraphics[scale=0.55]{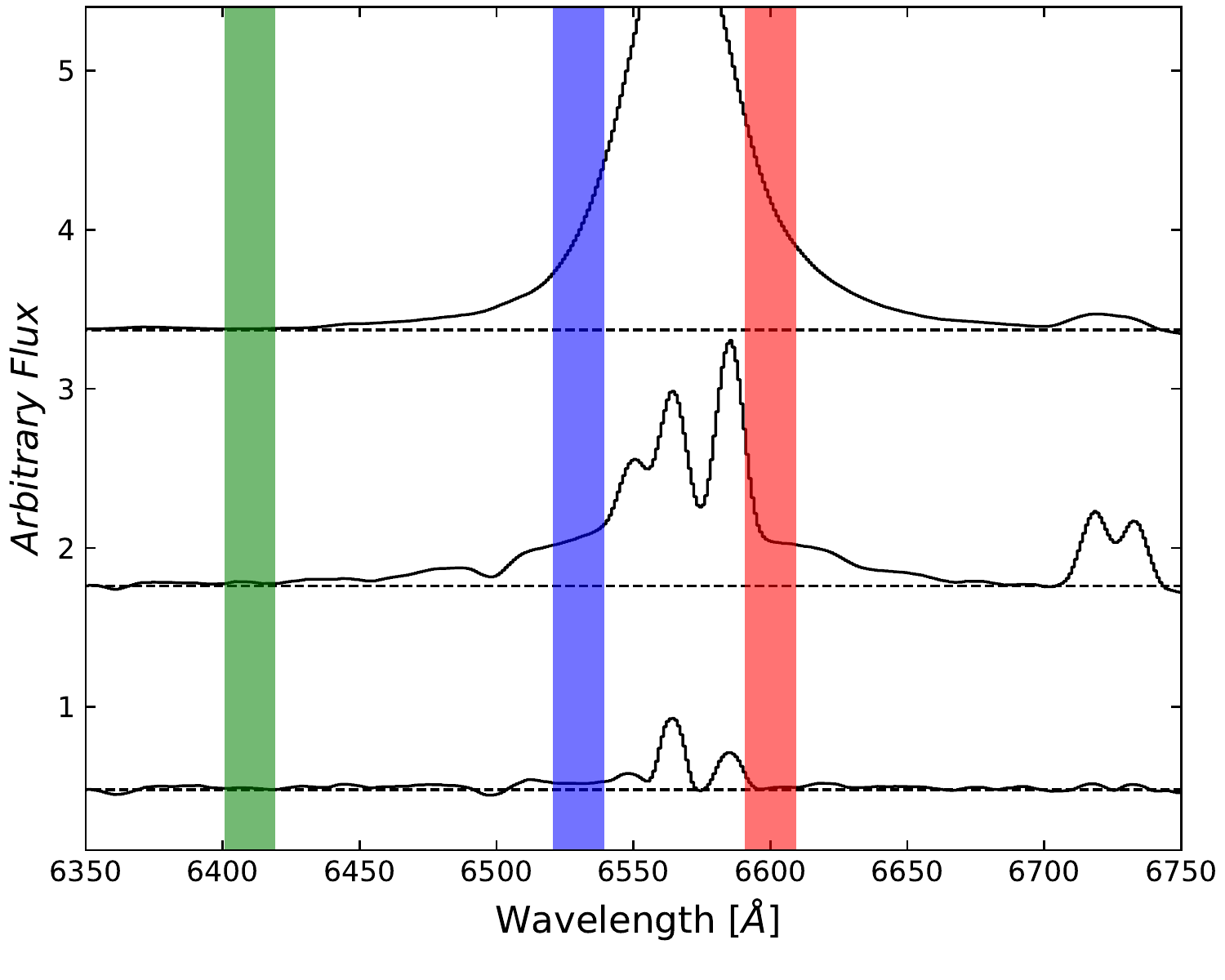}
  \caption{Three different spectra in \ha\ region. The upper panel corresponds to the spectrum of a type-1 AGN, the intermediate panel to a Seyfert 1.9, and the lower panel to a galaxy without active nuclei. The bands illustrate the spectral regions and positions we adopt for the continuum (green) and the broad component (blue and red).}
\label{fig:halphamethod}
\end{figure}

\begin{figure*}
  \centering
    \hspace*{-0.2cm}
  \includegraphics[scale=0.56]{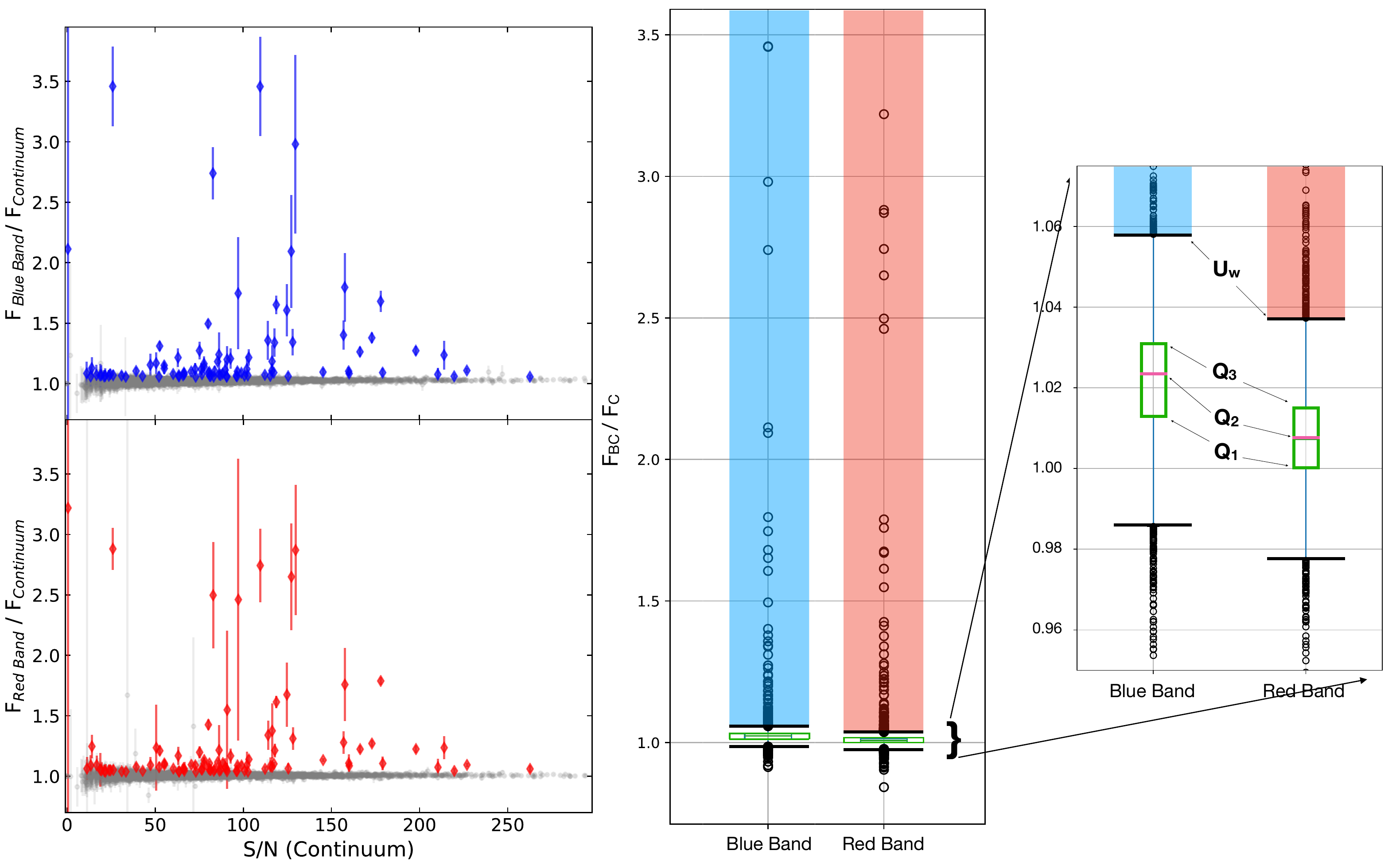}
  \caption{Left: Flux ratio distribution compared with the continuum S/N at 6400-6420 \AA. Colored diamonds represent the upper outliers ($\sim$3.6\% red band,  $\sim$2.7\% blue band) while gray dots are the non-active galaxies, which flux ratio is around or lower than 1. Middle: Boxplots used for the selection of AGN. The outliers of each band are the black circles upside the upper whiskers (blue and red bands). Right: Zoom-in of the quartiles region. The first and third quartiles are shown as green boxes, upper and lower whiskers as black lines. Below the upper limits live $\sim$96\% and $\sim$97\% of the MPL-7 sample in the red and blue bands, respectively (grey dots in the left diagram). }
\label{fig:bands}
\end{figure*}

In contrast to previous works that use only one spectral region near \ha, we propose two flanking spectral regions to take into account the range of wavelengths and asymmetries of the \ha\ broad emission line and the presence of stellar absorption lines. We define three spectral regions or bands that help us characterize the \ha\ region. One adjacent region is selected such that no emission or absorption lines are present and where the continuum flux (\fcont) is defined. Two additional regions were selected to flank \ha, defined as the blue and red (\fb\ and \fr) band fluxes measured at both sides of a region where the broad emission component is expected. To select the bands, the location of various absorption stellar lines as well as the widths of the narrow emission lines were taken into account. In type-1 AGN, the width of the narrow lines covers a range of FWHM from a few hundred \kms\ up to 1,000 \kms\, while for the broad \ha\ line, the FWHM could be as broad as 10,000 \kms\ or more \citep{Antonucci1993}. Regarding the stellar absorption lines, in the region around \ha, the deepest one is TiO at 6498 \AA\ \citep{BicaAlloin1986}. Considering the maximum FWHM value that the \niill\ can have as upper limits, and avoiding the TiO stellar absorption at 6498 \AA, we define wavelength intervals of 20 \AA\ width, at 
\begin{itemize}
    \item 6400-6420 \AA\ for the continuum band (\fcont)
    \item 6520-6540 \AA\ for the blue band (\fb), and 
    \item 6590-6610 \AA\ for the red band (\fr)
\end{itemize}
The width was chosen similarly to \cite{Oh2015} because it permits us to get reliable values of flux for a given S/N ratio. Due to the position bands, the FWHM that we can detect is limited by FWHM(\habc) $\gtrsim$ 1500 \kms, considering that the position of the blue band is 2$\sigma$ ($\sim$ 33 \AA) away from the \ha\ rest frame for a Gaussian with the minimum FWHM, and 1690 \kms\ for the red band (2$\sigma \sim$ 37 \AA).

Figure \ref{fig:halphamethod} shows three examples of spectra illustrating the location of the blue and red bands with respect to the broad \ha\ line: a type-1 AGN (top), a Seyfert 1.9 (middle), and a galaxy without a broad line (bottom). The \fcont\ band is shown in green, and \fb\ and \fr\ bands are shown in blue and red, respectively. Notice that \fcont\ is located in a feature-less region, far enough of the \ha\ broad component (to reach this continuum band, \ha\ should have a FWHM > 10 000 \kms). On the other side, \fb\ and \fr\ bands are as close as possible to \ha\ to detect the weakest broad components, if any. The location was selected outside the \niill\ width set above to avoid flux contamination of this line. 

The band fluxes were estimated using the task \textit{splot} from IRAF \citep{Tody1986}. The spectral flux was normalized along the entire observed range by fitting and subtracting a fifth-order spline function. In galaxies where the AGN power law is dominant, the normalization does not fit well around the \hb\ region. However, the slope around the \ha\ region is not so steep \citep[see. e.g.][]{VandenBerketal2001}, so the fit with the spline function was enough to model it. The \textit{splot} task returns values of the average flux and the RMS over each band and the S/N of the continuum region. 
With the flux values, we compute the ratios  \fb/\fcont\ and \fr/\fcont, between the blue and red bands (\fbr) over the continuum. 

Left panels of Figure \ref{fig:bands} show the values of the \fbr/\fcont\ ratios in the red and blue bands as a function of the S/N. It can be seen that both flux ratios can be separated into two populations: 
\begin{itemize}
    \item \fb/\fcont\ and \fr/\fcont\ $\sim$ 1. A first population containing the great majority of the galaxies (grey dots) suggesting the absence of a broad line around \ha, and 
    \item \fb/\fcont\ > 1 and \fr/\fcont\ > 1. A second population having higher values of \fb/\fcont\ and \fr/\fcont\ (blue and red dots) suggesting the possible presence of the \ha\ broad line.
\end{itemize}
While objects in the first population are mainly composed of non-active galaxies and type-2 AGN, objects in the second population are proposed as our best candidates to show a broad \ha\ line. 
The number of objects with values lower than 0.95 are 15 and 4 for the red and blue bands, respectively. In those spectra, we find star-forming galaxies with negative slopes and no broad \ha\ line.

To separate both populations, we made use of a boxplot-whiskers diagram (Figure \ref{fig:bands}, middle and right panels). This statistical tool help us 
to visualize the distribution of the computed flux ratio values and 
to establish a quantitative criterion for the broad \ha\ identification. While for the first population, the boxplot and their whiskers shows mainly flux ratio values around 1, the candidate sample shows a much broader distribution of flux ratio values 
placed out of the box and beyond of the whiskers (black circles in middle and right panels of Figure \ref{fig:bands}). These outliers are our type-1 AGN candidates. To analyze the distribution of the flux ratio \fbr/\fcont\ values, we developed a python code based on the \textit{Pandas} package \citep{pandas2010} to estimate the median, average and the quartiles\footnote{The flux ratio values at which 25, 50 and 75\% of the sample is contained.} of the distribution, used to 
build the boxplots for each blue and red bands.

The green area in the middle panel of Figure \ref{fig:bands} (also shown in the zoom in the right panel), is known as the box, 
which is delimited by the first ($Q_1$) and third quartiles ($Q_3$) (the 25\%\ and 75\%\ percentiles of the cumulative distribution of the sample, respectively). 
The median value of the distribution (or $Q_2$) lies within this area. The computed $Q_3$ values are 1.031 and 1.015 for the blue and red bands. For both blue and red distributions, the $Q_3$ value is near one, which means that in at least 75\% of the galaxies, we do not expect to find a broad \ha\ component. However, among the remaining 25\%\ objects, we still have many galaxies that belong to the "non broad-line" distribution. Then, we compute the upper whisker defined as $U_W = 1.5 \times  IQR \, + \, Q_3$, where $IQR = (Q_3-Q_1)$ is the interquartile range (for a symmetric distribution, $IQR = Q2$, which is not our case). We will not consider the lower whisker since it has values near and below one, which indicates the absence of a broad-line feature. 
The $U_W$ computed values are 1.037 and 1.056 for \fr\ and \fb\ respectively. Finally, we will consider as type-1 AGN candidates those outliers objects that are beyond the upper whisker. Above the $U_W$ limits, we found 171 outliers for \fr, and 126 for \fb\ ($\sim$4\% and $\sim$3\% of the total sample).
The first two Columns of Table \ref{tab:boxplot} resumes the values of the quartiles, the upper whisker, the mean and the maximum values for each band. 

\begin{figure}
  \centering
   \includegraphics[width=\columnwidth]{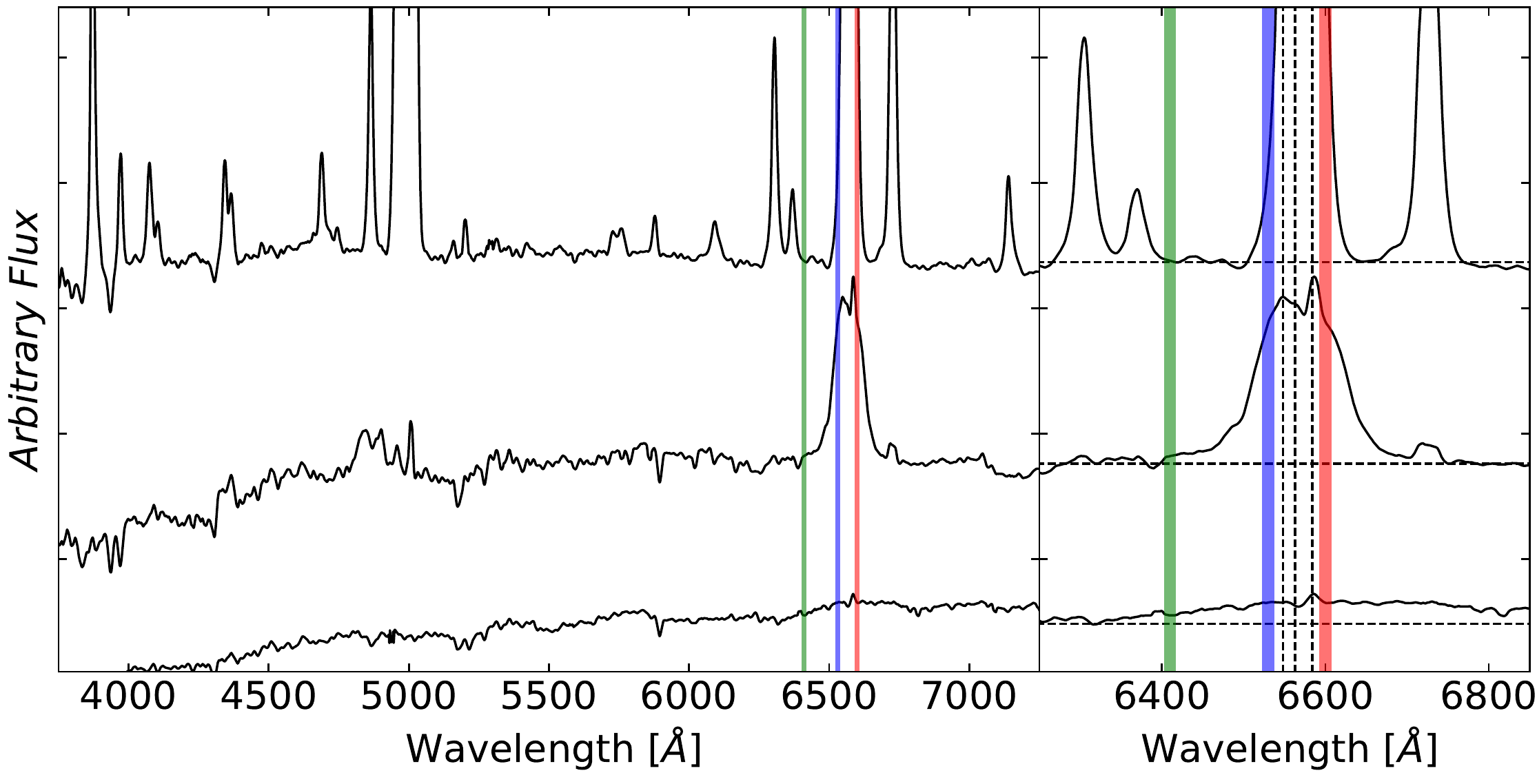}
  \caption{Three different spectra were found in the 93 candidates samples. The upper panel corresponds to the spectrum of a type-2 AGN, the intermediate panel to a type-1, and the lower panel to that of an M-type star. The left panel shows the overall spectra, while the right panel is a zoom-in to the \ha\ alpha region.}
\label{fig:candidates}
\end{figure}

The next step is to consider only the spectra with a broad-line feature that stands out over the \ha\ region continuum. To do that, we select only the candidates cataloged as outliers in both boxplots, reducing the sample to 94 candidates ($\sim$2\% of the total sample). After considerably reducing the initial sample, it was possible to visually inspect the spectra to verify the presence of the broad emission line component. Examples of candidate spectra are shown in Figure \ref{fig:candidates}, where the left panels show the overall spectra and the right panels zoom in into the \ha\ region. Two galaxies, 9031-12705 and 9036-6101, are the same object (same MaNGA-ID 1-210186), so it reduces the sample to 93 candidates. In addition, another 18 objects seem to show optical spectra resembling those of M-type stars \citep[see for example Figure 8, bottom row of][]{Oh2015}. The right lower panel in Figure \ref{fig:candidates} shows an example spectra of such objects, where the shape of the stellar continuum resembles a faint broad line in the \ha\ region. For our purposes, these types of objects were considered false positives. 

\begin{figure}
  \centering
   \includegraphics[scale=0.41]{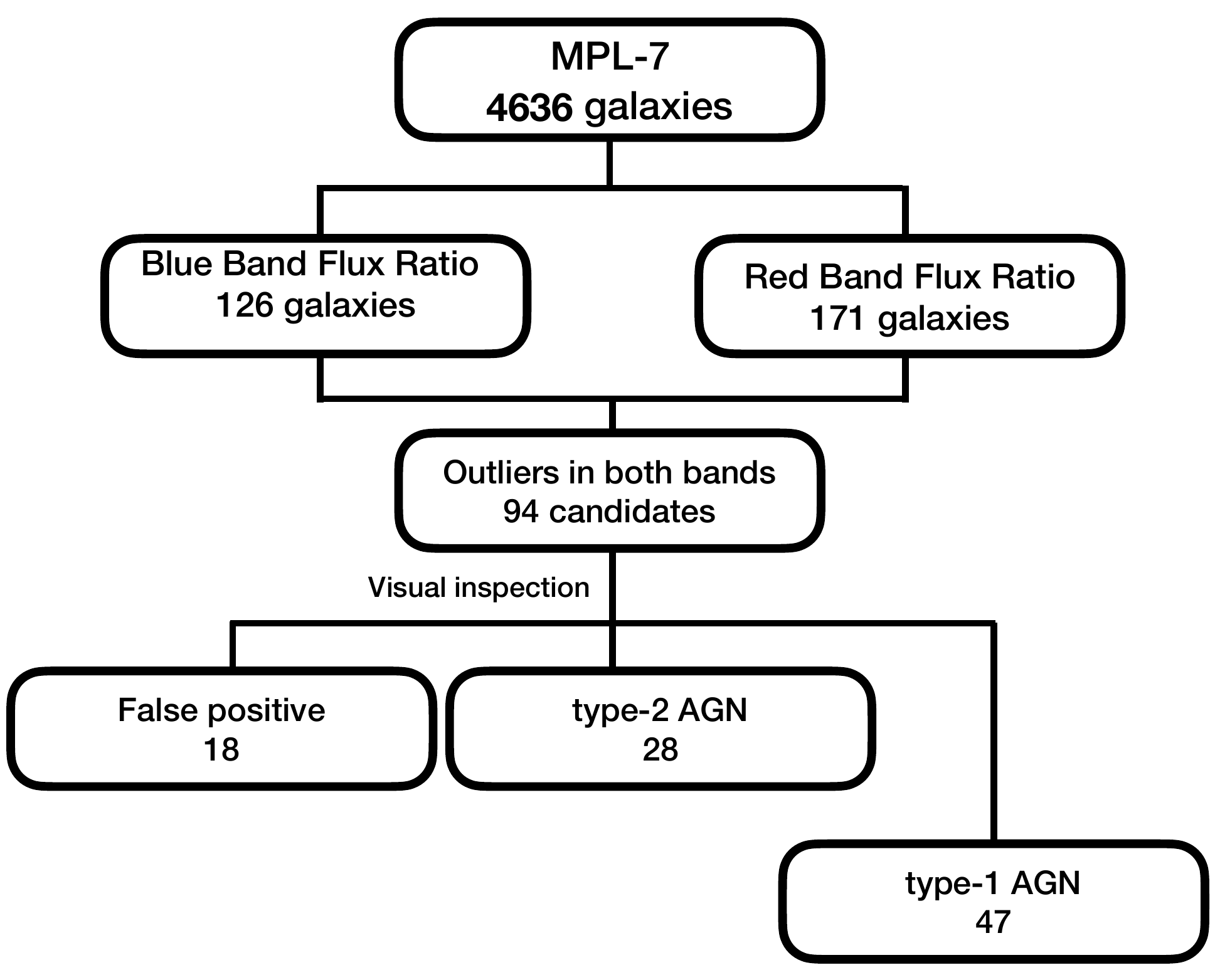}
  \caption{Summary of the methodology performed to find the type-1 AGN sample.}
\label{fig:summary}
\end{figure}

The remaining 75 candidates have broad and/or narrow emission lines. All of them were classified as AGN due to the strength and FWHM of the emission lines (according to the minimum width given by the bands position), consistent with having high \fbr/\fcont\ values. In the visual inspection, we separated them into type-1 and 2 AGN. The upper panel in Figure \ref{fig:candidates} illustrates a type-2 AGN spectrum showing only narrow emission lines (which were also found on the Seyfert region in the BPT diagrams, section \ref{sec:type2}), while the middle panel shows a type-1 AGN spectrum, where the broad component stands out on the continuum. This final cut leaves us with 28 type-2 AGN and \textbf{47 type-1 AGN} (1\% objects of the total sample). These 47 type-1 AGN are considered our final sample. 
Figure \ref{fig:summary} summarizes the different steps followed in our methodology. Sec. \ref{sec:sample} presents a detailed description of these objects. 

Table \ref{tab:47type1} list our type-1 AGN sample and some of their global properties. Column (1) is the SDSS Object ID, Column (2) shows the plate-IFU of MaNGA, Columns (3) and (4) are the RA and DEC in epoch 2000.0, Column (5) is the g-band magnitude from SDSS, Columns (6) and (7) are the redshift and the stellar mass of the host galaxy, respectively, both from NASA-Sloan Atlas, Column (8) is the optical luminosity of the \oiii\ emission line, extracted from Pipe3D, Column (9) is the estimated luminosity of the \ha\ broad component extracted from our emission lines fitting (results will be published in a forthcoming paper), and Column (10) is the empirical classification of the AGN described in section \ref{sec:sample}.

The fraction of type-1 AGN found with our method (1\%) in the MaNGA DR15 sample is consistent with the expected  fraction of high-ionization AGN (type-1 and 2) among local galaxies (2\%\ to 10\%), given a redshift and SMBH mass galaxy range \citep[e.g.,][and references therein]{Huchra1992,Ho1997,Netzer2013}. In turn, the type-1 AGN fraction is around 10\%\ of the total AGN population. This is also consistent with the results of \cite{Stern2012} and \cite{Oh2015}, which report type-1 AGN fractions of about 1\%\ from their initial samples.

\subsubsection{Rejected candidates}
We also visually inspected all the rejected candidate objects appearing in only one band flux ratio (33 and 78 for the red and blue boxplots). The reason is that the BLR line profiles may show a broad component displaced with respect to the restframe due to wind-driven outflows or particular kinematics of this region \citep[e.g., Figure 9.8 of][]{Netzer2013}. 
However, our results indicate that all these rejected candidates could be classified as type-2 AGN  or objects with M-type stellar spectra. We also found galaxies with more than one narrow emission line that could be related to multiple star-forming regions. All these objects were discarded from our final sample since we selected only the outliers in both boxplots.

\subsubsection{The role of the S/N in the flux ratio method}
\begin{figure}
  \centering
   \includegraphics[width=\columnwidth]{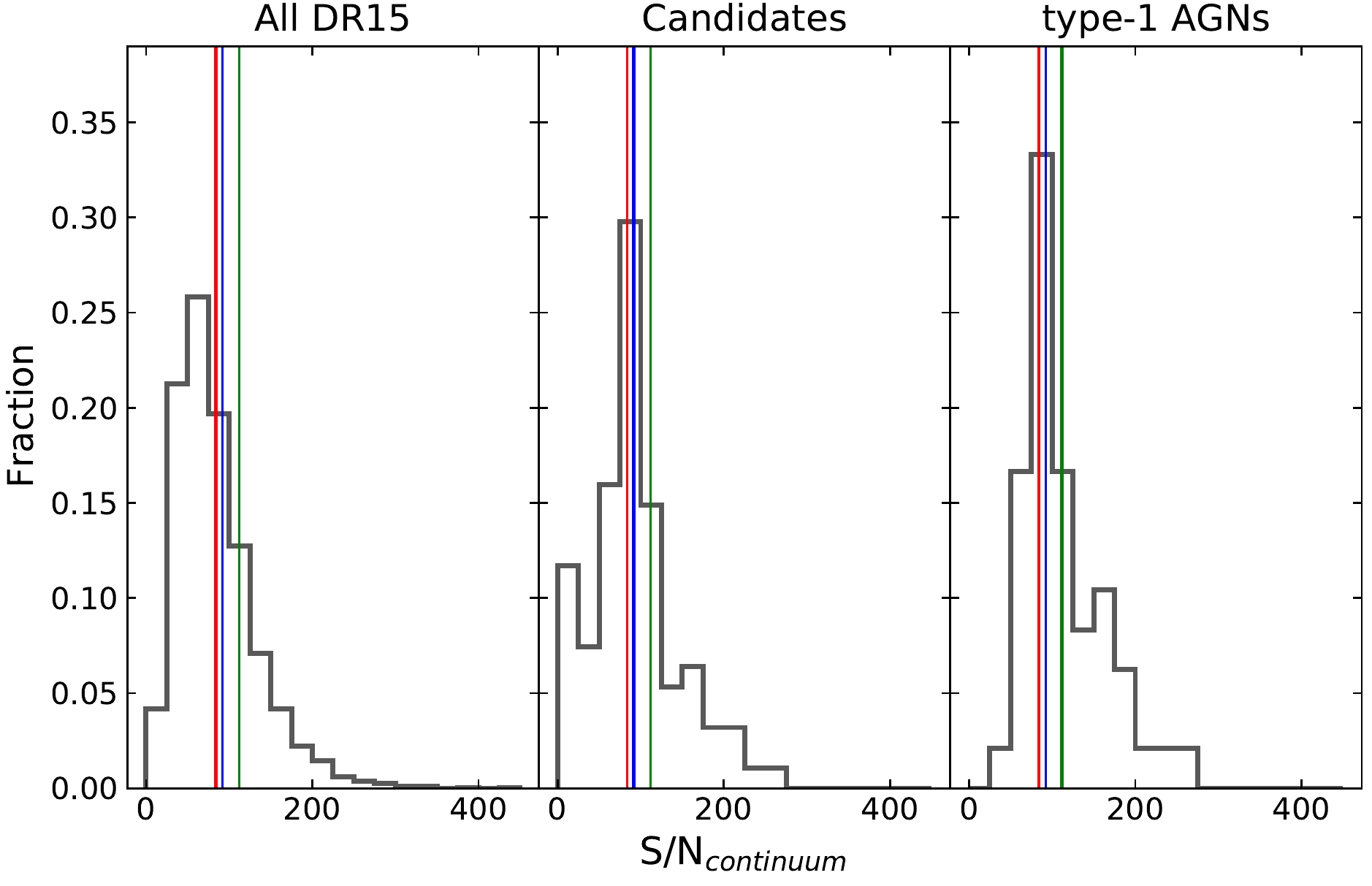}
  \caption{Histograms of the S/N of \fcont\ for the DR15 samples. Each panel shows the distribution of S/N for all DR15 sample (left), the candidates' sample (middle), and the type-1 AGN sample (right). Vertical lines indicate the average S/N for the respective sample: red for all DR15, blue for the candidates, and green for the type-1 AGN. The median S/N increases towards the type-1 AGN sample.}
  
\label{fig:SN_STL}
\end{figure}

An advantage of our method is the high S/N of the integrated spectra associated with our final sample. On average, the \fcont\ S/N computed for the MPL-7 spectra is 84, increasing to 92 for the 93 candidates and to around 112 for our 47 type-1 AGN (Table \ref{tab:N_snr}, and 
Fig. \ref{fig:SN_STL}). 

Looking at the distribution of the S/N in the complete sample, we have 4206 spectra with S/N < 150 and 430 with S/N > 150. An estimate of the fraction of type-1 AGN objects in the range S/N < 150 is 0.9\%, a lower value than 2.6\% corresponding to that in the range S/N > 150, so this is not an apparent decrease of AGN candidates as the S/N ratio increases but instead, a decrease in the number of galaxies as the S/N ratio is increasing. In the particular case of our type-1 AGN sample, 12 galaxies have S/N > 150, however, the averages flux ratios and redshift are not different from the galaxies with S/N < 150. The increase of the S/N is expected because the contribution of the nuclear luminosity is higher for the active galaxies with respect to quiescent ones. The high S/N achieved in the integrated central 3 arcsec MaNGA spectra is useful to avoid spurious contamination in the red and blue bands when searching for the broad line component. Equally important, it allows us to identify \ha\ broad line component by a direct inspection of the observed spectra, without the need of a host stellar component subtraction. In the next section, we carry out a first-order Host Galaxy subtraction to the spectra and compare the results obtained before and after host subtraction in order to test the robustness of these conclusions. 


\subsection{Host Galaxy subtraction proof}
\label{sec:HGproof}

Previous authors like \cite{Stern2012} and \cite{Oh2015} have shown that when considering spectra from the SDSS survey, the fiducial 3 arcsec aperture may contain a significant contribution from the host galaxy. Their work suggests that it is needed to remove that contribution to identifying 
the broad emission line component in low-luminosity AGN. 
However, with our present method, after synthesizing the MaNGA spectra in a similar 3 arcsec aperture, we could proceed with identifying 
the broad component, including the stellar contribution of the host galaxy. Thus, it is important to test our results and estimate the influence of the stellar contribution in our identification method. 

To take the stellar contribution into account, we apply a first-order stellar subtraction using \textit{Starlight} \citep[STL; ][]{CidFernandes2005}.  \textit{Starlight} is a program that uses stellar population synthesis: an observed spectrum is decomposed in terms of a superposition of stellar populations of various ages and metallicities to model the underlying stellar galaxy continuum. The spectral base used consists of 150 simple stellar populations, with six metallicities and 25 different ages \citep[using the][libraries]{Bruzual2003}. To consider the AGN contribution, we added 6 power laws of the form $f_\nu = \nu^{-\alpha}$ within the range of 0.5 < $\alpha$ < 3.0, to model the AGN continuum (if present). The result is a decomposition in a synthetic stellar model, a power-law contribution, and the emission line spectra after subtracting the original and the stellar model, including the resulting power law contribution. In cases with strong or dominant AGN emission, the STL model could not converge to an appropriate solution. In a forthcoming paper of this series, the problem of decoupling the power-law + host galaxy contributions in the AGN spectra will be addressed from the point of view of two independent methods (Cortes-Suarez in prep). 

The flux ratio method and the statistical analysis described in Section \ref{sec:fluxratio} were similarly applied to the resulting emission line spectra + power-law contribution after subtracting the stellar contribution 
to the whole DR15 sample. The middle columns in Table \ref{tab:boxplot} report the boxplot numbers computed as described in the previous section. The upper whisker values are $U_{W,STL}$, of 1.028 and 1.016 for \fr\ and \fb, respectively. 
The new $U_{W,STL}$ values are slightly lower compared to the ones originally estimated from the observed spectra. A plausible reason for this is that in the absence of the stellar component, the continuum flattens, so the flux ratios are closer to unity. We found 298 and 143 galaxies for the \fr/\fcont\ and \fb/\fcont\ ratios, above $U_{W,STL}$. 

Proceeding similarly and matching the results of our red and blue object selection, we identified 93 candidates. Even when the resulting number of candidates is the same as in Sec. \ref{sec:fluxratio}, 20 of them are different objects. After a visual inspection, ten false-positive were detected (a lower quantity than without the stellar subtraction), and 83 candidates showing narrow or broad emission lines were identified. Looking for type-1 AGN, the same objects were recovered as in the case of no stellar contribution subtraction except one, 1-71872. This object was found above the upper limit only in \fr/\fcont\ ratio. Its \fb/\fcont\ ratio lies just below the limit, 1.015. 


Comparing the results of the type-1 AGN identification before and after subtracting the host stellar contribution, we do not find an effect of the stellar subtraction in the selection method. It seems that for the MaNGA sample, with an average S/N of 84 (see Table \ref{tab:N_snr}), we obtain comparable results.
Under these conditions it may not be necessary to subtract the stellar component in the target spectra. We also find that the stellar subtraction is useful to reduce the number of false positives and increase the quantity of type-2 AGN (we found 37 in this sample), 
although one of our type-1 AGN is missed. Summarizing, given a set of spectra having significantly high S/N ratios, our direct identification method appears enough for the recognition of broad emission line 
AGN in the MaNGA survey, avoiding the introduction of possible degenerations when considering the stellar subtraction.

In Appendix \ref{appendix:A} we report our identification of type-2 AGN candidates and of objects showing M type spectra. Although our method is focused on the detection of broad emission lines, it allowed for the identification of 28 type-2 AGN (12\% of the candidates) when no subtraction is considered and of 37 (16\%) after subtraction of the stellar component. This sample of type-2 AGN shows some particular characteristics in their associated spectra like strong narrow emission lines with FWHM high enough to reach the blue or red bands, 
and narrow emission lines profiles that seem to be Lorentzian (or double Gaussian).


\subsection{Selection based on BPT and WHAN diagrams}
\label{sec:type2}

\begin{figure}
\centering 
\includegraphics[width=\columnwidth]{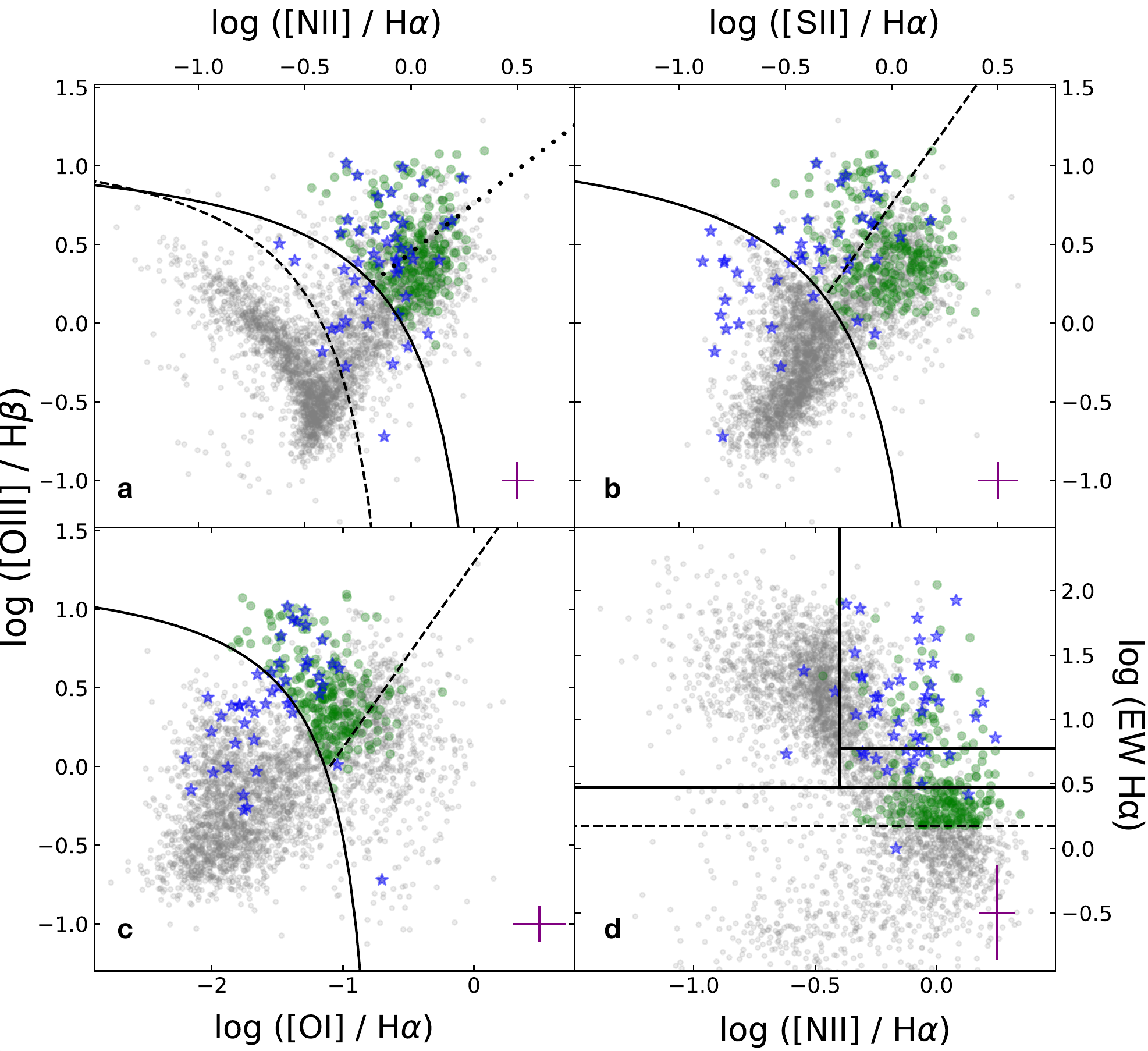}
  \caption{BPT and WHAN diagrams of the MPL-7 sample. Data were obtained from the Pipe3D VAC 
  Type-1 AGN detected using the line ratio method are shown in blue stars, while type-2 AGN are shown in green stars. The WHAN diagram is shown at the bottom-right with the EW(\ha) $>$ 1.5\AA\ threshold as a dashed line. Error bars are shown as purple crosses.}
\label{fig:MPL7_BPT}
\end{figure}

We have built the non-resolved BPT diagrams \citep{Baldwin1981} using the VAC generated through the Pipe3D pipeline \citep{Sanchez2016a} for the central 2.5 arcsec integrated spectra of the MaNGA DR15 sample\footnote{The 
Pipe3D VAC were obtained using integrated spectra of the spaxels contained in the central 2.5 arcsec, 1 effective radius and all the Field of View. }. Although these routines are mainly optimized for the identification of narrow lines, they also use an approximation for the identification of broad emission lines. We decided to use the Pipe3D VAC to compare our results with other works discussed in Section \ref{sec:comparison}. For the AGN identification we follow the criteria outlined in \cite{Sanchez2018}, by considering galaxies located above the \cite{Kewley2001} line in the three independent BPT diagrams, and having EW(\ha) > 1.5 \AA\ (panels (a) to (c) of Figure \ref{fig:MPL7_BPT}). The \cite{Kewley2006}, \cite{Kauffmann2003} and \cite{Schawinski2007} demarcation lines are also shown in these diagrams.  

We found 242 AGN candidates ($\sim$ 5\% of the total sample), marked with green starry symbols in Figure \ref{fig:MPL7_BPT}. 
Of them, three galaxies have more than one datacube in common J110431.08+423721.2: 8256-12704, 8274-12704, 8451-12701 (MaNGA- ID 1-558912); J155953.98+444232.4: 9036-2703, 9031-3704 (MaNGA ID 1-209772); and J160436.23+435247.3: 9031-12705, 9036-6101 (MaNGA-ID 1-210186). A cross-match of these results with our 
type-1 AGN sample (blue stars in Figure \ref{fig:MPL7_BPT}) let us find 16 matches. 
The low type-1 AGN detection means that applying standard stellar population synthesis and placing the results in BPT diagrams even with additional restrictions on the EW(\ha) values, 66\%\ of the already identified type-1 AGN were missed. 
A characteristic in common in these 31 missed objects is 
that the narrow component of the Balmer lines is stronger than the other narrow lines explaining their location in the composite and star forming region of the BPT diagrams. We also noticed that narrow emission lines are partially or entirely embedded into the \ha\ broad component in some galaxies, something that was not considered in the Pipe3D emission line fitting procedures, and consequently, the emission of the narrow emission lines are being underestimated or lost. In this way, the \nii\ diagram show 46 of 47 type-1 AGN, the \oi\ diagram 45 AGN, and the \sii\ diagram 44 AGN. So it is important to note that these diagrams do not detect the spectra dominated by the AGN emission. 

The remaining candidates were considered as type-2 AGN since they only show narrow emission lines. Combining the candidates emerging in the BPT diagrams with those from the flux ratio method (including the type-2 AGN), we identified \textbf{283 AGN: 47 type-1, and 236 type-2}. Imposing  EW(\ha) $>$ 3 \AA\ to avoid ionization processes due to post-AGB stars or shocks \citep{cid-fernandes2010}, the sample amounts to 125 AGN: 45 type-1, and 80 type-2. Other works \citep[e.g.,][]{CanoDiaz2016,Lacerna2020} consider values of EW(\ha) > 6 \AA\ to assure that the dominant ionization process comes from non-stellar nuclear activity. With this restriction, we find 77 AGN: 33 type-1 and 44 type-2. Since more restrictive EW(\ha) conditions translate into a loss of broad emission line AGN candidates, we adopt EW(\ha) > 1.5 \AA. 

Panel (d) of Figure \ref{fig:MPL7_BPT} shows the AGN sample in the WHAN diagram \citep[log(\nii/\ha) vs. EW(\ha), ][]{CidFernandes2011}. AGN are identified if log(\nii/\ha) > -0.4 and EW(\ha) is between 3 and 6 \AA\ for weak AGN, and 
larger than 6 \AA\ for strong AGN (solid lines). 
The EW(\ha) > 1.5 \AA\ threshold is the dashed line. Although all the candidates in the BPT diagrams are above the Kewley line (green stars), in the WHAN diagram, most of them have EW(\ha) less than 3 \AA, which suggests that ionization processes could not be associated only with the AGN \citep{CidFernandes2011}. 
It is important to note that this diagram is able to detect most of our already identified broad-line AGN: 11 in the weak AGN region and 31 in the strong AGN region, missing only 5 (11\%) of them. Two of them have EW(\ha) < 3 because the narrow component is entirely embedded onto the broad one. The other three are located in the SF region due to the intensity of the \ha\ narrow component.


BPT diagnostic diagrams based on narrow emission lines, although useful, can identify only 34\% of our type-1 AGN sample. Interestingly, the WHAN diagram recovers a higher (89\%) fraction, despite not being optimized for detecting the broad components, suggesting that this method could be helpful in their identification.



\section{Comparison with other Catalogs}
\label{sec:comparison}

Although BPT and WHAN diagrams are 
helpful to detect 
AGN in large optical surveys, they identify type-2 AGN principally. Another way to identify them is by cross matching with AGN catalogs at wavelengths such as X-rays, infrared (IR), or radio, where they can also be detected \citep[e.g.][]{Coffey2019,Comerford2020}. 
In this section, we compare the results of our selection method for broad-line AGN with previous works that also identify not only type-1 but also type-2 AGN. We mainly consider 
works based on spectral data from 
the SDSS DR7 and the MaNGA survey.

\subsection{SDSS DR7 AGN Catalogs}
\label{sec:SDSS-DR7comparison}
\defcitealias{Stern2012}{SL12}
\defcitealias{Oh2015}{Oh15}
\defcitealias{Liu2019}{Liu19}

\cite{Stern2012}, \cite{Oh2015} and \cite{Liu2019} (hereafter \citetalias{Stern2012}, \citetalias{Oh2015} and \citetalias{Liu2019}, respectively) carried out a systematic search of type-1 AGN, based on the detection of broad \ha\ emission using data coming from the SDSS DR7 database \citep{Abazajianetal2009}. The results of these works have provided substantial numbers of newly discovered type-1 AGN in the local universe setting the basis for a more robust statistical analysis to test and constrain the physical properties of the BLR 
and the AGN kinematics. 
 
\citetalias{Stern2012} searched for the \ha\ broad emission line in the SDSS DR7 spectroscopic data, restricting the redshift range to 0.005 $>$ z $>$ 0.31 (232 837 objects). They subtracted the host galaxy contribution using a galaxy eigenspectra derived from a PCA of SDSS galaxies \citep{Yipetal2004}, with a power law component (L$_\lambda \propto \lambda^{-1.5}$) representing the AGN continuum, to derive a featureless continuum. Then, they interpolated a local continuum around \ha\ to fit and subtract 
the narrow emission lines. The residual flux ($\Delta$F) around \ha\ was summed to look for a broad component. 
They considered a broad line candidate if the ratio of the residual flux over the dispersion $\Delta$F / $\sigma$ $>$ 2.5 (3\%\ of the initial sample). 
For those objects, they fitted high order Gauss–Hermite functions for the broad \ha\ profile. They considered broad-line AGN candidates those objects with a line width $\Delta$v $>$ 1000 \kms, and after a visual inspection, their final sample of candidates consists of 1.5\%\ of the initial sample (3579 objects).

\citetalias{Liu2019} looked for AGN in 866,302 objects cataloged as “galaxies” or “quasars” in the SDSS DR7 at z < 0.35. Using the EL-ICA algorithm \citep{Lu2006}, which considers the library of simple stellar population of \citet{Bruzual2003}, they decompose the spectra into stellar and nonstellar nuclear components. Then, they fitted the narrow and broad lines in the \ha\ and \hb\ regions, including the FeII multiplets and the AGN continuum. Their broad-line selection criteria is based on: (i) \habc\ flux above 10$^{-16}$ erg s$^{-1}$ cm$^{-2}$, (ii) S/N(\habc) $\geq$ 5, (iii) the height of the best fit of the \habc\ above 2 RMS of the spectrum with no stellar contribution, measured in a region free of emission lines near \ha, 
and (iv) FWHM of the broad component higher than the FWHM of the narrow line. They find 14,584 sources that meet those criteria, amounting to the biggest type-1 AGN catalog of SDSS DR7.


\citetalias{Oh2015} looked for type-1 AGN in galaxies with z $<$ 0.2 also using spectroscopic data from the SDSS DR7 database (664 187 galaxies). \habc\ was identified by computing a ratio between the mean fluxes at 6460–6480 \AA, defined as a pseudo-continuum interval, and 6523–6543 \AA\ to highlight the \habc. 
Considering a threshold of 1 $\sigma$, they selected 17\% of the objects as type-1 AGN candidates, for which they simultaneously fit the stellar continuum and the emission lines. They selected objects with a FWHM(\habc) $>$ 800 \kms, and an amplitude-over-noise ratio (A/N) of \habc\ larger than 3 (8.6\%\ of the 1$\sigma$ demarcation). 
Then they measured the total flux of the \ha\ broad component at the red side of \niib\, and the averaged dispersion of the continuum in four specific windows between 4500\AA\ and 7000\AA, considering as candidates those with total fluxes two times greater than the continuum dispersion. Their final sample consists of 5553 sources (0.8\% of the initial sample).

In spite that our method is based on flux ratios, it departs from \citetalias{Oh2015} method in three important aspects:

\begin{itemize}
    \item New appropriate spectral windows for the continuum and blue bands.
    \item The introduction of the red band.
    \item The avoiding of the stellar subtraction and line fitting. 
\end{itemize}

\textit{Band positions.-} The methodology of \citetalias{Oh2015} did not successfully detect spectra with multiple or very broad \ha\ components, which overlap emission lines from \oib\ to \siia, as shown in their Figure 8. An example of a multiple component broad emission line AGN, that extends up to 6450 \AA\ is J075244.19+455657.4 (Figure \ref{fig:8714}). 
Another example can be seen in Figure A.3 of \cite{Lacerna2016}. 
For this reason, our continuum band is shifted 60 \AA\ to the blue with respect to \citetalias{Oh2015}, maintaining the same range width.

\textit{Red Band.-} 
Besides the identification based on the estimate of a flux ratio and conditions on A/N and FWHM, \citetalias{Oh2015} carried out a spectral fitting and decomposition to estimate the properties of the broad \ha\ line. Based on that, they could 
compare the area of the broad \ha\ component beyond [NII] $\lambda$6584 as an alternative measure to FWHM for the broad \ha\ line, thus generating an additional condition for the identification of type-1 AGN. In contrast, we took advantage of the higher S/N in our integrated spectra. The introduction of the red band in our identification criterion is a good alternative to track the extent of the \habc. 
Thus, there is no need to invoke a fitting of the broad and/or narrow components in our direct method. 


\textit{Stellar subtraction.-} Modeling the stellar component and subsequent subtraction in the observed spectra of galaxies is fraught with difficulties. In the case of galaxies hosting AGN, there is no unique solution representing the host galaxy and the power-law contributions. Furthermore, in powerful or dominant AGN cases, the stellar fitting and subtraction may be non-sense due to the lack of stellar absorption lines. For these reasons, and because the flux ratio selection process using the boxplot-whiskers statistical method allows us to recover all type-1 AGN, we decided to implement a direct method on the observed spectra without stellar subtraction. 

\begin{figure}
    \includegraphics[width=\columnwidth]{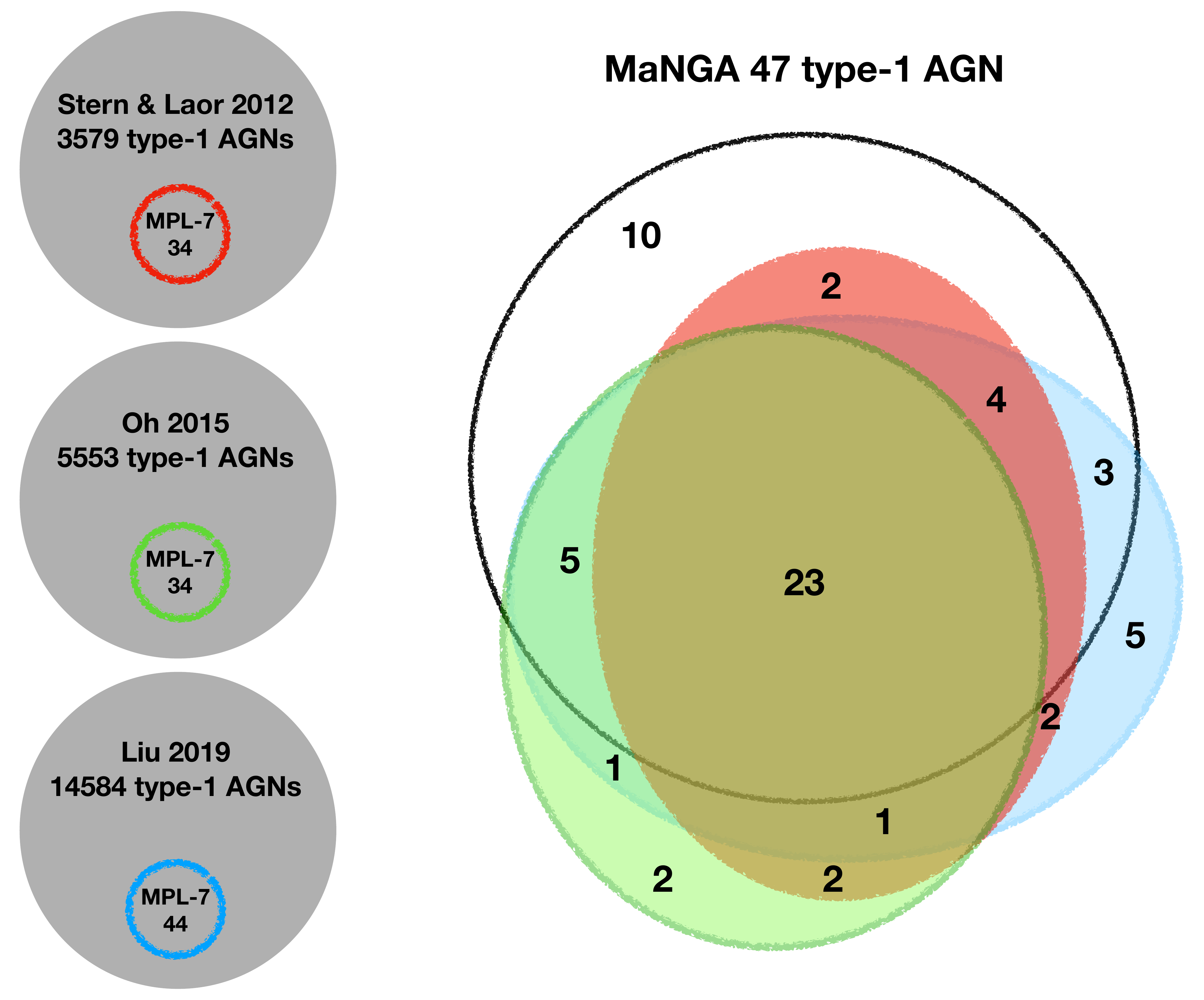}
    \caption{Venn Diagram of the MPL-7 cross-matched samples with our type-1 AGN sample (Black circle). The Red circle is the \citetalias{Stern2012}  sample, Green Circle the \citetalias{Oh2015} sample, and Blue Circle the  \citetalias{Liu2019} sample. The objects outside our type-1 AGN sample are in our type-2 AGN catalog (see section \ref{sec:type2}). Three objects in the \citetalias{Liu2019} sample do not have classification 
}
    \label{fig:venn}
\end{figure}

\subsubsection{SDSS DR7 Type-1 AGN detections}
After a cross-match with the full MaNGA DR15 sample, we find 44 galaxies in common with \citetalias{Liu2019} (35 type-1 AGN), 34 in common with \citetalias{Oh2015} (28 type-1 AGN), and also 34 in common with \citetalias{Stern2012} (29 type-1 AGN), with 23 type-1 AGN in common among the four samples ($\sim$49\%). 
The distribution and coincidences are illustrated in the Venn Diagram of Figure \ref{fig:venn}. A large number of coincidences were found despite the differences between each method.                                    

Objects that are not in our sample were inspected visually, identifying some changing-look AGN. In Appendix \ref{appendix:B} we report a list of changing-look candidates in the MaNGA galaxies. We started a follow up and the results will be reported in a forthcoming paper. 

\begin{figure}
    \includegraphics[width=\columnwidth]{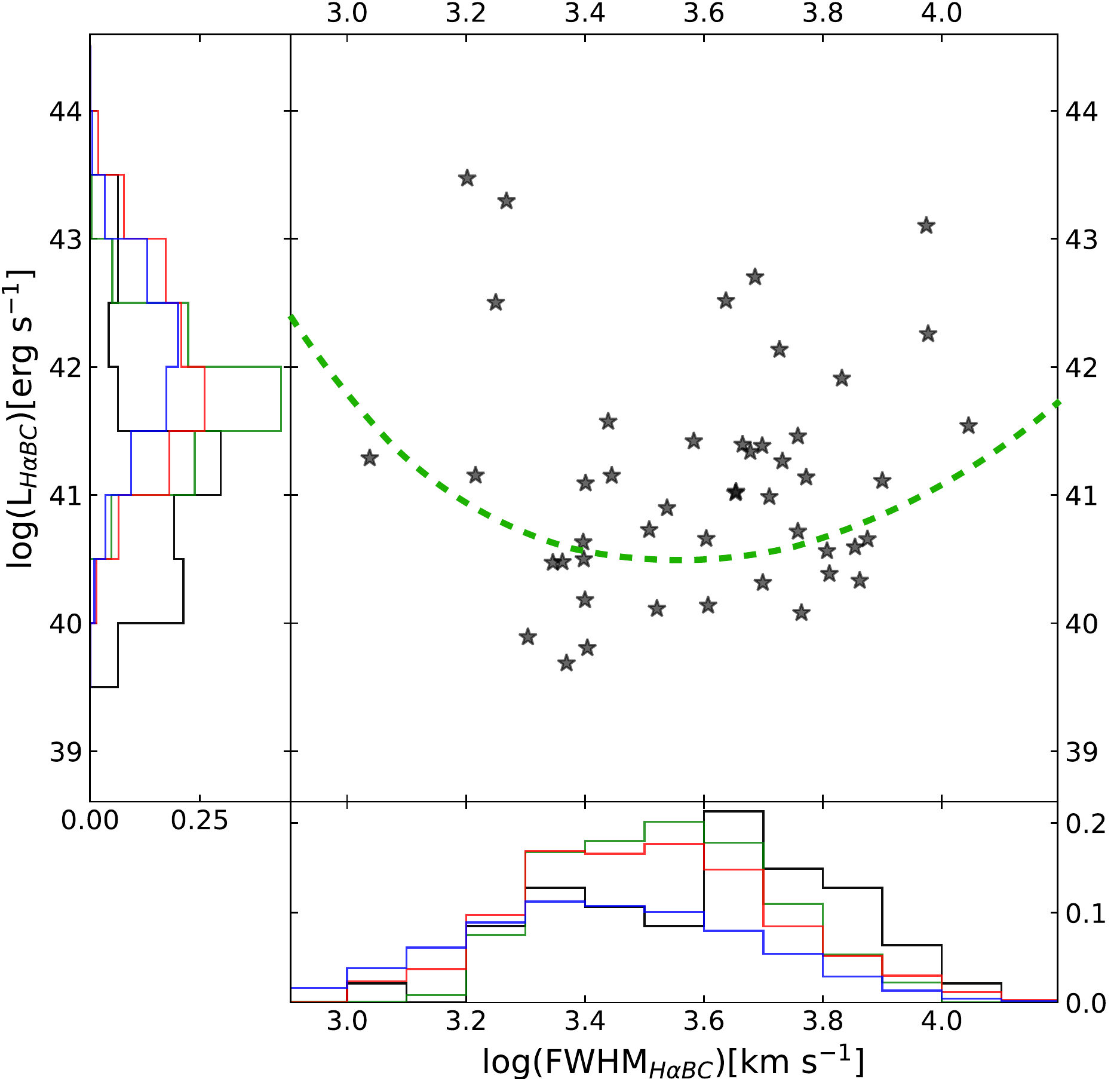}
    \caption{ The FWHM and luminosity of the \ha\ broad component for our type-1 AGNs. The axes are shown as histograms to see the distribution of both values (black lines). We also include the distributions of \citetalias{Stern2012} (red lines), \citetalias{Oh2015} (green lines), and \citetalias{Liu2019} (blue lines). The green dashed line is the demarcation line of \citet{Oh2015} that uses to select their type-1 AGN sample.}
    \label{fig:fwhmvslumha}
\end{figure}

Figure \ref{fig:fwhmvslumha} shows the FWHM(\habc) - L(\habc) luminosity relation and their corresponding distributions for the \habc\ 
for our type-1 AGN sample (black lines) and for the \citetalias{Stern2012} (red lines), \citetalias{Oh2015} (green lines), and \citetalias{Liu2019} (blue lines) samples. We compute both quantities for our 47 object sample after carefully subtracting the host galaxy contribution and fitting the emission lines plus the power law contribution (Cortes-Su\'arez et al. in prep.). The green dashed line shows the 90\% completeness limit for the identification of broad emission line AGN proposed by \citetalias{Oh2015} (see their Figure 4). As already emphasized by \citetalias{Oh2015} below this threshold, it is possible to find low-luminosity type-1 AGN, and in fact, we find 17 objects below that limit. The lowest luminosity value found is log L(\habc) $\sim$ 39.69 in an interval of log FWHM $\sim$ 3.0 - 4.0, with a median of 3.65. 
As a comparison, the type-1 AGN sample of \citetalias{Stern2012} has a luminosity interval of log L(\habc) $\sim$ 40-44 that peaks at 41.75. In contrast, our sample extends to lower values with a median value of L(\habc) $\sim$ 41.03, being in fact dominated by low luminosity AGN (77\% have log L(\habc) < 41.5). 
In the case of the \citetalias{Liu2019} sample, their log L(\habc) has values in the interval 38.5-44.3 with a median value near 42. 

The lower luminosity values reported by \citetalias{Liu2019} are associated with low-luminosity AGN \citep[located below the selection curve of][]{Oh2015} and correspond to FWHM \habc\ $\sim$ 500 km s$^{-1}$. 
In contrast, due to the position bands of our method, the FWHM that we can detect is limited by FWHM(\habc) $\gtrsim$ 1600 km s$^{-1}$ in average (Section \ref{sec:fluxratio}), i.e., we identify candidates only with a broad visible component, which is the definition of type-1 AGN in the optical range. 
When cross matching our objects with the \citetalias{Liu2019} sample, we lose 3 of 44 galaxies. However, in a visual inspection we do not detect a visible broad component. The remaining 35 objects are in our type-1 catalog or were classified as type-2 AGN. This could suggest that our flux ratio method is able to recover these low-luminosity AGN, as long as they have a visible broad component.  

\begin{figure}
  \includegraphics[width=\columnwidth]{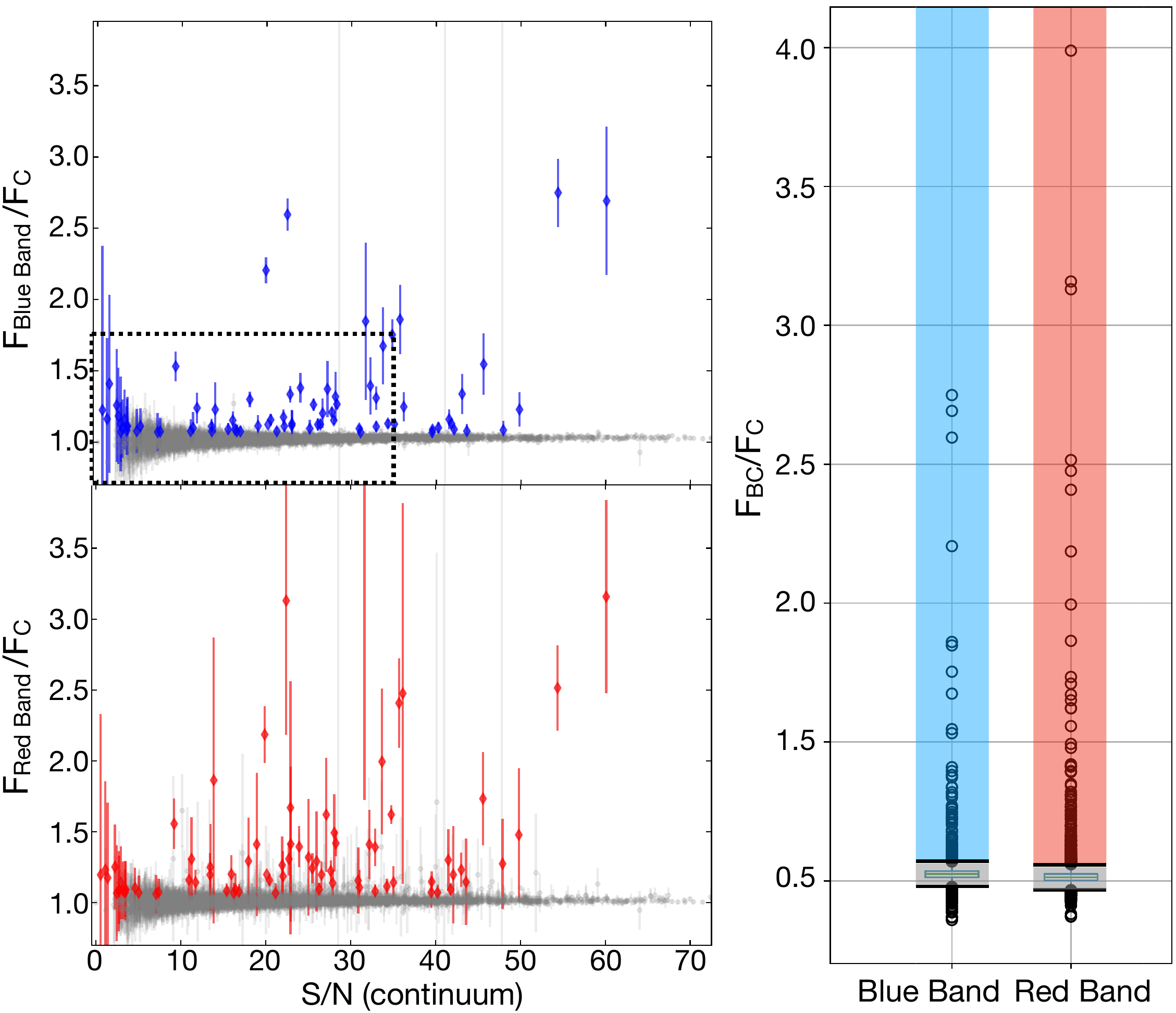}
  \caption{Left: Flux ratio distribution as shown in figure \ref{fig:bands} but using the DR7-SDSS spectra. Black box in top figure illustrate the area of the Figure 1 of \citet{Oh2015}.  Right: Boxplots used for the AGN selection. Outliers found were 108 and 245 for the blue and red bands, respectively. From them, 76 were classified as AGN candidates.}
  \label{boxplot_dr7}
\end{figure}

\subsubsection{Flux ratio method applied in the SDSS DR7}
We further applied our direct flux ratio method to the 
SDSS DR7 spectra that matched the actual DR15 MaNGA sample, finding 4300 MaNGA objects in common. Figure \ref{boxplot_dr7} shows the resulting distribution of the flux ratios \fr/\fcont\ and \fb/\fcont\ versus the S/N in the continuum (left panels), and the boxplot (right panel) for these objects. 
The more scattered distribution with respect to Figure \ref{fig:bands}, is 
possibly due to a lower S/N of the DR7 data compared to our sample data. Despite of using a similar aperture (3 arcsec), we got lower S/N values, with a fraction of 2.8\% objects having a S/N lower than 10 (Figure \ref{fig:SN_STL}), in comparison with a fraction of only 0.2\% objects for the MaNGA sample.

For that reason, the upper whisker limits $U_{W,DR7}$ increased to 1.059 and 1.069 for red and blue bands, respectively, compared to the MaNGA values (last two columns of Table \ref{tab:boxplot}). Repeating our procedures and considering again only the superior outliers in both bands (108 blue and 245 red, Figure \ref{boxplot_dr7}), 76 candidates were identified, including all the \citetalias{Stern2012} and \citetalias{Oh2015} AGN candidates and 88\%\ of the \citetalias{Liu2019} sample. From those candidates, 39 galaxies match our sample of 47 MaNGA type-1 AGN. Among the eight missing objects, two have no DR7 spectra. Two more do not 
show a broad emission line in the DR7 spectra but, contrary to those reported in Appendix \ref{appendix:B}, they do in the MaNGA spectra. The other four objects were lost because the upper whisker limits increased and 
the broad \ha\ component in these objects is weak. 

Finally, we further found five objects having a broad emission line in the DR7 spectra but that were not reported in \citetalias{Stern2012}, \citetalias{Oh2015}, and \citetalias{Liu2019} catalogs. 
In the case of \citetalias{Liu2019} work, the missing candidates correspond to one of our eight type-1 AGN not found with our method, and one type-2 AGN. Other three objects are not in our catalog because they show a type-2 AGN profile. 15 objects were identified as false positives ($\sim$19\% of the candidates).

Table \ref{tab:N_snr} summarizes the number of objects obtained for the MaNGA observed spectra, the spectra after subtracting the Host Galaxy with Starlight, and for the SDSS DR7 spectra.
Column (2) is the size of each sample. Columns (3) and (4) are the number of outliers above each $U_W$, blue and red. Column (5) shows the number of candidates above both $U_W$. Column (6) gives the number of false positives, and Column (7) reports the fraction of type-1 AGN recovered from our final sample. The last three Columns show the average S/N for all the galaxies (Col. 8), the candidates (Col. 9), and the type-1 AGN (Col. 10). 

The difference of the S/N between the MaNGA extracted spectra, and the SDSS DR7 sample (Col. 9 of Table \ref{tab:N_snr}) is that the latter uses a single aperture while we integrated over several spaxels. As we mentioned above, the sum of spaxels helps to increase the S/N. The decrease of the S/N increases the value of the $U_W$. This anticorrelation is due to the fact that, when the noise increases, the flux ratio with values close to one increases as well, since the noise dilutes the weak broad components. However, we have shown that, although the values of $U_{W,B\& R}$ depend on the S/N of the sample, our method can find more type 1 AGN in nearby galaxies than other methods, following fewer steps. 

In the next section, we compare our type-1 sample with the ones obtained with different methodologies using the MaNGA data.

\begin{table}
  \begin{center}
    \begin{tabular}{l|cc|cc|cc}
      \hline \hline
 & \multicolumn{2}{|c|}{Observed Spectra} & \multicolumn{2}{|c|}{Starlight Spectra}& \multicolumn{2}{|c|}{DR7 Spectra}\\
  \hline
Value & BB & RB  & BB & RB & BB & RB  \\
\hline
$Q_1$ & 1.014 & 1.000 & 0.999 & 1.004  & 1.012 & 1.001\\
$Q_2$ & 1.023 & 1.007 & 1.002 & 1.008  & 1.025 & 1.012\\
$Q_3$ & 1.031  & 1.015 & 1.005 & 1.013 & 1.035 & 1.024 \\
$U_W$ & 1.056  & 1.037  &  1.028 & 1.016 & 1.069 & 1.059 \\
Mean & 1.026 & 1.012 & 1.004 & 1.013 & 1.026 & 1.021 \\
Max & 3.459 & 3.220 & 2.267 & 2.344 & 2.749 & 3.989 \\
\hline \hline
    \end{tabular}
    \caption{Boxplot values obtained in the Blue Band (BB) and Red Band (RB) for each sample studied in this work, the \textit{Observed Spectra} from MaNGA, \textit{Starlight Spectra} after subtracting the Host Galaxy contribution to the MaNGA sample, and the \textit{DR7 Spectra} from the SDSS survey.}
    \label{tab:boxplot}
  \end{center}
\end{table}

\begin{table*}
\centering
\begin{tabular}{l|cccccc|ccc}
      \hline \hline
 & \multicolumn{6}{c|}{Number of objects} & \multicolumn{3}{c}{Average S/N}\\
 \hline
Sample & N & $U_{W,B}$ & $U_{W,R}$  & $U_{W,B\&R}$ & False & Type-1  & S/N(full) & S/N(candidates) & S/N(type-1 AGN) \\
 & & & & (candidates) & positive & AGN &   &  & \\
 (1) & (2) & (3) &(4) &(5) &(6) &(7) &(8) &(9) &(10) \\
\hline
MaNGA Observed & 4636 & 126 & 171 & 93 & 18 & 100\% & 84 & 92 & 112\\
MaNGA Starlight & 4628 & 143 & 298 & 93 & 10 & 98\% & - & - & - \\
SDSS DR7 & 4300 & 108 & 245 & 76 & 15 & 83\% & 51 & 46 & 61\\
\hline \hline
    \end{tabular}
    \caption{Summary of the boxplot results applied in the observed spectra (MaNGA Observed sample), after subtracting the HG (MaNGA Starlight sample) and SDSS DR7 spectra. 
    The second Column is the size of each sample. The third and four Columns are the outliers above each $U_W$ while the fifth Column shows the candidates that are above both $U_W$. The sixth Column is the number of false positives and the seventh the percentage of type-1 AGN found. The last three Columns show the \fcont\ S/N average for the full sample, the candidates sample, and for the type-1 AGN sample.}
    \label{tab:N_snr}
\end{table*}

\subsection{AGN Catalogs in MaNGA}
\label{sec:manga_catalogs}
\defcitealias{Sanchez2018}{S18}
\defcitealias{Rembold2017}{R17}
\defcitealias{Wylezalek2020}{W20}
\defcitealias{Comerford2020}{C20}

We next make a non-extensive review of some works that have tried the identification of type-1 and 2 AGN within the MaNGA survey, paying particular attention to their methods and results to look for differences and coincidences with our method and final sample. 
 
\cite{Rembold2017} used the MPL-5 data sample and cross-matched it with the SDSS DR12 \citep{Alam2015} to obtain line fluxes and equivalent widths of \hb, \ha, \oiiib, and \niib\ of the integrated nuclear spectrum \citep{Thomas2016}. 
Then, they built the \nii\ BPT and the WHAN diagrams considering as AGN candidates those galaxies located simultaneously in the Seyfert or LINER region of both diagrams. They found 62 objects fulfilling this criterion but did not make a 
distinction between type-1 or 2 AGN. 

Of the 47 type-1 and 236 type-2 AGN found in the MPL-7 (Sec. \ref{sec:method}), 23 type-1 and 132 type-2 objects were found in the MPL-5 sample. Of them, 5 type-1 and 32 type-2 AGN were found in the \cite{Rembold2017} AGN list. 
Applying the criteria of the BPT and WHAN diagrams of Figure \ref{fig:MPL7_BPT}, we can identify 14 type-1 AGN. 
It is worth mentioning that the same aspects that affect the AGN identification described above needs to be considered. These are the difference of observational epoch between MaNGA and previous SDSS observations, the re-reduction of the DR12 data (Section \ref{sec:intro}), and that the BPT diagrams do not consider the BEL contribution for the strongest AGN emitters. Although the results may be different depending on the pipeline used, in this case, Pipe3D helped us to recover a higher AGN fraction, despite its gross approximation for the identification of type-1 AGN.

\cite{Sanchez2018} processed the data in the MPL-5 sample adopting the GSD156 library of stellar populations \citep{CidFernandes2013} using Pipe3D. Then they extracted the spectra of the central region (3 arcsec of diameter) of the MaNGA galaxies to look for AGN using the three BPT diagrams. They imposed two criteria, an emission line ratio above the Kewley demarcation line in the three diagnostic diagrams, and a conservative criterion for the EW(\ha) > 1.5 \AA\ to include the weakest AGN. Their final AGN sample consists of 97 type-1 and type-2 objects. They 
tried to isolate the type-1 objects by fitting narrow Gaussians for \ha\ and 
\niill\ (with a FWHM < 250 \kms), plus an additional broad component for \ha\ (1000 < FWHM < 10 000 \kms). 
However, they did not consider the power law contribution of the AGN emission. They obtained 36 type-1 AGN candidates, of which 10 objects coincides with our sample (21\%). 
The remaining 26 objects were identified as false positives as they do not show any broad component in the visual inspection. This result is somewhat expected, as the method applied is not optimized to detect broad components. As described in Section \ref{sec:type2}, Pipe3D does not consider the contribution of the AGN power law nor that the narrow components are immersed in the broad components of the Balmer lines.

Another AGN sample gathered from the MaNGA survey is that of \cite{Wylezalek2018,Wylezalek2020}, who also worked on the MPL-5. 
Unlike the present work and others like \citet{Rembold2017} and \citet{Sanchez2018}, they looked for the AGN signatures 
considering the entire IFS spectral data cube. 
They used data from the MaNGA Data Analysis Pipeline \citep[DAP;][]{Yan2016a,Westfall2019} 
to generate spatially resolved (spaxel by spaxel) \nii\ and \sii\ BPT diagrams for each galaxy. Their AGN candidates were selected for having 
a fraction larger than 10\% or 15\% of the spaxels in the \nii\ and \sii\ BPT diagrams. 
Then they imposed various conditions; 1) spaxels having  S/N > 5, to avoid those on which the Pipeline failed to reconstruct the spectra; 2) EW(\ha) > 5\AA; 3) a distance connecting the spaxel measurement and the star formation demarcation line in the \sii\ BPT diagram lower than 0.3, and 4) the \ha\ surface brightness SB(\ha) > 10$^{37.5}$ erg s$^{-1}$ kpc$^{-2}$, in the spaxels selected with the above criteria, as a tracer for diffuse ionized gas. They identified 308 AGN candidates. 

\citet{Wylezalek2018} also compared the MaNGA observations with the SDSS single fiber spectra from different epochs, detecting AGN candidates that single nuclear spectra methods could not, 
especially galaxies that may have recently turned their nuclear activity off, or AGN emission-like regions overshadowed by contamination from diffuse ionized gas, extra-planar gas, and photoionized by hot stars. 
To identify the type-1 AGN they cross-matched the MaNGA MPL-5 sample with the catalog from \citet{Oh2015} finding 67 coincidences. Of these 67 objects, 12 are in our sample, which were catalogued by \cite{Wylezalek2020} as Seyfert (8 galaxies), star forming galaxies (3 objects), and LINER (1 object). 

More recently, \cite{Comerford2020} have compiled a catalog of type-1 and 2 AGN using the MaNGA MPL-8 survey (6261 galaxies). To identify them, they used various criteria based in different wavelengths. 
1) Mid-infrared WISE colors. 
They used the \cite{Assef2015} criteria considering  W1 (3.4 $\mu$m) and W2 (4.6 $\mu$m) colors, finding 67 AGN. 
2) Swift observatory's Burst Alert Telescope (Swift/BAT) of ultra hard X-ray detections (14 - 195 keV). They used the \citet{Oh2018} AGN catalog, which is based on the cross-match of the 105-month BAT catalog and the SDSS DR12 \citep{Alam2015}, finding 17 matches. 
3) NVSS \citep{Condon1998} and FIRST \citep{Becker1995} radio observations at 1.4 GHz. 
They cross match the MaNGA objects with the \cite{Best2012} AGN radio catalog, based on the NVSS and FIRST detections of the SDSS DR7 galaxies. They found 325 radio AGN in MaNGA, catalogued as high-excitation radio galaxies (HERGs 3 objects), low-excitation radio galaxies (LERGs 143 objects), and no classification (260 objects).
4) Broad emission lines in SDSS spectra. To look for type-1 AGN, they cross-matched the MPL-8 sample with the \citet{Oh2015} catalog, founding 55 broad-line objects. In total, they present a sample of 406 AGN; of them, 309 (76\%) were detected by their radio emission. 

A cross-match with our type-1 AGN sample, yields 34 coincidences: 16 in WISE, 7 in Swift/BAT, 7 with radio detection (2 HERG, 1 LERG and 4 unclassified), and 27 broad-line AGN. For our type-2 AGN candidates, 36 coincidences were found: 14 in WISE, 3 in Swift/BAT, 23 with radio detection (0 HERG, 12 LERG), and 4 broad-line AGN. After a visual inspection of these 4 AGN labeled as broad-line by \citet{Oh2015}, one was catalogued as variable (1-460288), two show Lorentzian emission line profiles in the narrow components (1-604907, 1-385099), and another one has low S/N showing no clear evidence of a broad component (1-72322).


The AGN selection by \citet{Rembold2017,Sanchez2018} and \citet{,Wylezalek2018,Wylezalek2020} is mainly based on the characterization of AGN using the BPT diagnostic diagrams, which, although very useful for detecting type-2 AGN, fail to detect type-1 AGN. On the other hand, the work of \cite{Wylezalek2018,Wylezalek2020} and \citet{Comerford2020} uses databases of previously detected type-1 AGN, i.e., they do not develop their own detection method to identify the broad line objects.

\section{AGN Multiwavelength Properties}
\label{sec:multiwavelength}
Galaxies with nuclear activity present different properties across a broad range of wavelengths, reflecting different contributions from the nuclear region and the host galaxy \citep[e.g.][and references there in]{Padovani2017}. Most ultraviolet (UV) to near-IR (NIR) emission emerging from an AGN is produced by the inner regions of the accretion disk \citep{Secrest2015,Assef2018}. For the most obscured objects, the emission from the active nucleus at wavelengths shorter than the optical range (including X-rays for the Compton Thick AGN) is hidden by the nuclear and circumnuclear dust. It is, therefore, very convenient to use the NIR to mid-IR (MIR) band because it allows identifying both unobscured (type-1) and obscured (type-2) AGN \citep{2012SternD,Mateos2012,Jarrett2017,Assef2018}. The emission is dominated by cold dust associated with star formation in the host galaxy at longer wavelengths. 

In the unified model \citep[UM; e.g.,][]{Antonucci1993}, the orientation of an optically-thick torus of dust and gas surrounding the central engine plays a central role in determining the observable features of an AGN. According to the UM, in Type 1 AGN, the accretion disk is oriented face-on, leaving an unobstructed view of the broad line region. In that orientation the emission of the ionizing photons coming from the accretion disk, can be used as indicator of intrinsic AGN luminosity. 

In contrast, the UM predicts that type-2 AGN are oriented edge-on, obstructing a direct view of the broad line region. These obscured AGN can be identified instead, by emission lines originated in the narrow line region (NLR), a region far from the central emission but still affected by the ionizing central continuum \citep{Hickox2018}. Therefore the emission lines coming from the NLR can also be used as indicators of intrinsic AGN luminosity. The flux of the \oiiib\ line is commonly used as such a diagnostic \citep[e.g., ][]{Bassani1999,Heckman2005} as it is one of the most prominent lines and suffers little contamination from star formation processes in the host galaxy. This line although attenuated by dust in the host galaxy, can be corrected by applying a reddening correction using the observed Balmer decrement (i.e. the observed ratio of the narrow \ha/\hb\ emission-lines compared to the intrinsic ratio) and a mean extinction curve for galactic dust. 

The X-ray emission is also another viable alternative to detect AGN since X-rays are produced in a corona of hot electrons, within a few gravitational radii from the central accreting disk \citep[e.g., ][]{Haardt1991,Kara2015}. This emission is less affected by obscuration and also by contamination from star formation processes \citep{Stern2012}, therefore it can be used as a reference measure of the nuclear emission power in type-1 AGN. Furthermore, since the soft X-ray emission is partially absorbed by the hot dust surrounding the central black hole, some correlations between MIR and  X-ray luminosities are expected \citep{Lutz2004,Gandhi2009,Suh2017,Suh2019}. 

A mixture of the nuclear activity emission with the host galaxy contribution is often seen in nearby AGN in the optical spectra. In this Section, we classified the observed spectra in our type-1 AGN MaNGA sample based on the relation of two spectral emission/absorption lines. Using the information in the IR and radio wavelengths, we locate our objects in the AGN classification diagrams in those bands. We further look for correlations of luminosity indicators at X-rays, optical, IR and radio wavelengths. That information, taking into account the presence of upper limits in the X-rays and Radio data, is useful to understand the impact of AGN activity on other properties of the host galaxies.

\subsection{An Empirical Type-1 AGN Optical Spectral Classification}
\label{sec:sample}
Type-1 AGN spectra show large spectral differences in the continuum shape and emission line strengths of the narrow and/or broad components \citep[e.g.,][]{Boroson1992,Sulentic2000,Shen2014,Hickox2018,Padovani2017}. 
The differences in the spectral profile is a consequence of the different SED and physical condition in the line emitting gas, ultimately related to the accretion disk orientation. To classify this diversity in terms of the Host Galaxy and Power Law (HG-PL) contributions, we propose to use spectral indexes similar to Lick indexes, designed for stellar population studies that go from CN$\lambda$4161 to TiO$_2\lambda$6233 \citep[e.g.][]{Wortheyetal1994}. We use particular EW measurements for spectral regions containing stellar absorption lines or AGN emission lines and their corresponding local nearby continua.  

We define an H-Index in the spectral interval 3950-3990 \AA\, where the CaII H absorption line is strong in spectra dominated by stellar absorption lines. When the AGN emission is dominant, the CaII H line is ``filled'' by the non-thermal continuum, and the emission lines of \neiii\ and \he\ can be seen. The size of this interval is broad enough to include the expected location of the CaII H absorption line for host galaxy dominated spectra \citep[that can reach velocities greater than 500 \kms;][]{Cherinka2011}, and the \neiii\ and \he\ emission lines in the power-law dominated spectra. A second spectral interval intends to measure \hb\ whether it is in emission or absorption. In the case of strong AGN contribution, part of the broad emission component is included. We choose the Lick \hb-Index defined in the spectral 4848-4878 \AA\ interval, where the \hb\ emission line could disappear if there is an intense stellar absorption line \citep{Greene2005}.

To compute the CaII H and \hb\ indexes, we used the continua around 4020 and 5100\AA, respectively, with a width of 50\AA. The continuum for estimating the CaII H-Index (H-index) is intended to consider the slope of the Balmer jump. For instance, \cite{Bruzual1983} characterized the amplitude of the discontinuity around 4000\AA\ (D4000) using the intervals 3750-3950 and 4050-4250 \AA. In the case of bluer galaxies, the amplitude of the discontinuities decreases. 

%

\begin{figure}
\centering
  \includegraphics[width=\columnwidth]{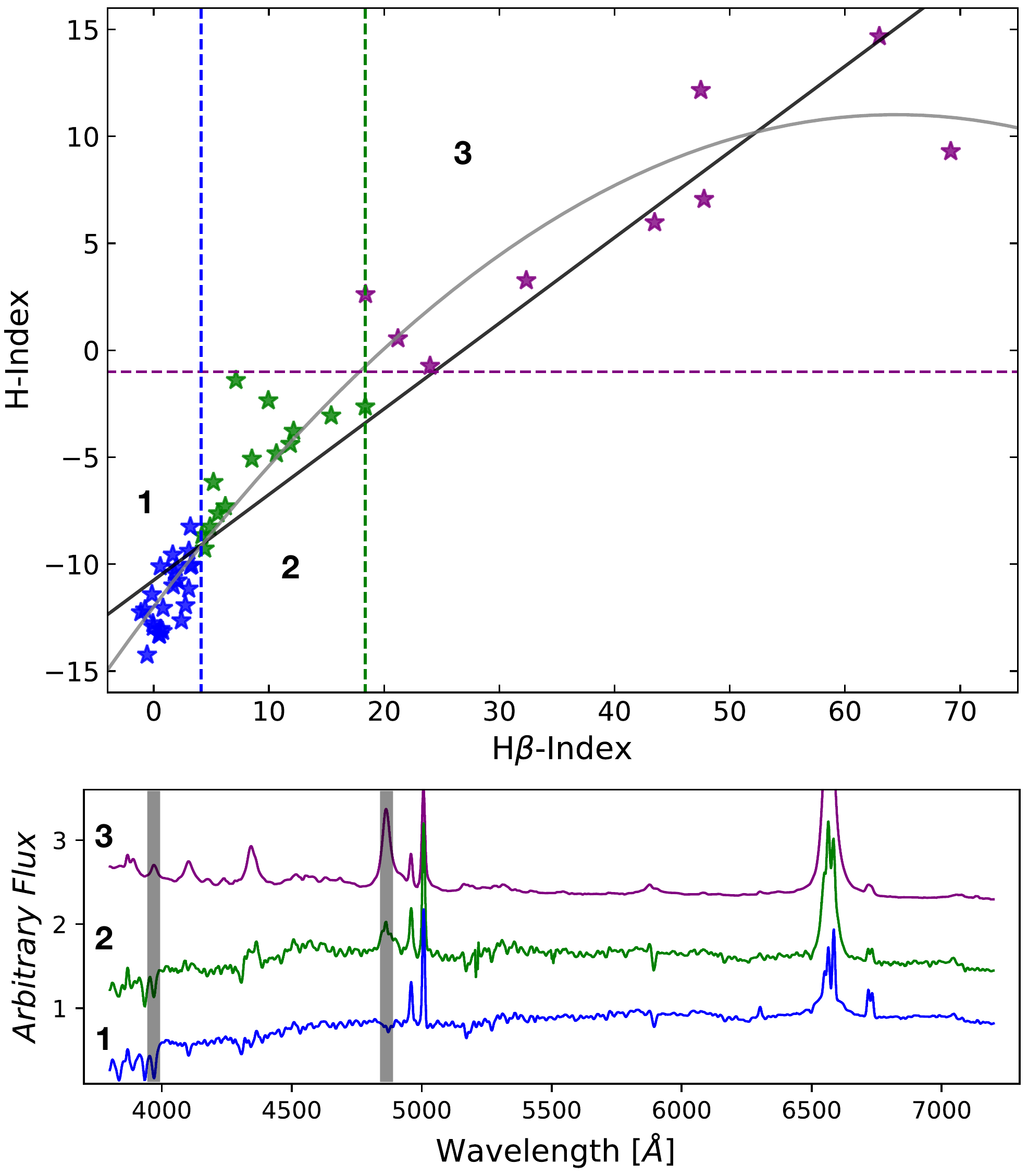}
  \caption{Top figure. Index values for all 47 AGN of the \hb\ (abscissa) and Blue interval (ordinate). The dashed lines indicate the limits to separate our AGN by the features in their spectra (see text). The color pattern indicates each group of AGN, blue for Dominant Galaxy (1), green for Intermediate type (2), and purple for Dominant AGN (3). Bottom figure shows an example spectrum for each group, where we can see the differences in the continua, the absorption /emission lines, and the shape of the broad lines.}
\label{fig:spectraAGN}
\end{figure}

The upper panel of Figure \ref{fig:spectraAGN} shows the resulting index values for the 47 type-1 AGN. In this plot, positive values are for emission lines. Since both quantities seem to be correlated, a linear relation is fit obtaining, 
\begin{equation}
    \mbox{H-Index} = (-10.75   \pm  0.44) + (0.40 \pm 0.02)\  \mbox{\hb-Index}
\end{equation}
with a relation value $r^2$ of 0.88. 

A quadratic fit,
\begin{equation}
\begin{split}
    \mbox{H-Index} = (-11.99 \pm 0.38) + (0.71 \pm 0.05)\  \mbox{\hb-Index} \\ - (0.006 \pm 0.001)\  \mbox{\hb-Index}^2
\end{split}
\end{equation}
seems to fit better the data, with a $r^2$ = 0.94, 
however, not a very different $r^2$ was found using a first order exponential fit. The plot shows that the H-Index values can be as negative as -14.2\AA\ in the case of dominant absorption lines, or positive, up to 14.7\AA, in the case of dominant emission lines. It is noticeable that 83\%\ of the objects in this distribution have negative values which is a signature of the presence of the stellar CaII absorption lines. In the case of the \hb-Index, the range values are between -1.1\AA\ and 69.2\AA, with only six objects measured in absorption. This empirical description intends to emphasize the different levels of nuclear activity in the observed spectra, where the most AGN dominant cases show positive values of both (EWs) indexes.

An inspection to Figure \ref{fig:spectraAGN} shows that the 47 type-1 AGN can be grouped into three specific regions of this diagram. 
\begin{itemize}
    \item \textit{AGN Dominant}. There are in the extreme region where the H-Index $>$ -1.0 \AA\ (9 galaxies above the purple line). The spectra of objects in this region show a Balmer broad emission line having various components, with an apparent absence of absorption lines, and the continuum dominated by the AGN power law.
    \item\textit{Galaxy Dominant}. There are located at the other extreme, where H-Index $<$ -8 \AA\ and \hb-Index $<$ 4.1 \AA\ (blue stars, 24 galaxies). The spectra of objects in this region show only a featureless \ha\ broad component with a continuum mostly dominated by the stellar contribution.
    \item \textit{Intermediate}. A third group can be described as in between of the previous two extremes, in the region H-Index $<$ -1 \AA\ and \hb-Index $<$ 18.7 \AA\ (green stars, 14 galaxies). The spectra of objects in this region show both the \ha\ and \hb\ broad emission lines, the AGN power law as well as absorption lines. 
\end{itemize}
The lower panel of Figure \ref{fig:spectraAGN} shows an example spectra for each group, evidencing different levels of the nuclear activity contribution in the observed spectra. The differences can be explained principally by the host galaxy contribution, dust contamination, obscuration of the broad line region and the state of nuclear activity \citep{Kauffmann2003,Laor2003,Hao2005,Donofrio2021}.
More detailed analysis of the observed properties of the broad emission lines for each group will be discussed in Cortes et al. (in prep).

\begin{table*}
\centering
\begin{tabular}{ccccccccccc}
\hline
\hline
\textbf{SDSS-ID}	& \textbf{MaNGA-ID}	&
\textbf{Plate-IFU} & \textbf{R.A.}		 & \textbf{Dec}	 & \textbf{m$_g^a$} & \textbf{z$^b$} & M$_*^c$	(M$\odot$) & \textbf{L[OIII]$^d$}	 & \textbf{L(H$\alpha_{BC}$)$^d$}	 & AGN	group$^e$		\\	
(1)&(2)&(3)&(4)&(5)&(6)&(7)&(8)&(9)&(10)&(11)\\
\hline	\\

J211646.33+110237.4 & 1-113712     & 7815-6104 & 319.1931 & 11.0437 & 16.67 & 0.081 & 10.49 & 42.13 & 42.14 & 1\\
J212851.19-010412.4 & 1-180204     & 7968-3701 & 322.2130 & -1.0701 & 15.29 & 0.052 & 10.74 & 39.59 & 40.47 & 2\\
J210721.91+110359.1 & 1-113405     & 7972-3704 & 316.8410 & 11.0664 & 17.5 & 0.042 & 9.91 & 40.03 & 40.32	 & 3\\			
J220429.49+122633.3 & 1-596598     & 7977-9101 & 331.1229 & 12.4426 & 15.26 & 0.027 & 10.49 & 39.93 & 41.15 & 2\\
J171411.63+575834.0 & 1-24092      & 7991-1901 & 258.5485 & 57.9761 & 16.3 & 0.093 & 10.18 & 42.22 & 43.47 & 1\\		
J171518.57+573931.6 & 1-24148      & 7991-6104 & 258.8274 & 57.6588 & 15.91 & 0.028 & 10.18 & 40.04 & 40.14 & 3\\
J072656.08+410136.0 & 1-548024     & 8132-6101 & 111.7337 & 41.0267 & 16.82 & 0.129 & 11.20 & 40.96 & 41.38 & 3\\
J073623.13+392617.7 & 1-43214      & 8135-1902 & 114.0964 & 39.4383 & 16.22 & 0.118 & 10.79 & 43.16 & 43.29 & 1\\
J073846.89+295328.5 & 1-121075     & 8144-3702 & 114.6950 & 29.8913 & 16.62 & 0.098 & 10.96 & 40.17 & 41.09 & 3\\
J040548.78-061925.8 & 1-52660      & 8158-3704 & 61.4533 & -6.3238 & 17.33 & 0.057 & 10.26 & 40.15 & 41.14 & 3\\				
J082840.99+173453.0 & 1-460812     & 8241-9102 & 127.1710 & 17.5814 & 16.42 & 0.067 & 10.68 & 40.23 & 40.66 & 3\\
J134630.60+224221.6 & 1-523004     & 8320-6101 & 206.6280 & 22.7060 & 15.73 & 0.027 & 10.07 & 39.11 & 39.81 & 3\\
J142004.29+470716.8 & 1-235576     & 8326-6102 & 215.0179 & 47.1213 & 16.17 & 0.070 & 10.76 & 40.93 & 41.57	 & 2\\			
J123651.17+453904.1 & 1-620993     & 8341-12704 & 189.2132 & 45.6512 & 14.69 & 0.030 & 10.44 & 40.39 & 40.73 & 2\\
J134300.79+360956.3 & 1-418023     & 8446-1901 & 205.7530 & 36.1657 & 16.32 & 0.024 & 9.49 & 39.67 & 40.48 & 2\\
J111803.22+450646.8 & 1-256832     & 8466-3704 & 169.5134 & 45.1130 & 16.43 & 0.107 & 11.30 & 41.28 & 41.91 & 2\\
J143031.19+524225.8 & 1-593159     & 8547-12701 & 217.6300 & 52.7072 & 15.23 & 0.045 & 10.76 & 40.23 & 40.56 & 3\\
J160505.15+452634.8 & 1-210017     & 8549-12702 & 241.2715 & 45.4430 & 15.19 & 0.043 & 10.81 & 40.05 & 41.39	 & 2\\
J153552.40+575409.5 & 1-90242      & 8553-1901 & 233.9680 & 57.9026 & 15.05 & 0.030 & 10.01 & 42.81 & 42.26 & 1\\
J153810.05+573613.1 & 1-90231      & 8553-9102 & 234.5420 & 57.6037 & 15.45 & 0.074 & 10.96 & 40.29 & 41.26 & 2\\
J162838.23+393304.4 & 1-594493     & 8603-6101 & 247.1590 & 39.5513 & 13.74 & 0.031 & 11.22 & 39.56 & 40.18 & 3\\
J170007.17+375022.2 & 1-95585      & 8606-12701 & 255.0300 & 37.8395 & 15.33 & 0.063 & 11.20 & 40.27 & 40.66 & 3\\
J212401.90-002158.7 & 1-550901     & 8615-3701 & 321.0080 & -0.3663 & 16.2 & 0.062 & 10.50 & 40.66 & 41.42 & 2\\
J075525.29+391109.8 & 1-71974      & 8713-9102 & 118.8554 & 39.1861 & 15.26 & 0.033 & 10.35 & 40.60 & 41.29 & 1\\
J075244.19+455657.4 & 1-604860     & 8714-3704 & 118.1842 & 45.9493 & 15.37 & 0.052 & 11.04 & 40.74 & 41.54 & 3\\
J075643.71+445124.1 & 1-44303      & 8718-12701 & 119.1820 & 44.8567 & 15.88 & 0.050 & 10.39 & 40.24 & 40.33 & 3\\
J082842.73+454433.3 & 1-574519     & 8725-9102 & 127.1781 & 45.7426 & 16.33 & 0.049 & 10.23 & 40.37 & 40.72 & 2\\
J080020.98+263648.7 & 1-163966     & 8940-12702 & 120.0870 & 26.6135 & 14.16 & 0.027 & 10.70 & 40.63 & 40.99 & 3\\
J164520.62+424527.9 & 1-94604      & 8978-6104 & 251.3360 & 42.7578 & 16.3 & 0.049 & 10.32 & 39.66 & 40.59 & 3\\
J134401.90+255628.3 & 1-423024     & 8983-3704 & 206.0080 & 25.9412 & 15.9 & 0.062 & 10.54 & 40.63 & 41.15 & 2\\
J113409.01+491516.4 & 1-174631     & 8990-12705 & 173.5380 & 49.2546 & 16.95 & 0.037 & 9.92 & 40.50 & 39.89 & 3\\
J112637.74+513423.0 & 1-149561     & 8992-3702 & 171.6570 & 51.5730 & 15.87 & 0.026 & 9.87 & 39.80 & 40.11 & 2\\
J112536.16+542257.1 & 1-614567     & 9000-1901 & 171.4010 & 54.3826 & 15.9 & 0.021 & 9.68 & 40.60 & 41.46 & 1\\
J160436.23+435247.3 & 1-210186     & 9036-6101 & 241.1510 & 43.8798 & 16.33 & 0.060 & 10.53 & 39.88 & 40.50 & 3\\
J162501.44+241547.4 & 1-295542     & 9048-1902 & 246.2560 & 24.2632 & 16.61 & 0.050 & 10.12 & 40.91 & 41.34 & 2\\
J075828.10+374711.6 & 1-71872      & 9181-12702 & 119.6170 & 37.7866 & 13.87 & 0.041 & 11.42 & 40.08 & 40.63 & 3\\
J075756.71+395936.0 & 1-71987      & 9182-6102 & 119.4860 & 39.9934 & 16.21 & 0.040 & 10.63 & 40.70 & 40.08 & 3\\
J030639.56+000343.1 & 1-37863      & 9193-12704 & 46.6649 & 0.0620 & 16.77 & 0.107 & 10.45 & 41.63 & 42.52 & 1\\				
J030510.60-010431.6 & 1-37385      & 9193-9101 & 46.2942 & -1.0755 & 15.66 & 0.045 & 10.82 & 40.09 & 41.11 & 3\\				
J030652.09-005347.5 & 1-37336      & 9194-6101 & 46.7171 & -0.8965 & 16.32 & 0.084 & 10.88 & 40.05 & 40.90 & 3\\					
J030834.31+003303.3 & 1-37633      & 9194-6103 & 47.1430 & 0.5509 & 16.05 & 0.031 & 10.28 & 39.21 & 40.39 & 3\\				
J172935.80+542940.0 & 1-24660      & 9196-12703 & 262.3990 & 54.4944 & 16.28 & 0.082 & 10.80 & 40.45 & 41.02 & 3\\
J081319.33+460849.6 & 1-574506     & 9487-3702 & 123.3310 & 46.1472 & 16.5 & 0.054 & 10.53 & 40.59 & 41.03 & 3\\
J081516.86+460430.8 & 1-574504     & 9487-9102 & 123.8203 & 46.0753 & 15.08 & 0.041 & 10.70 & 41.78 & 42.50 & 1\\
J075217.84+193542.2 & 1-298111     & 9497-12705 & 118.0744 & 19.5951 & 15.81 & 0.117 & 10.90 & 43.35 & 43.10 & 1\\
J084654.09+252212.3 & 1-385623     & 9500-1901 & 131.7254 & 25.3701 & 16.10 & 0.051 & 10.16 & 41.68 & 42.70 & 2\\
J134245.70+243524.0 & 1-523211     & 9881-1902 & 205.6900 & 24.5901 & 15.7 & 0.027 & 10.04 & 38.96 & 39.69 & 3\\			
 &  &  &  &  &  & 
\\
\hline
\hline
\end{tabular}
\vspace{4pt}
\caption{Main properties of the 47 galaxies with active nuclei according to our selection criteria. ($^a$)The magnitude corresponds to the g band provided by SDSS (DR14). ($^b$) The redshift was obtained by Pipe3D \citep{Sanchez2016a,Sanchez2021}. ($^c$) Stellar masses from NSA catalog derived from Sersic fluxes. ($^d$) The \habc\ and \oiii\ luminosities were obtained from our emission lines fitting (Cortes-Suarez et al. in prep.) ($^e$) Empirical classification of the type-1 AGN described in section \ref{sec:sample}, 1 for AGN Dominant, 2 for Intermediate, and 3 for Galaxy Dominant.}
\label{tab:47type1}
\end{table*}

\subsection{Multiwavelength Data}


We have collected information at different wavelengths 
from the WISE catalog \citep{Wright2010} at IR wavelengths, in the radio continuum from FIRST \citep{Becker1995} and NVSS \citep{Condon1998}, and in X-ray catalog using ROSAT \citep{Boller2016}.  

The detection fraction for our 47 type-1 AGN sample is complete in the WISE NIR and MIR bands (W1=3.4$\mu$m, W2=4.6$\mu$m, W3=12$\mu$m, and W4=22$\mu$m). In the case of X-ray data the detection fraction is 55\%\ coming from the ROSAT catalog.  The matched radio continuum detection fraction is 51\% coming mainly from the FIRST and NVSS surveys. Sections \ref{sec:wise} and \ref{sec:radio} provide a more detailed description of the IR and radio properties of the current sample, while section \ref{sec:MultiwavelengthLuminosities} presents a correlation analysis between optical and the multiwavelength luminosities by taking into account the presence of upper limits in the 20 cm and X-ray data.

Table \ref{tab:multifrecuencia} reports the values found, Column (1) is the MaNGA identification number, Columns (2-5) are the WISE fluxes, 
Column (6) is the flux in the soft X-rays range from ROSAT, Column (7) is the radio-continuum luminosity (L$_{\rm Radio}$) coming from NVSS/FIRST, Column (8) reports the radio-loudness classification 
using the X-ray radio-loudness parameter \citep[R$_K$,][]{Terashima2003} with the values of the FIRST and NVSS fluxes (Sec. \ref{sec:MultiwavelengthLuminosities}), 
Column (9) shows the excitation index characterization following \cite{Buttiglione2010}. Column (10) shows the radio source characterization following \cite{Best2005,Best2012,Mingo2016}. Column (11) is the radio characterization (LERG/HERG) following \cite{Best2012}.

\begin{table*}
\centering
\begin{tabular}{ccccccccccc}
\hline
\hline
\textbf{MaNGA-ID} & w1 & w2 & w3 & w4 & {0.1-2.4 KeV} & {20 cm} & {\textbf{Log R$_X$}}  & {HE/LE} & {Radio Source}& {HERG/LERG}\\ 
(1) & (2) & (3) & (4) & (5) & (6) & (7) & (8) & (9) & (10) & (11)\\
\hline \\
1-113712  & 43.75 & 43.66 & 43.60 & 43.82 & 42.72$\pm$0.14 & 38.70$\pm$0.03 & -4.92 & HE & SF & -\\
1-180204  & 43.30 & 42.99 & 43.25 & 43.24 & \textit{42.47} & \textit{37.98} & - & HE & - & -\\
1-113405  & 42.73 & 42.60 & 42.72 & 42.78 & \textit{42.13} & \textit{37.79} & - & HE & - & -\\
1-596598  & 43.04 & 42.81 & 42.41 & 42.44 & 41.86$\pm$0.15 & \textit{37.22} & - & LE & - & -\\
1-24092   & 44.22 & 44.19 & 44.17 & 44.32 & 43.51$\pm$0.04 & \textit{38.51} & - & - & - & -\\
1-24148   & 42.99 & 42.56 & 42.24 & 42.33 & \textit{41.57} & \textit{37.47} & - & HE & - & -\\
1-548024  & 44.03 & 43.81 & 43.87 & 43.95 & \textit{43.02} & \textit{38.83} & - & HE & - & -\\
1-43214   & 44.81 & 44.81 & 44.64 & 44.69 & 44.15$\pm$0.04 & \textbf{39.30$\pm$0.05} & -5.75 & - & SF & -\\
1-121075  & 43.56 & 43.31 & 43.33 & 43.21 & \textit{42.75} & \textit{38.55} & - & LE & - & -\\
1-52660   & 42.92 & 42.66 & 42.37 & 42.65 & \textit{42.41} & - & - & HE & - & -\\
1-460812  & 43.67 & 43.46 & 43.27 & 43.36 & \textit{42.54} & \textbf{38.96$\pm$0.04} & -4.47 & HE & AGN & HERG\\
1-523004  & 42.74 & 42.40 & 42.40 & 42.43 & \textit{41.68} & \textit{37.37} & - & LE & - & -\\
1-235576  & 43.73 & 43.57 & 43.50 & 43.52 & 43.06$\pm$0.05 & 38.68$\pm$0.02 & -5.27 & HE & SF & -\\
1-620993  & 42.95 & 42.69 & 42.91 & 43.05 & 43.15$\pm$0.03 & \textbf{38.11$\pm$0.04} & -5.94 & HE & SF & -\\
1-418023  & 42.50 & 42.33 & 42.28 & 42.47 & 42.21$\pm$0.06 & \textit{37.30} & - & HE & - & -\\
1-256832  & 44.07 & 43.88 & 43.84 & 43.91 & 43.98$\pm$0.04 & \textbf{39.86$\pm$0.02} & -5.02 & HE & AGN & HERG\\
1-593159  & 43.53 & 43.24 & 43.53 & 43.71 & 42.13$\pm$0.12 & \textbf{38.59$\pm$0.04} & -4.43 & HE & SF & -\\
1-210017  & 43.39 & 43.11 & 42.98 & 42.91 & \textit{42.30} & \textit{37.81} & - & HE & - & -\\
1-90242   & 43.61 & 43.60 & 43.62 & 43.76 & 43.33$\pm$0.01 & 38.25$\pm$0.01 & -5.58 & HE & SF & -\\
1-90231   & 43.56 & 43.28 & 43.53 & 43.55 & 42.83$\pm$0.08 & \textbf{38.68$\pm$0.07} & -5.05 & HE & SF & -\\
1-594493* & 43.45 & 43.07 & 42.20 & 42.09 & 43.58$\pm$0.02 & \textbf{41.12$\pm$0.01} & -3.36 & - & AGN & LERG\\
1-95585   & 43.47 & 43.17 & 43.00 & 43.01 & \textit{42.11} & \textbf{38.70$\pm$0.06} & -4.30 & LE & AGN & HERG\\
1-550901  & 43.43 & 43.17 & 43.44 & 43.68 & 42.86$\pm$0.08 & 38.12$\pm$0.05 & -5.64 & HE & SF & -\\
1-71974   & 43.25 & 43.13 & 43.41 & 43.51 & 43.12$\pm$0.03 & \textit{37.61} & - & - & - & -\\
1-604860*  & 43.75 & 43.54 & 43.29 & 43.31 & 43.11$\pm$0.05 & \textbf{40.60$\pm$0.01} & -3.40 & LE & AGN & LERG\\
1-44303   & 43.01 & 42.66 & 42.72 & 42.87 & \textit{42.18} & \textit{37.94} & - & HE & - & -\\
1-574519  & 43.14 & 42.91 & 43.03 & 42.96 & \textit{42.23} & \textit{37.92} & - & HE & - & -\\
1-163966  & 43.62 & 43.51 & 43.52 & 43.72 & 42.02$\pm$0.12 & \textbf{38.42$\pm$0.02} & -4.50 & HE & SF & -\\
1-94604   & 42.97 & 42.70 & 42.63 & 42.48 & 42.22$\pm$0.12 & \textit{37.91} & - & HE & - & -\\
1-423024  & 43.48 & 43.26 & 43.45 & 43.51 & 42.62$\pm$0.11 & \textit{38.14} & - & HE & - & -\\
1-174631  & 42.66 & 42.44 & 42.41 & 42.55 & \textit{41.99} & \textit{37.69} & - & HE & - & -\\
1-149561  & 42.64 & 42.42 & 42.44 & 42.75 & \textit{41.92} & \textit{37.37} & - & HE & - & -\\
1-614567  & 43.13 & 43.06 & 42.89 & 42.88 & 42.08$\pm$0.07 & 37.37$\pm$0.04 & -5.61 & HE & SF & -\\
1-210186  & 43.12 & 42.83 & 42.79 & 42.57 & \textit{42.84} & \textit{38.12} & - & HE & - & -\\
1-295542  & 43.21 & 42.96 & 43.21 & 43.48 & \textit{42.27} & \textbf{38.37$\pm$0.08} & -4.79 & HE & SF & -\\
1-71872* & 43.72 & 43.35 & 42.64 & 42.57 & 42.21$\pm$0.11 & \textbf{41.22$\pm$0.01} & -1.88 & LE & AGN & LERG\\
1-71987   & 42.91 & 42.72 & 43.00 & 43.62 & \textit{42.61} & \textbf{40.22$\pm$0.01} & -3.30 & HE & AGN & HERG\\
1-37863   & 44.15 & 44.10 & 44.10 & 44.24 & 43.35$\pm$0.11 & \textbf{39.31$\pm$0.04} & -4.93 & HE & SF & -\\
1-37385   & 43.24 & 42.91 & 42.55 & 42.38 & \textit{42.32} & \textit{37.87} & - & LE & - & -\\
1-37336   & 43.61 & 43.35 & 43.50 & 43.42 & \textit{43.16} & \textit{38.43} & - & LE & - & -\\
1-37633   & 42.95 & 42.58 & 42.03 & 41.86 & \textit{41.97} & \textbf{38.03$\pm$0.05} & -4.84 & LE & AGN & LERG\\
1-24660   & 43.83 & 43.67 & 43.71 & 43.90 & 42.30$\pm$0.17 & 38.72$\pm$0.03 & -4.48 & HE & SF & -\\
1-574506  & 43.26 & 43.03 & 42.90 & 42.98 & 42.92$\pm$0.06 & \textit{38.06} & - & HE & - & -\\
1-574504  & 43.60 & 43.48 & 43.77 & 44.07 & 42.85$\pm$0.05 & \textbf{38.67$\pm$0.03} & -5.07 & LE & SF & -\\
1-298111  & 44.55 & 44.62 & 44.66 & 44.91 & 43.14$\pm$0.17 & \textbf{40.14$\pm$0.02} & -3.90 & HE & AGN & HERG\\
1-385623  & 43.27 & 43.12 & 43.05 & 42.89 & 43.39$\pm$0.05 & \textit{37.95} & - & HE & - & -\\
1-523211  & 42.57 & 42.22 & 41.67 & 41.68 & \textit{41.65} & \textit{37.39} & - & LE & - & -\\
\\
\hline
\hline
\end{tabular}

\vspace{4pt}
\caption{The multiwavelength $\lambda$ L$_{\lambda}$ luminosities (in log erg s$^{-1}$) of the AGN type-1 sample. Columns 2-5 are the WISE luminosties for each Near-IR and MID-IR band. Column 6 is the X-Ray luminosity at 0.1-2.4 KeV from the Second ROSAT all-sky survey (2RXS) source catalogue. Column 7 is the Radio luminosity (L$_{\rm Radio}$) at 20 cm from FIRST catalogue. Bold values are from the NVSS catalogue. Column 8 is the X-ray radio-loudness parameter from \citet{Terashima2003}. Column 9 is the Excitation Index classification by \citet{Buttiglione2010}. Column 10 is the radio source classification obtained from figure \ref{fig:Best-Mingo}. Column 11 is the \citet{Best2012} classification for galaxies with AGN radio detections.  Italic values are the upper limits estimated from the catalog detection limit (including CLEAN bias) at source position. AGN with jets are indicated with an asterisk.}
\label{tab:multifrecuencia}
\end{table*}


\subsection{The WISE color-color diagram}
\label{sec:wise}
The selection criteria using one \citep{2012SternD,Assef2018} or two \citep{Mateos2012,Jarrett2017} WISE colors are based on the fact that the SEDs of stars and star-forming galaxies in the MIR are different from that of AGN. The reason is due to the fact that stars have a blackbody SED that drops at wavelengths longer than a few microns, and the reprocessed photons from very hot dust around star-forming regions peak at around a few tens of microns. For the case of AGN, the radiation from the accretion disk heats the dust of the surrounding torus to the dust sublimation temperature (1000-1500 K), so that the AGN spectral power law generated is nearly flat, which is clearly distinguishable from stellar ones.

As described in \citet{Caccianiga2015}, WISE colors may be helpful to study the effect of the combination of the AGN emission with the host galaxy contribution. 
The W1 and W2 bands trace the continuum emission from low-mass evolved stars. The W3 band is dominated by both the 11.3 $\mu$m PAH (polycyclic aromatic hydrocarbon) and the 12.8 $\mu$m [NeII] emission features. The W4 band traces the dust continuum that is reprocessed from star formation and AGN activity \citep{Jarrett2017}.

In this context, \citet{Stern2012} studied the WISE colors of AGN in the COSMOS Survey \citep{Scoville2007}, selecting the galaxies with the active nucleus based on W1 and W2 magnitudes. They show that down to a W2 magnitude of 15.05, 78\% of Spitzer-identiﬁed AGN have W1 – W2 $>$ 0.8 and 95\% of the objects with such red WISE colors are bona-ﬁde AGN. Further on, \cite{Assef2018} probed the AGN selection on significantly deeper WISE magnitudes by extending and improving the WISE AGN selection. These tests have provided different selection criteria separately optimized for reliability and completeness (90\% for W1 – W2 $>$0.5).

We use the AllWISE Source Catalog \citep{Wright2010,Mainzer2014} to retrieve fluxes for the complete MPL-7 MaNGA sample. We found information for 4277 galaxies, including all our type-1 and type-2 AGN objects except for three type-2 AGN. 
We have adopted the magnitudes measured with a profile-fitting photometry.

\begin{figure}
    \centering
    \includegraphics[width=\columnwidth]{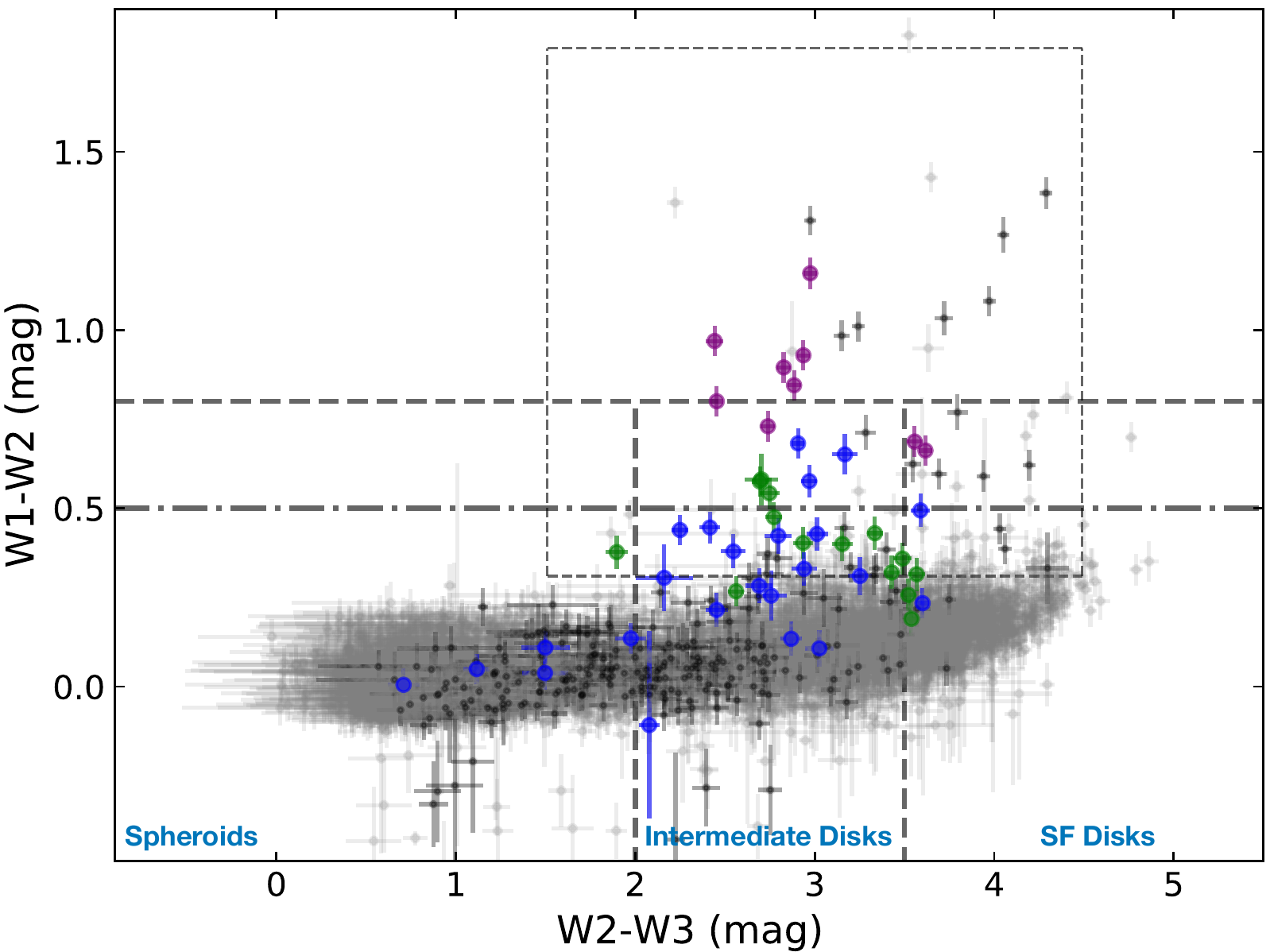}
   
    \caption{The W1-W2 vs W2-W3 color diagram. Our type-1 AGN sample is shown as AGN dominant (purple dots), Host dominant (blue dots) and Intermediate (green dots). 3992 galaxies of MPL-7 are shown as grey dots. We also add the type-2 AGN population from section \ref{sec:type2} as black dots. The inner box is the characteristic region for the AGN, including the demarcation line of $W1-W2=0.8$ \citep{2012SternD} and $W1-W2=0.5$ \citep{Assef2018}. The vertical dashed lines separate the galaxies between spheroids (left), intermediate disks (middle), and star-forming disks (right) from \citet{Jarrett2017}.}
    \label{fig:Wise}
\end{figure}

Figure \ref{fig:Wise} shows the distribution of all the MPL-7 MaNGA 
sample in the color-color W1-W2 vs W2-W3 diagram \citep{Jarriett2011,Caccianiga2015}. The color dots correspond to the type-1 AGN empirical classification described in Section \ref{sec:sample}: AGN dominant (purple), host dominant (blue), and intermediate (green). 
Type-2 AGN are shown in black, and the complement (non-AGN) DR15 sample in grey dots. The short-dashed black box encompasses the region of different types of AGN presented in \citet[][see its Fig. 1]{Caccianiga2015}. Among the AGN types included in this box are the blazar ‘WISE Gamma-ray Strip’ (WGS) for BL Lacs, Flat Spectrum Radio Quasars (FSRQ), defined by \citet{Massaro2012}, and the AGN wedge defined by \citet{Mateos2012,Mateos2013} for X-ray-selected AGN. The horizontal dashed and dot-dash lines are the AGN limit proposed by \citet[][at W1-W2=0.8]{2012SternD} and \citet[][at W1-W2=0.5]{Assef2018}, respectively. Vertical dashed lines show the regions occupied by spheroidal (early-type), intermediate-disks and SF disks (late-type) galaxies as presented in \citet{Jarrett2017}. 

Figure \ref{fig:Wise} shows that about two-thirds (66\%) of our type-1 AGN sample lies within the short-dashed black box enclosing different types of AGN in \citet{Caccianiga2015}, 
14 of our type-1 AGN meet the revised infrared \cite{Assef2018} selection criteria, and only five lie above the original \cite{2012SternD} criterion. The rest of the objects are located below any of the specified AGN regions. If our empirical classification of AGN is considered, the AGN dominant galaxies lie above the AGN \cite{Assef2018} line. In contrast, the intermediate and host dominant groups remain below this limit, indicating that an increasing contribution of the host galaxy is placing the AGN below this limit. However, segregation between host-dominated and intermediate AGN types can still be appreciated in the sense that below W1-W2$\sim$0.25, there are no intermediate types, although above and below this limit and within the 1.9 > W2-W3 > 3.6 interval, it is possible to find host dominated ones, which extend to values of W2-W3 as low as 0.8. For comparison, in the case of the type-2 AGN population, $\sim$10\%\ is located in the AGN region, 3\% is above the \cite{2012SternD} limit and $\sim$ 6\% of the \cite{Assef2018} line. Most ($\sim$90\%) of type-2 objects are outside this region, suggesting that the source of the emission is highly contaminated or due to another kind of source in the host galaxy (e.g., post-AGB or shocks, see Section \ref{sec:type2}). 

\subsection{FIRST/NVSS Radio Properties and the Identification of Radio-jets}
\label{sec:radio}

Radio observations are an alternative for detecting AGN. The dominant emission process in the radio band is of non-thermal origin, due to synchrotron emission. The intensity of this emission has been associated with fundamental physical differences in nuclear activity, mainly due to differences in the accretion efficiency \citep{Padovani2017}. As a result, two intrinsically different populations of AGN radio emitters, emerge. 
The first one is associated with quasars in high-accretion (above 1\% of Eddington) high-excitation regime showing strong emission lines. In this scenario, the material is accreted in a radiatively efficient mode through an optically thick and geometrically thin accretion disk. These are known as high-excitation radio galaxies (HERGs) or radiative-mode AGN \citep[e.g.,][]{Best2012}.
The second mode refers to objects showing highly energetic radio jets and weak lines due to inefficient radiation, low accretion (below 1\% of Eddington) and low excitation regime. The radiation of these AGN is due to advection-dominated accretion flows (ADAF). These are optically thin, geometrically thick accretion flows, emitting the bulk of their energy in kinetic form through radio jets. They are called low-excitation radio galaxies (LERGs) or jet-mode AGN.
In this subsection, we intend to carry out a characterization of our type-1 AGN sample in terms of their radio properties. Radio detections were have found for 23 of 47 AGNs (21 from FIRST, and 17 from NVSS, with 15 coincidences between them). The radio data is reported in Table \ref{tab:multifrecuencia} (Column 7), bold-facing the ones from NVSS. Three objects in our sample show higher density fluxes in NVSS when compared to those reported in FIRST. A quick look at the FIRST maps show that these are extended sources also with filamentary emission (see Figure \ref{fig:Radiojets}). We recall the reader that the 5$\sigma$ detection limits of FIRST (with a resolution $\theta$ = 5.4 arcsec) and NVSS (with $\theta$ = 45 arcsec) are 1 mJy beam$^{-1}$ and 2.3 mJy beam$^{-1}$ respectively, so that FIRST is more than twice as sensitive as the NVSS to sources with angular diameters < 5.4 arcsec. However, in terms of brightness, their RMS image noises are $\sigma$ = 3.21 K and 0.14 K respectively, so the NVSS is about 20 times more sensitive than FIRST to extended sources. Thus we decided to report the NVSS values for these extended sources (stars inside circles in Figure \ref{fig:LoiiivsLRadio}). For objects without FIRST detections, we report upper limits for the radio emission by considering the catalog detection limit (including the CLEAN bias). \footnote{The reader is referred to http://sundog.stsci.edu/first/catalogs/readme.html for more details). }



\subsubsection{AGN Vs. Star-forming emission}
\label{sec:agn-sf}

The origin of the radio emission in galaxies can emerge from both nuclear activity and intense star formation. The radio emission from star forming regions is mostly due to the synchrotron emission of particles accelerated in supernova shocks. In this scenario, the radio luminosity should be correlated with the star formation rate (SFR). High SFR leads to correspondingly high radio power \citep[e.g.,][and references therein]{Condon1992,SandersMirabel1996}. Particularlly, intense star formation activity in radio quiet AGN would give rise to radio emission. In terms of the Eddington ratio, higher values of Eddington ratio are associated with concomitant high SFR \citep[e.g.,][]{Sani2010,Ganci2019}.

\begin{figure}
    \centering
    \includegraphics[width=\columnwidth]{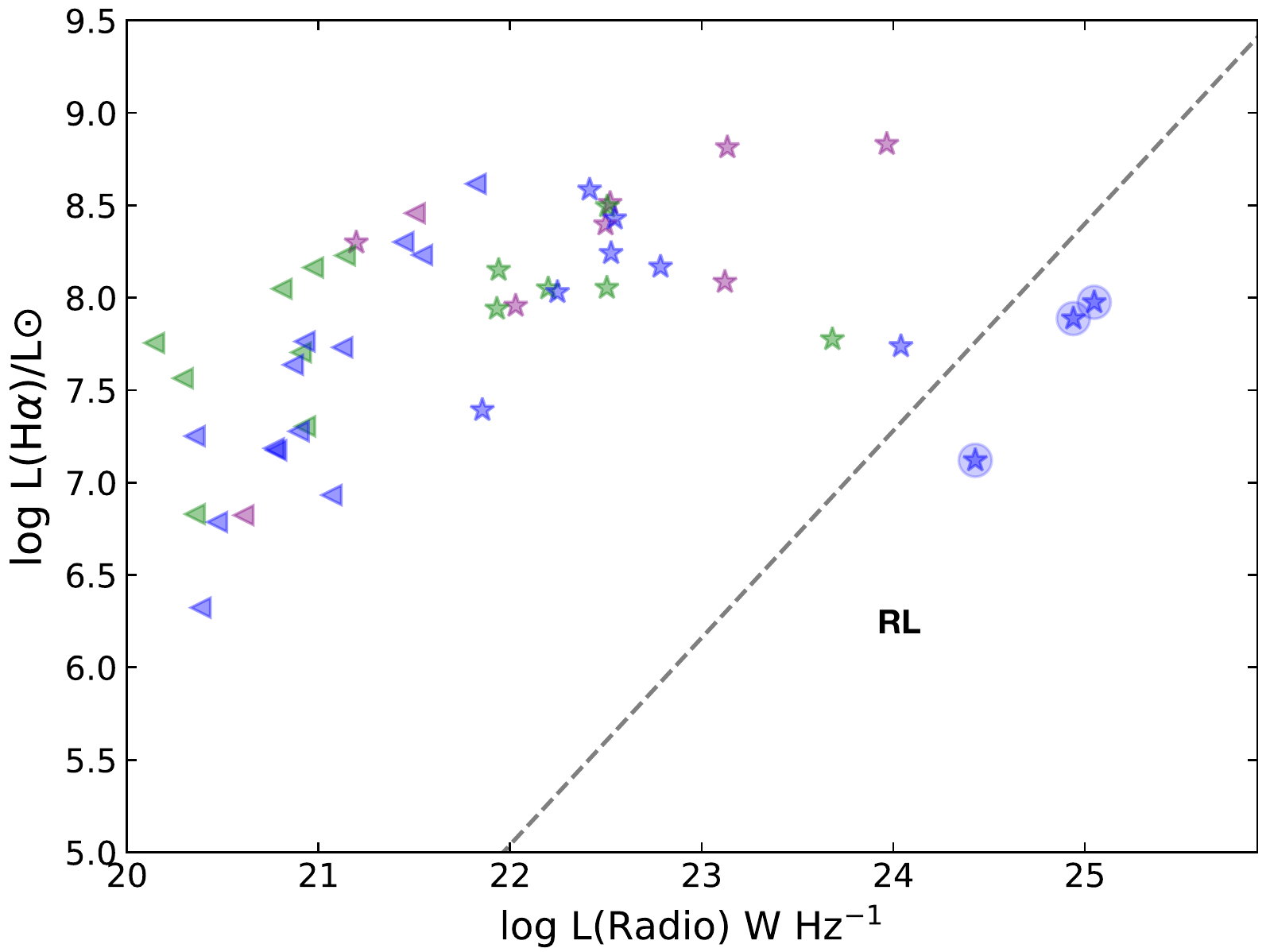}\\
    \includegraphics[width=\columnwidth]{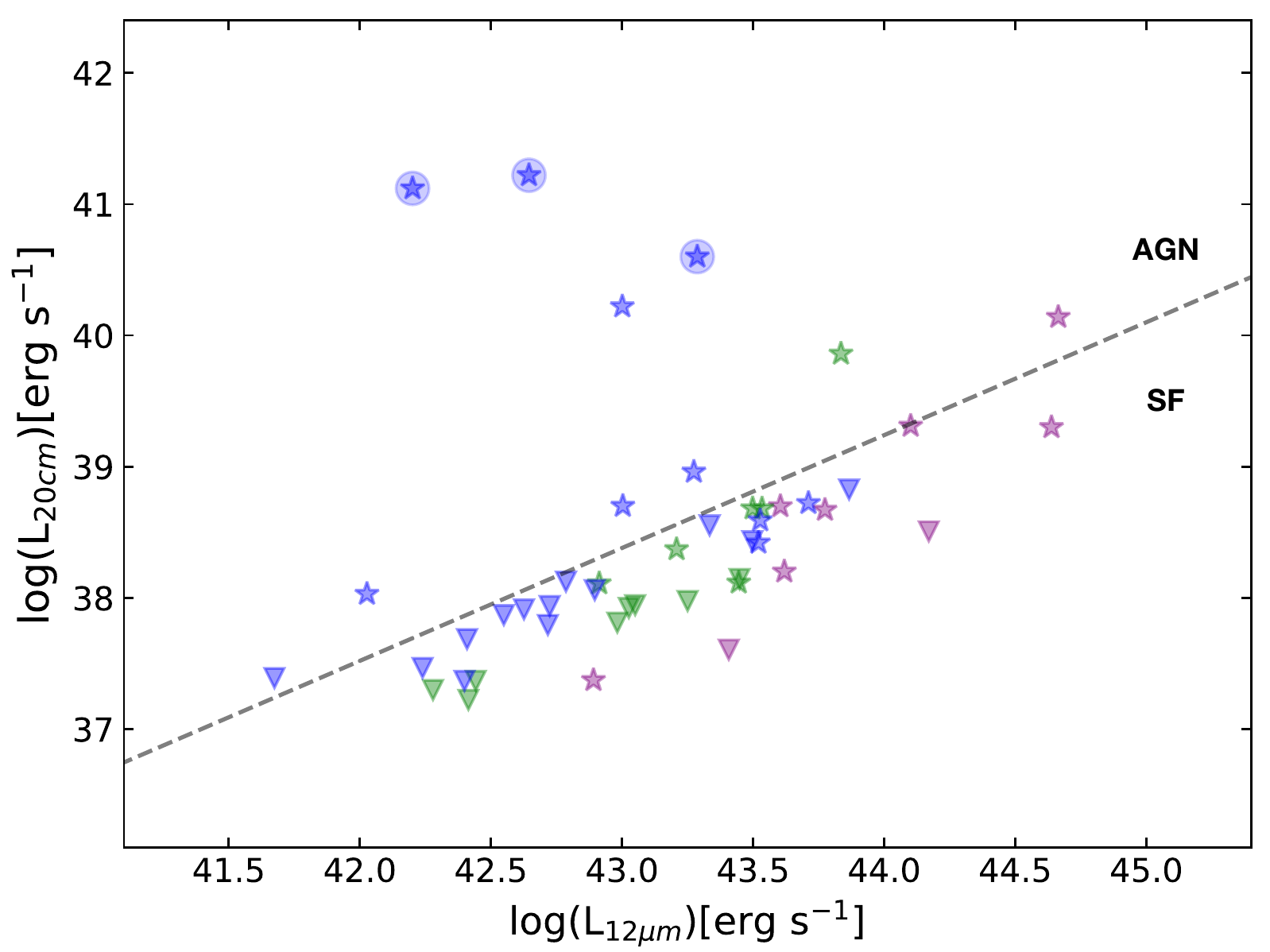}\\

    \caption{The classification of radio sources for our type-1 AGNs. The top plot is the relationship between \ha\ and radio luminosity. Dashed line is the limit between Radio Loud sources and Star Forming sources \citep{Best2012}. Bottom plot is the radio/mid-IR star formation correlation. Dashed line is the best fit of Star Forming sources from \citet{Mingo2016}. Objects above this linear are classified as AGN source.}
    \label{fig:Best-Mingo}
\end{figure}

To establish the origin of the radio emission in our sample, whether thermal from the SF regions or non-thermal from radio jets, we can use diagnostic diagrams. Using optical data, the BPT diagram is useful to locate objects in the AGN-composite-SF regions (Section \ref{sec:method}). In the log\oiii/\hb\ Vs. log\nii/\ha\ plane of Figure \ref{fig:MPL7_BPT}, blue stars shows our type-1 AGN sample, of which one object lies in the SF region, while another 14 are placed in the composite region. In the case of the WHAN diagram, we have two type-1 objects having EW(\ha) < 3 \AA, suggesting that for those objects, the ionization processes could not be associated only with the AGN \citep{CidFernandes2011}. However, as discussed previously, BPT or WHAN diagrams are not optimized for broad-line AGN.  

Taking into account the radio emission, several authors have use different criteria to separate radio-emitting AGN from starbursting galaxies \citep[e.g.][]{Best2012,Mingo2016,Hardcastle2019}. The log(L$_{\rm radio}$) versus log(L\ha) diagram proposed by \citet[][see its Figure A1]{Best2012} shows that for star-forming galaxies both luminosities are correlated, which provides a direct measure of the SF rate. On the other hand, the location of radio-loud AGN (RL) in this diagram is due to their much higher radio luminosity to \ha\ luminosity ratio. 
The adopted division line (considering the observed \ha\ luminosity) is log(L\ha/L$\odot$) = 1.12 $\times$ (log(L$_{\rm radio}$/W Hz$^{-1}$) - 17.5). Galaxies placed above the theoretical model are considered star-forming galaxies. We build the diagram (Figure \ref{fig:Best-Mingo} upper panel) using the radio flux (column 7, Table \ref{tab:multifrecuencia}) and the flux of \habc\ estimated from our emission line fitting (Cortes-Suarez et al. in prep.) using our host galaxy subtracted data (column 10, Table \ref{tab:47type1}). As expected, we find our three AGN with extended radio emission in the AGN radio-loud region.

Other option is to use the correlation between radio and MIR emission at 12$\mu$m for SF galaxies \citep[][see also the Fig. \ref{fig:Wise} for the location of the SF disks in the WISE diagram]{Mingo2016}. The correlation found by \citet{Mingo2016} for SF galaxies (log(L$_{20cm}$) = (0.86 $\pm$ 0.04) log(L$_{12\mu m}$) + (1.4 $\pm$ 1.5)) can be used to set the division between star-forming sources and radio emitting AGN. 
The lower panel of Figure \ref{fig:Best-Mingo}, locates our objects in this diagram, stars are for radio measurements and inverse triangles show the radio upper limits. We find 9 radio emitting objects in the AGN region: 7 of them classified as host dominated AGN (3 of them with extended radio emission), one intermediate and one AGN dominated object. 
In the SF region, the majority of the objects are intermediate and dominated AGN (5 and 6 objects respectively). Looking at the radio upper limits, all except one host dominated object lie in the SF region. Column 10 of Table \ref{tab:multifrecuencia} is a summary of the AGN-SF radio source classification.

\subsubsection{Radio classification}

For objects whose radio emission has been labeled as coming from an AGN in the previous section, we proceed to determine to which AGN population (HERG/LERG) belongs depending on their radio intensity following three criteria proposed by \citet{Best2012}.

\begin{itemize}
    \item[ ] \oiiib \textit{Equivalent Width.--} Objects with EW(\oiii) values lower than 5 \AA\ are considered LERGs, while objects with EW(\oiii) greater than 5 \AA\ are considered HERGs. Since the \oiiib\ line was detected for all our sources, we compute the EW(\oiii) from our emission line fitting (Cortes-Suarez et al. in prep.) using our stellar subtracted data (see section \ref{sec:type2}) finding in all the cases values higher than 5 \AA, with 9.1 \AA\ as the lower value, suggesting that with this single criterion, our 9 objects are all HERGs. For the remaining type-1 AGN, the lowest EW(\oiii) value is 7.6 \AA. 
    \item[] \textit{\loiii\ vs L$_{\rm Radio}$.-- }
    A second criterion is based on the \oiiib\ line luminosity (\loiii) vs radio luminosity (L$_{\rm Radio}$) diagram. Figure \ref{fig:LoiiivsLRadio} shows the distribution of our 9 objects in filled stars. For comparison, we also show the rest of the type-1 AGN sample in empty stars for the objects with radio detection, and in empty left triangles for the upper limits.
    We use the \loiii\ obtained from our emission line fitting. Pipe3D does not report a direct measure of the \oiii\ flux, and the derived one from other emission lines associated with broad components may introduce large uncertainties. 
    The dashed line (black) is the lower limit for the distribution of HERGs, such that with these criteria, sources below this limit can be classified as LERGs. 
    The different colors emphasize our empirical classification (purple for AGN-dominated, green for intermediate and blue for host-dominated spectra).

    This diagram identifies 6 HERGs and 3 LERGs. In the HERG domain, we find one AGN-dominated, one intermediate type, and four host-dominated types. In the LERG domain, we find 3 host-dominated types, two of them showing radio jets (Figure \ref{fig:Radiojets}). 
        
    \begin{figure}
    \centering
    \includegraphics[width=\columnwidth]{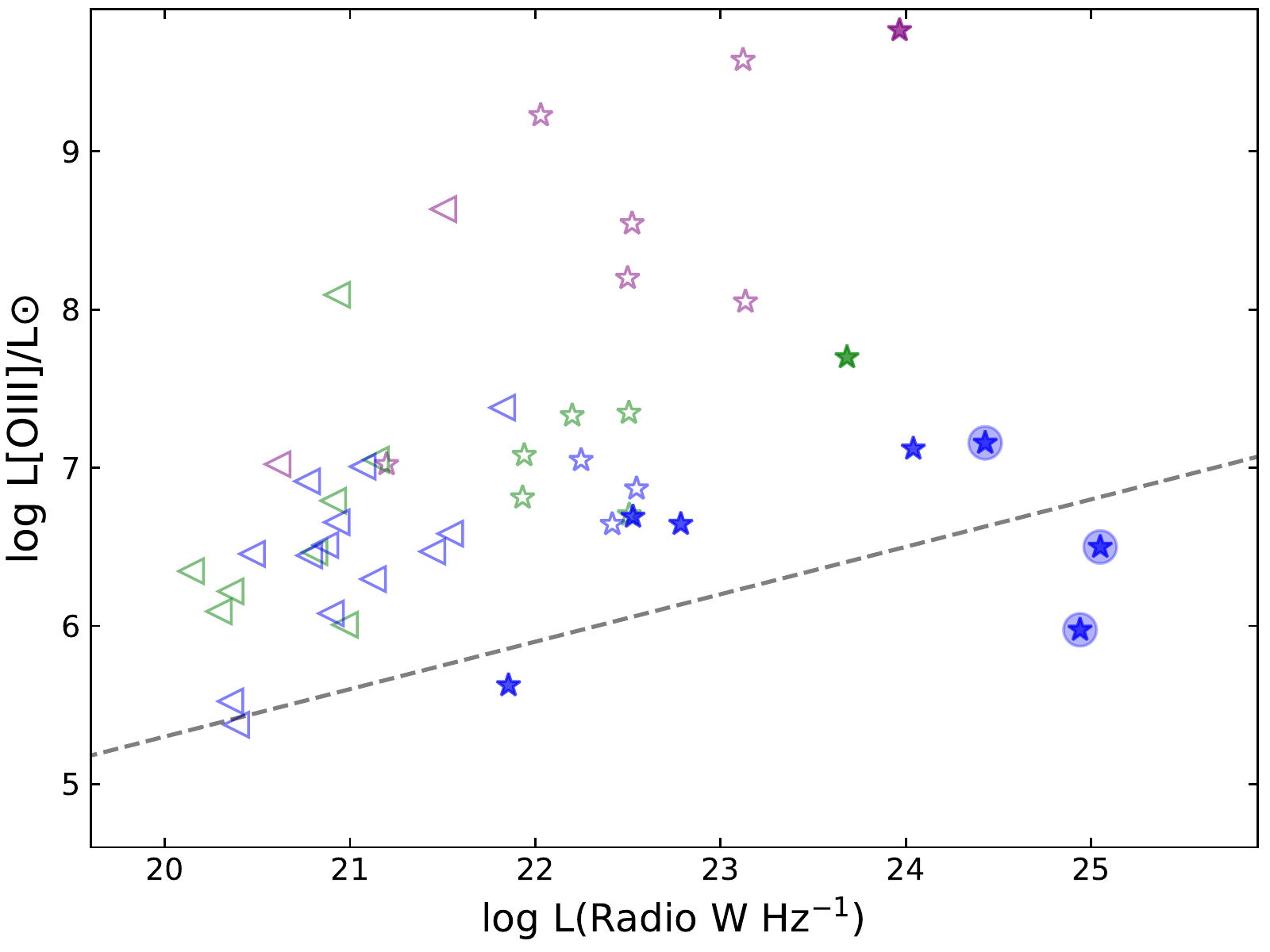}
    \caption{\oiii\ emission line luminosity versus radio luminosity. Type-1 AGN are shown as AGN Dominant (purple), Galaxy Dominant (blue), and Intermediate (green). We used the dashed line of \citet{Best2012} to separate HERGs and LERGs for our 9 objects in filled stars. We also add the upper limits as left triangles.}
    \label{fig:LoiiivsLRadio} 
    \end{figure}

    \item[] \textit{Eddington ratio.-- }
    A third criterion takes into account the accretion rate. Differences between LERGs and HERGs seems to be related to the Eddington-scaled accretion rate, or equivalently, the Eddington ratio, with the ADAF mode occurring when the accretion rate is well below the Eddington limit \citep[e.g.][]{Narayan1995,Kollmeier2006,Trump2011}. \cite{Best2012} estimated the Eddington-scaled accretion rate by comparing the total energetic output of the black hole, calculated as the sum of the radiative luminosity and the jet mechanical luminosity, with the Eddington luminosity. On the other hand, the Eddington ratio, defined as the ratio between the bolometric and Eddington luminosities (\lbol/\ledd), considers only the radiative luminosity. 
    HERGs typically accrete at a rate between of 1-10 per cent of Eddington rate, while LERGs predominantly accrete at rates below one per cent of Eddington rate \citep{Best2012}. The dependence between the radio emission and the accretion rate has also been studied in quasars at low and high z, up to 3, showing similar results \citep[][and references therein]{Sulentic2000,Shen2014,Marziani2018}. In particular, for low z quasars (z $<$ 0.8) lower black hole mass objects in the range 10$^6$-10$^7$ \msun\ showing optical spectra dominated by low ionization emission lines (such as FeII or the 
    Balmer lines), almost no high ionization emission lines (such as \oiii), have the highest Eddington ratios, with no radio dominance  \citep{Marziani2003,Ganci2019}. In the other extreme, quasars with black hole masses $\sim$ 10$^{9}$\msun\ are dominated by high ionization emission lines, showing broadest permitted lines, lowest Eddington ratios and radio emission. 
\end{itemize}


The bolometric luminosity of each radio source was estimated from the observed luminosity of the \oiiib\ emission line, using the bolometric correction \lbol\ = C$_{\rm [OIII]}$ \loiii, where  $C_{\rm [OIII]}$ = 142 for AGN with log \loiii\ = 40-42 \citep{Lamastra2009}. 
We adopt the uncertainty $\sim$ 0.4 dex on individual estimates of the bolometric radiative luminosity, from the scatter around the \lbol\ versus \loiii\ relation \citep{Heckman2004}.

The Eddington luminosity is \ledd\ = 1.3 $\times 10^{31}$ \mbh/ M$_{\odot}$ W, with an intrinsic scatter less than 0.3 dex. The black hole mass was estimated from the velocity dispersion of the galaxy $\sigma$, as measured by \citet{Aquino-Ortiz2018} and then using the well-established \mbh\ – $\sigma_{\star}$ relation of \citet{Tremaine2002},  log(\mbh/M$_{\odot}$ ) = $\alpha$ + $\beta$ log$(\sigma_{\star}$/200 \kms where ($\alpha,\beta$= 7.67 ± 0.115, 4.08 ± 0.751) for barred active galaxies and ($\alpha,\beta$= 8.19 ± 0.087, 4.21 ± 0.44) for non-barred active galaxies \citep[e.g.,][]{Graham2008,Gutekin2009} . We do not consider \mbh\ computations for seven AGN since the reported 
velocity dispersion values are below 70 \kms, the resolution limit of MaNGA data. A visual inspection of those seven objects shows that their spectra are AGN-dominant type where the absence of prominent absorption lines can lead to underestimate the velocity dispersion. 


\begin{figure}
    \centering
    \includegraphics[width=\columnwidth]{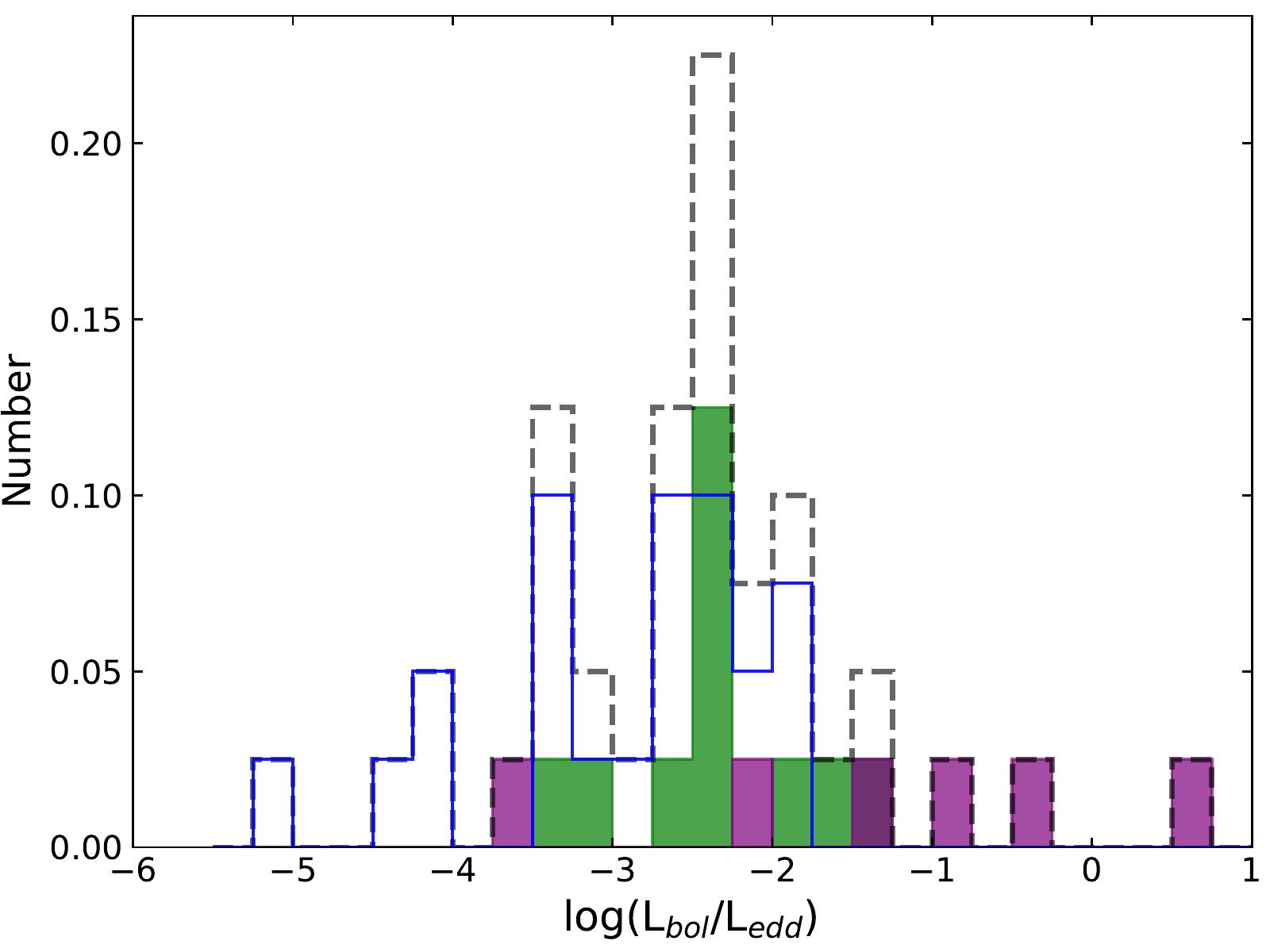}
    \caption{ The distribution of Eddington-scaled accretion rates for the type-1 AGN sample (black dashed line). We also show the distribution of each type-1 AGN group defined in section \ref{sec:sample}, purple bars for AGN Dominant, Green bars for Intermediate and Blue bars for Galaxy Dominant.}
    \label{fig:LbolLedd}
\end{figure}

Figure \ref{fig:LbolLedd} shows the Eddington ratio distributions for the complete type-1 AGN sample. 
The segregation in terms of our Host, Intermediate and AGN dominated empirical classification is also clear and gradual. On one extreme the AGN-dominated hosts (purple bars) show the higher Eddington-scaled accretion rates, while on the other, the Host-dominated types (blue bars) show the lower accretion rates, with the Intermediate types (green bars) in between these extremes. For the case of the 9 objects, one of them could not be classified for not having \mbh\ estimation. The resulting division between the two populations is: 2 HERGs and 6 LERGs, separated by log(\lbol/\ledd) $\sim$ -3.  We used this limit considering the observed separation in our distribution together with the LERG/HERG distribution shown in the Figure 6 of \citet{Best2012}. 

Since the results of each criterium shows different fraction of HERG/LERG populations, we decided to include a fourth criterium.  

\citet{Buttiglione2010} proposed a classification based on the emission line excitation properties of galaxies by defining an “excitation index” (EI) parameter, combining four emission line ratios: EI = log([OIII]/\hb) -  $\frac{1}{3}$[log([NII]/\ha) + log([SII]/\ha) + log([OI]/\ha)]. 
They demonstrate that this parameter is bimodal and use it to classify galaxies, dividing the Low Excitation (LE) and High Excitation (HE) populations at a value of EI = 0.95. 

\begin{figure}
    \centering
    \includegraphics[width=\columnwidth]{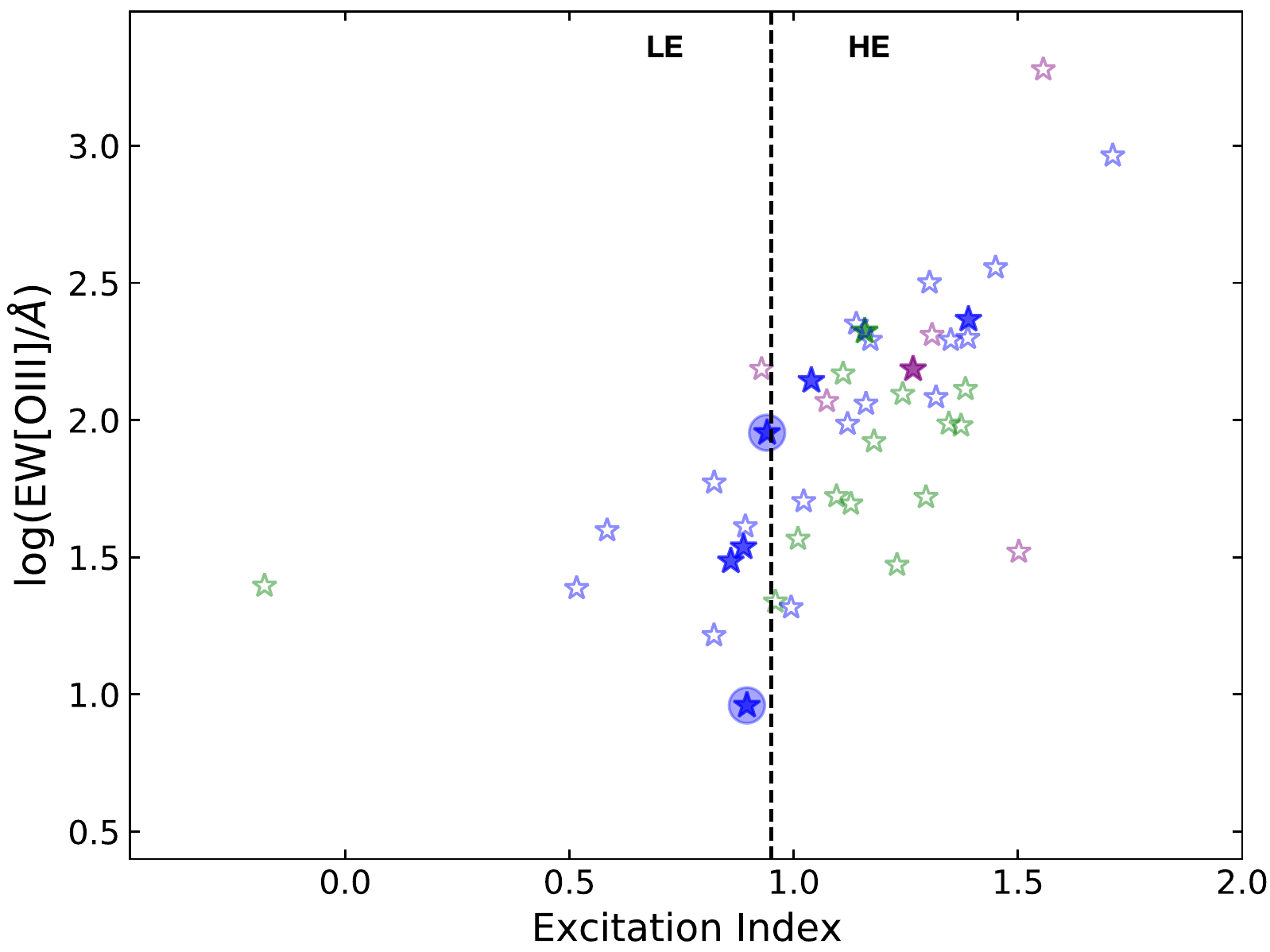}\\
    \caption{\oiiib\ versus the excitation index (EI) plane. Dashed line is the \citet{Buttiglione2010} limit between High Excitation Galaxies (>0.95) and the Low Excitation Galaxies (<0.95). Type-1 AGN are shown as AGN Dominant (purple), Galaxy Dominant (blue), and Intermediate (green).}
    \label{fig:butiglionini}
\end{figure}

According to this diagram and using the emission line estimations from Pipe3D, we found 4 High Excitation Galaxies and 4 Low Excitation Galaxies. Figure \ref{fig:butiglionini} shows the location of these objects and, for illustration, we also show the rest of the type-1 AGN sample in empty stars. All the LE objects are Host Dominated AGN. 
Among the LE objects, we identify two extended radio sources (stars inside circles in Fig.\ref{fig:LoiiivsLRadio}). 
A third object with extended radio emission
it is expected to share the same LE region, however not all of its emission line data is reported. 


\subsubsection{HERGs and LERGs objects}
\label{sec:herg-lerg}

Using the four criteria described in the last section, we proceed to characterize our 9 objects as HERGs or LERGs. A galaxy is assigned as HERG if it meets two of the three criteria of \cite{Best2012} and it is classified as HE in Figure \ref{fig:butiglionini} otherwise, it is assigned as LERG. According to this, we identify 5 HERGs and 4 LERGs (column 11, Table \ref{tab:multifrecuencia}). 

\begin{figure*}
    \centering
    \includegraphics[scale=0.67]{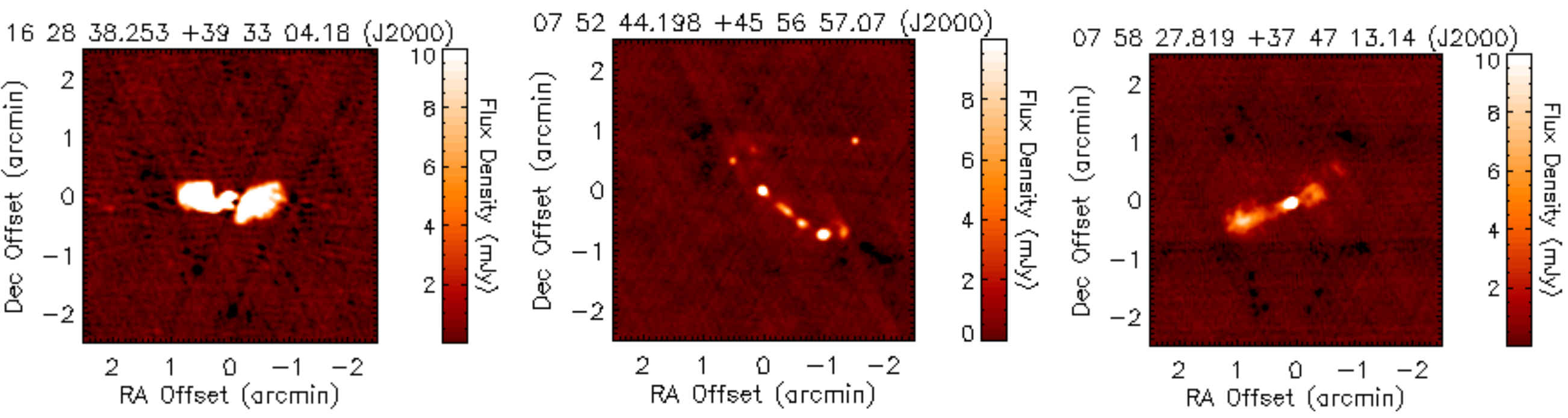}
    \caption{FIRST images of the radio sources with observable jets. From left to right, 1-594493, 1-604860, and 1-71872.}
    \label{fig:Radiojets}
\end{figure*}

The HERG/LERG fraction found agrees with the number of galaxies that shows radio jets. The three AGN with the highest NVSS radio flux density shown in Figures \ref{fig:LoiiivsLRadio}-\ref{fig:butiglionini} as blue stars in circles are 1-594493, 1-604860, and 1-71872, were classified as LERGs. Figure \ref{fig:Radiojets} is a mosaic showing their corresponding FIRST images. The images show clear evidence radio jet structures powered by the active nucleus, in agreement with the definition of LERGs. The three galaxies with radio jets are of the host-dominated type according to our empirical classification.
The identification of an additional LERG candidate (1-37633) in Figure \ref{fig:LoiiivsLRadio} may also suggest the presence of radio jets in its radio-continuum images, however we do not find visual evidence on the FIRST image. 
This candidate also belongs to the host-dominated type.

According to the radio luminosity functions, LERGs dominate at radio luminosities below L$_{\rm 20cm}\sim$10$^{26}$W Hz$^{-1}$ and HERGs become dominant above this value. However, \cite{Best2012} found that both classes can be found in all luminosities. We also agree with this result since, in Figure \ref{fig:LoiiivsLRadio}, we show that our higher value of L$_{\rm 20cm}$ is around 10$^{25}$W Hz$^{-1}$ indicating that our HERGs are also found in the low luminosity regime. Moreover, the number of HERGs could increase if we consider the 14 SF radio sources since all can classify as HERGs with the same criteria of radio characterization. 

Comparing our results with those found by \cite{Comerford2020}, who in turn used the \cite{Best2012} catalog as described in Section \ref{sec:manga_catalogs} we find five coincidences: 2 HERGs, 1 LERG, and two unclassified. One of their HERGs is cataloged as LERG by us. This reclassification corresponds to 1-6048960 and is in agreement with the presence of jets (middle panel, Figure \ref{fig:Radiojets}). The two objects unclassified by them, we considered them as HERGs. 
Since we could classify 9 of the 23 radio-emitting galaxies (including the seven coincidences), our type-1 AGN sample adds information for 6 radio sources to the \citet{Comerford2020} radio MaNGA catalog.


\subsubsection{Radio loudness}
 
One of the common criteria to classify the radio loudness is the Kellerman's ratio \rk\ = f$_{\nu,radio}$/f$_{\nu,opt}$ \citep{Kellermann1989}, defined as the ratio of the radio flux density at 5GHz and the optical flux density in the B band. The radio-loud (RL) versus radio-quiet (RQ) limit set by \cite{Kellermann1989} is \rk\ $=$ 10.

Since the optical flux is estimated at 4400\AA\ and our AGN sample is mostly Intermediate and Host-Dominated types (see section \ref{sec:sample}), optical flux estimations could be biased by the stellar contribution to the AGN continuum. But, if we look at the upper panel of Figure \ref{fig:Best-Mingo}, were \habc\ was used (instead the AGN continuum at 4400 \AA), the three objects with extended radio emission of Fig. \ref{fig:Radiojets} are well separated.

Other option is to classify RL/RQ objects using the X-ray radio-loudness parameter \rx\ = L$_{\nu,radio}$/L$_{\nu,2-10keV}$ \citep{Terashima2003}, defined as the ratio of the radio luminosity at 5GHz and the X-ray luminosity in the 2-10 keV band. Indeed, the X-ray luminosity is ideal to avoid extinction problems that normally occur in the optical band. Objects with log(\rx) $>$ -4.5 are considered RL galaxies, while RQ galaxies have log(\rx) values lower than -4.5.
Table \ref{tab:multifrecuencia} shows the estimated log(\rx) (Column 8) values from FIRST,  NVSS and ROSAT data. Applying this criterion to the objects whose radio emission was classified as AGN in section \ref{sec:agn-sf}, we find six radio-loud galaxies (and one in the borderline), with the higher values being coincident with the LERGs showing radio jets. In both cases, whether using optical or X-ray data, results indicate that our sample is dominated by radio quiet high excitation AGN.

\subsection{Optical vs Multiwavelength correlations}
\label{sec:MultiwavelengthLuminosities}

With fluxes available at different wavelengths, we estimate their corresponding luminosities and explore the correlations of various indicators of AGN emission power in our sample of type-1 AGN. To this purpose, we estimate specific $L_{\lambda}$ luminosities in \habc\ and \oiii\ emission lines as well as in the WISE, Radio Continuum, and X-Ray wavelengths as follows: 
\begin{equation}
 L_{\lambda}=4\pi\ D_{L}^2 \lambda\ F_{\lambda} 
 \label{eq:luminosity}
\end{equation}
where D$_{L}$ is the luminosity distance and F$_{\lambda}$ is the flux retrieved at different wavelengths from each catalog.

As mentioned above, the \oiii\ and  \habc\ fluxes were estimated from our emission line fitting (Cortes-Suarez et al. in prep.) using our host galaxy subtracted data. Values of L$_{\rm [OIII]}$ and \lhabc\ are reported in Table \ref{tab:47type1} (Columns 9 and 10).

The WISE data is reported in Vega magnitudes ($m_{vega}$), so the flux density (in Jansky) is:
\begin{equation}
    F_{\nu}[Jy]=F_{\nu0}\times\ 10^{-m_{vega}/2.5}
\end{equation}
where $F_{\nu\ 0}$ is the zero magnitude flux density of Vega for each WISE band (309.540 (W1), 171.787 (W2), 31.674 (W3) and 8.363 (W4)). The IR spectral slope was considered equal one\footnote{In the slope range from -1 to 2, the differences in the derived flux are below 0.7\%.}. 
The FIRST and NVSS data are reported in flux density units (mJy). The ROSAT data is reported in count rates so we applied a factor of 1.08$\times$10$^{-11}$ erg cm$^{-2}$ to convert them into fluxes \citep{Boller2016}. Our estimates to the corresponding luminosities are reported in Table \ref{tab:multifrecuencia} (Columns 2 to 7).

\begin{figure}
    \centering
    \includegraphics[width=\columnwidth]{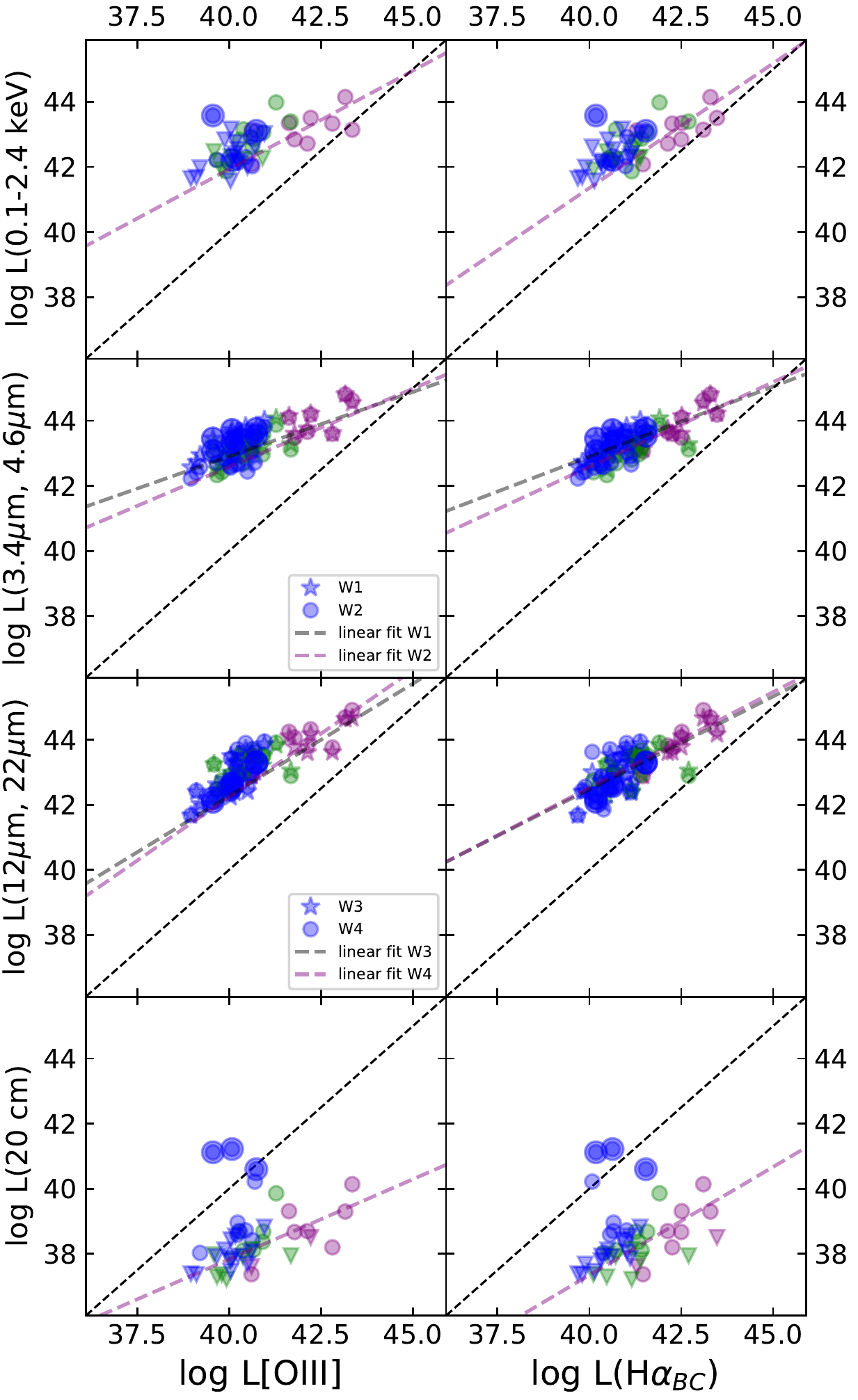}
    \caption{Optical luminosities versus other multiwavelength luminosities. The sample was separated in AGN Dominant (purple), Intermediate (green), Galaxy Dominant (blue circles), 
    and with observable jet (with shadow circle). In the case of non-detection in the radio and X-ray data, upper limits were estimated and shown as downwards triangles. Black dashed lines is the unity relation. Purple dashed line is a linear fit (see text for details).}
    \label{fig:OptvsMV}
\end{figure}

Figure \ref{fig:OptvsMV} shows the correlations of the optical indicators of the AGN power, log \lhabc\ and log \loiii\ with the continuum luminosity log L$\lambda$ ($\lambda$ = NIR and MID-IR, x-Ray, and NVSS/FIRST).
In all panels the black dashed line represents the one-to-one correlation. The sample is divided again in AGN-Dominant (purple marks), Intermediate (green marks) and Galaxy-Dominant (blue marks). 

\begin{itemize}
    \item[] \textit{X-rays.- } The fourth panel (from bottom to top) shows the relation of the optical indicators of AGN luminosity versus X-ray continuum luminosity from 0.1 to 2.4 KeV. Given the physical region where the emission from these indicators come from, a tight correlation is expected. The non-detections in the observations of the emission properties of galaxies gives valuable information, and it is important to take them into account in order to avoid biasses in the interpretation of possible relations.  
   We have adopted X-ray upper limits coming from the X-ray upper limit server\footnote{ \url{http://xmmuls.esac.esa.int/upperlimitserver/}}  and use survival analysis methods combining detections and non-detections \citep[censored data, see][]{Isobe1986,Feigelson2012}. More precisely, we used the Buckley-James (B-J) method for estimating the regression parameters \citep{Buckley1979}, which is robust when the distribution of points around the regression line is not Gaussian with no formal restriction about the random distribution of censored points. As shown by \cite{james1984}, the Buckley-James estimator of slope is asymptotically correct under a wide variety of conditions. 

   The \lhabc\ - X-ray correlation ($\rho$ = 0.70) is consistent with that expectation despite of the low X-ray detection fraction (about 55). We point out that (i) most of the non-detections correspond to the Host-dominated types, (ii)  loiii\ luminosities also show a good linear correlation ($\rho$ = 0.66) despite this narrow emission line could be contaminated by other sources (post-AGB, shocks, SF) and (iii) the X-ray detections corresponding to the AGN-dominated types tend to follow closer the one-to-one relations. The three galaxies with radio-jets now share the observed trend with the rest of the galaxies. 
   
    \item[] \textit{Near infrared.- } The third panel (from bottom to top) shows the correlation of the optical indicators of the AGN power with the NIR (W1 and W2) WISE luminositites. Given the aperture sizes in the W1 and W2 flux measurements, and their sensitivity to the continuum from old stellar populations, these luminosities should be proportional to the host stellar mass. On the other hand, our optical luminosity indicators are good tracers of the AGN power and thus proportional to the mass of the black hole. Therefore, in principle a correlation should be expected, reflecting a scaling relation between central black hole mass and host galaxy stellar mass \citep[e.g., ][]{Reines2015}. We find good and tighter correlations with \lhabc\ ($\rho > 0.78$) than those with \loiii\ ($\rho >0.74 $). This may be due to the fact that the BLR 
    is closer than the \oiii\ emitting region (NLR) to the ionizing source. We also find a clear segregation in terms of our empirical classification with AGN-dominated types lying in the high luminosity 
    region and the Host-dominated types lying in the corresponding low luminosity region, with the Intermediate types in between. The three galaxies with radio-jets also follow the trend dictated by the rest of the galaxies.

    \item[] \textit{Mid infrared.- } The second panel (from bottom to top) of Figure \ref{fig:OptvsMV} shows the correlation of the optical indicators of the AGN luminosity and the MIR (W3 and W4) WISE luminosities. Since the optical emission emerging from the broad and narrow line regions can be partially absorbed by the hot dust surrounding the central region and re-emitted in the infrared a correlation is also expected. As shown in Table \ref{tab:correlacionpearson}, the Pearson correlation coefficients after linear fittings confirm strong correlations, exceeding $\rho > 0.75$ in the case of \lhabc\ and tighter ($\rho > 0.81$) in the case of \loiii, also with less scatter that the other correlations. This result confirms that the MIR luminosity is also a good indicator of the intrinsic AGN power. 
    A clear segregation is seen with the AGN-dominated types lying in the region of higher optical and WISE luminosities, while the Host-dominated types lying, in contrast, in the region of low optical and WISE luminosities, with the Intermediate types in between. The three galaxies with radio-jets follow the observed trend dictated by the other galaxies. The outlier in the \loiii\ Vs L(MIR) diagram is 1-43214, an AGN-dominated object that seems to have a nuclear starburst \citep{Lakicevic2017}.

    \item[] \textit{Radio.- } The first lower panel shows the relation of the optical indicators of AGN luminosity versus radio-continuum luminosity at 20cm (L$_{\rm 20cm}$ or L$_{\rm radio}$). Filled circles are for radio-detections, downwards triangles indicate upper limits and the larger filled circles with an outer ring are three galaxies showing radio-jets (Figure \ref{fig:Radiojets}). The fit to the observed data  with the (B-J) method is shown (purple dashed line) and reported in Table \ref{tab:correlacionpearson}.  
    Notice that, if the three galaxies with extended radio emission are omitted in both optical indicators, the (B-J) fit provides  a correlation $\rho$ = 0.47 for \lhabc\ and $\rho$ = 0.60 for \loiii. 
    As shown in the previous Section, the position of sources in this diagram reflects the variation of the radio properties \citep[e.g., ][]{Best2012}.  
    The amplitude of log L$_{\rm 20cm}$ range expands towards the sources with radio-jets that significantly depart from the rest of the sources. In these plots, we can appreciate a trend (although with some scatter) of AGN-dominated galaxies having higher optical and radio luminosities, in contrast to Host-dominated AGN having lower optical and radio luminosities, with the Intermediate types in between these.

\end{itemize}

Table \ref{tab:correlacionpearson} shows the results of the Pearson correlation coefficient after fitting the observed distributions of optical and multiwavelength luminosities. We also include the slopes and intercepts of our best-fitting linear regression lines.
We can see that all slopes are positive and lower than 1, which is indicative that the luminosity of \oiii\ or \habc\ increases faster than the other wavelengths. Other authors have previously found these types of correlations  \citep[e.g.,][]{LaMassa2010}. On the other hand, we observe that the slopes of the X-ray and radio correlations are higher for \habc. In comparison, the slopes of the correlations with \oiii\ are higher for the MIR. The difference in the slopes can be explained in terms of the distance from the BLR or NLR to the emission regions of the other wavelengths (the hot corona, the dusty torus or the jets). For the NIR, we have found that the increment of these luminosities is almost half of the increase in the optical luminosities.

From Figure \ref{fig:OptvsMV}, we can see that for X-ray and IR relations, the AGN Dominant group is closer to a 1:1 relation than the Host Galaxy Dominant or Intermediate groups. This can be explained by the stellar contamination since \oiii\ could be strongly affected by the host galaxy emission. When we compare both upper panels, the \oiii\ - X-ray relation is shallower than the \habc\ - X-ray due that the nuclear \oiii\ could be contaminated with emission from 
the host galaxy light or star formation processes. 
The IR relations are stronger than the other wavelengths. However, 
we can see that NIR relations are shallower than MIR, despite the use of \oiii\ or \habc, since W3 and W4 are more related to the nuclear emission of the AGN. In the case of the radio emission, only jetted galaxies (6\% of our sample) are closer to the 1:1 slope. 
We consider that the main differences in the slopes found in this work for each wavelength lie both in the group of AGNs used (Host Dominant, Intermediate, and AGN Dominant) and in the accuracy of the subtraction of the stellar component.

\begin{table}
\scriptsize
\centering
\begin{tabular}{l|cccc}
\hline
\hline
\textbf{Variables}	&	\textbf{$\rho$-value}	&	\textbf{a}	&	\textbf{b}	\\
\hline
log(L$_{\rm{XR}}$) versus log(L$_{\rm{[OIII]}}$) &	0.66	&	0.61$\pm$0.11	&	17.71	\\
log(L$_{\rm{W1}}$) versus log(L$_{\rm{[OIII]}}$) &	0.74	&	0.40$\pm$0.05	&	27.07$\pm$2.12	\\
log(L$_{\rm{W2}}$) versus log(L$_{\rm{[OIII]}}$) &	0.81	&	0.48$\pm$0.05	&	23.37$\pm$2.08	\\
log(L$_{\rm{W3}}$) versus log(L$_{\rm{[OIII]}}$) 	&	0.81	&	0.69$\pm$0.06	&	14.64$\pm$2.38	\\
log(L$_{\rm{W4}}$) versus log(L$_{\rm{[OIII]}}$) 	&	0.82	&	0.78$\pm$0.06	&	11.03$\pm$2.53	\\
log(L$_{\rm{radio}}$) versus log(L$_{\rm{[OIII]}}$) 	&	0.33(0.60)	&	0.50$\pm$0.23	&	17.71	\\
log(L$_{\rm{XR}}$) versus log(L$_{\rm{H\alpha_{BC}}}$)	&	0.70	&	0.77$\pm$0.14	&	10.70	\\
log(L$_{\rm{W1}}$) versus log(L$_{\rm{H\alpha_{BC}}}$)	&	0.78	&	0.43$\pm$0.05	&	25.69$\pm$2.14	\\
log(L$_{\rm{W2}}$) versus log(L$_{\rm{H\alpha_{BC}}}$)	&	0.83	&	0.52$\pm$0.05	&	21.74$\pm$2.13	\\
log(L$_{\rm{W3}}$) versus log(L$_{\rm{H\alpha_{BC}}}$)	&	0.79	&	0.57$\pm$0.07	&	19.61$\pm$2.69	\\
log(L$_{\rm{W4}}$) versus log(L$_{\rm{H\alpha_{BC}}}$)	&	0.75	&	0.59$\pm$0.08	&	19.06$\pm$3.18	\\
log(L$_{\rm{radio}}$) versus log(L$_{\rm{H\alpha_{BC}}}$)	&	0.26(0.47)	&	0.83$\pm$0.29	&	3.07	\\
\hline
\hline
\end{tabular}
\caption{Pearson correlation coefficients for figure \ref{fig:OptvsMV} of the best-fitting linear regression line, $Y$ = $a$X + $b$. L$_{\rm XR}$ is the luminosity at 0.1-2.4 keV,  L$_{\rm W1}$ and L$_{\rm W2}$ are the NIR luminosities, L$_{\rm W3}$ and L$_{\rm W4}$ are the FIR luminosities, and L$_{\rm radio}$ is the luminosity at 20 cm.  The parenthesis values do not include the 3 AGN with observable jets. For the cases of L$_{\rm XR}$ and L$_{\rm radio}$ we calculated using Buckley–James linear regression method to consider the upper limits. This method only gives the error of the slope. }
\label{tab:correlacionpearson}
\end{table}

\section{Summary and Conclusions}
\label{sec:conclusions}

A main goal of the MaNGA survey has been to map the detailed composition and kinematic structure of 10,000 nearby galaxies using integral field unit (IFU) spectroscopy. Its unprecedented detail allows us for a more detailed census of active galactic nuclei located at the centers of galaxies over a wide range of physical parameters. 

We present a method for identifying type-1 AGN, using flux ratios to look for the \ha\ broad emission line. Considering the central three arcsec integrated spectra of the MaNGA DR15 galaxies, we estimated the fluxes in two bands placed at the blue and red side near the expected position of the broad \ha\ line. The line flux ratios are between the bands and the adjacent continuum. The position of the bands and the continuum take into account possible \habc\ asymmetries and the presence of stellar absorption lines. Then using the statistical tool boxplot-whiskers diagram, we identified and separated type-1 AGN candidates from quiescent galaxies.


Our selection method was tested, on one side, in the host galaxy subtracted spectra from the MaNGA DR15, and on the other side, in the observed spectra from the SDSS DR7 sample. For the case where the HG was subtracted, it was proved that even for the lowest S/N 
of our sample (S/N=26, and 11 for the candidate objects), our method is good enough to detect the broad \ha\ component, so we can omit the HG-subtraction, avoiding the bias introduced with such procedures. It seems that the identification of the objects could depend on the line intensity, as considered by \citet{Liu2019} where they propose a lower limit for the \habc\ flux of 10$^{-16}$ erg s$^{-1}$ cm$^{-2}$. Another possibility is that the minimum S/N of the sample affects the upper whisker limit, as in the case of SDSS DR7, where the minimum S/N=1.
The results of the broad AGN identification were also compared with those from other methods applied to the SDSS DR7 and MaNGA spectroscopic data. 

Our main results are as follows:

\begin{itemize}
    \item We identified 47 broad emission line AGN from a sample of $\sim$4700 galaxies in the DR15 sample, amounting to 1\% up to z $\sim$ 0.13. The broad \habc\ luminosity of this sample spans from  10$^{38}$ < L\ha\ < 10$^{44}$ erg s$^{-1}$, and logFWHM(\habc) $\sim$ 3--4 
    covering a range of Eddington ratio 
    -4.18 to -0.34 in logarithm values. The details of the broad component measurements will be presented in Cortes-Su\'arez et al. in prep. 

    \item The result of our selection method was compared with those from other methods based on the BPT and WHAN diagrams using MaNGA data as well as the SDSS DR7 spectroscopic data. We lose 66\%\ of our sample using the BPT diagrams, and 11\%\ using the WHAN diagram (considering a threshold of 3\AA). 
    Applying our method in the DR7 spectra that match with the MPL-7 objects, we missed 17\%\ of the galaxies. The missing 8 objects, two do not have DR7 spectra, two are changing look objects, and the other four have a weak \habc.
    
    \item In the comparison with type-1 AGN catalogs built with the SDSS DR7 database and with different MPL versions of MaNGA, we found that 
    our method selects more efficiently the broad line galaxies. The percentage of detected objects for different authors on the basis of equal samples using SDSS DR7 are: \citet{Liu2019} 81\%, \citet{Oh2015} 65\%, \citet{Stern2012} 67\%. Using MaNGA data: \citet{Rembold2017} 36\%, \citet{Sanchez2018} 21\%, \citet{Wylezalek2018} 26\%, and \citet{Comerford2020} 72\%. 
    
    \item We classify the AGN-HG contribution empirically using spectral indexes into 
    AGN-dominated (19\%), Intermediate (30\%) and Host-dominated types (51\%), evidencing different levels of nuclear activity contributing in the observed spectra. 
    
    \item Using the \oiiib\ vs excitation index plane, we find that our sample is composed of 4 HE and 4 LE galaxies.
    
    \item From the photometric information in the IR and radio wavelengths, we identify our type-1 AGN sample in the WISE color-color diagram finding a good agreement between AGN-HG dominance and WISE colors, bluer for AGN dominated and redder for HG dominated objects. In the radio emission characterization, we find 5 HERGs and 4 LERGs, three of them
    showing visible evidence of radio-jets in the FIRST images.  

    \item In the multiwavelength analysis, luminosity indicators of the AGN power in the X-ray, optical, IR and radio ranges were confronted, finding that L(\habc) provides similar correlations than L[OIII]. The correlation with the broad emitting region is expected, given that the proximity of this region is best related to nuclear activity. Our type-1 AGN sample also shows segregation in these diagrams when our empirical classification is considered.   
\end{itemize}

The identification of this type-1 AGN sample, the analysis of 
their most basic properties combined with the multiparametric space of the MaNGA data and their properties in different wavelengths, will enable further comprehensive investigations of the properties of this sample AGN, and their connection with the host galaxies.


\section*{Acknowledgments}

ECS acknowledges the fellowship 825458 from CONACyT. CAN acknowledges support form grant IN111422 PAPIIT UNAM, and CONACyT project Paradigmas y Controversias de la Ciencia 2022-320020. HMHT acknowledges support from UC MEXUS-CONACYT grant CN-17-128.

Funding for the Sloan Digital Sky Survey IV has been provided by the Alfred P. Sloan Foundation, the U.S. Department of Energy Office of Science, and the Participating Institutions. SDSS acknowledges support and resources from the Center for High-Performance Computing at the University of Utah. The SDSS web site is www.sdss.org.

SDSS is managed by the Astrophysical Research Consortium for the Participating Institutions of the SDSS Collaboration including the Brazilian Participation Group, the Carnegie Institution for Science, Carnegie Mellon University, Center for Astrophysics | Harvard \& Smithsonian (CfA), the Chilean Participation Group, the French Participation Group, Instituto de Astrof\'isica de Canarias, The Johns Hopkins University, Kavli Institute for the Physics and Mathematics of the Universe (IPMU) / University of Tokyo, the Korean Participation Group, Lawrence Berkeley National Laboratory, Leibniz Institut f\"ur Astrophysik Potsdam (AIP), Max-Planck-Institut f\"ur Astronomie (MPIA Heidelberg), Max-Planck-Institut f\"ur Astrophysik (MPA Garching), Max-Planck-Institut f\"ur Extraterrestrische Physik (MPE), National Astronomical Observatories of China, New Mexico State University, New York University, University of Notre Dame, Observatorio Nacional / MCTI, The Ohio State University, Pennsylvania State University, Shanghai Astronomical Observatory, United Kingdom Participation Group, Universidad Nacional Aut\'onoma de México, University of Arizona, University of Colorado Boulder, University of Oxford, University of Portsmouth, University of Utah, University of Virginia, University of Washington, University of Wisconsin, Vanderbilt University, and Yale University.  This project makes use of the MaNGA-Pipe3D \textbf{VAC}.  
We thank the IA-UNAM MaNGA team for creating this catalogue. 

\section*{Data Availability}

The data underlying this article are available at the MaNGA-Pipe3D Valued Added Catalog at \url{https://www.sdss.org/dr17/manga/manga-data/manga-pipe3d-value-added-catalog/} and \url{
http://ifs.astroscu.unam.mx/MaNGA/Pipe3D_v3_1_1/tables/
}. 
The datasets were derived from sources in the public domain using the SDSS-IV MaNGA public Data Release 15, at \url{https://www.sdss.org/dr15/}



\bibliographystyle{mnras}
\bibliography{example} 



\appendix
\section{Type-2 AGN list}
\label{appendix:A}
In section \ref{sec:type2} is described the selection of AGN using the classical BPT diagrams \citep{Baldwin1981} and WHAN diagram \citep{CidFernandes2011}. Here we present the remaining list of AGN classified as type-2 found with the flux ratio method and using the \citet{Sanchez2018} criterion.

\begin{table*}
\centering
\begin{tabular}{cccccccc}
\hline
\hline
\textbf{SDSS-ID} &\textbf{MaNGA-ID} &
\textbf{Plate-IFU} & \textbf{R.A.}   & \textbf{Dec}  & \textbf{m$_g^a$} & \textbf{z$^b$} & M$_*^c$ (M$\odot$) \\ 
(1)&(2)&(3)&(4)&(5)&(6)&(7)&(8)\\
\hline \\

J162640.78+274639.1 & 1-561297  & 9025-6101 & 246.6700 & 27.7775 & -20.97 & 0.079 & 10.89  \\
J162926.70+293228.0 &  1-272639    & 9025-6104 & 247.3613 & 29.5411 & -20.97 & 0.053 & 10.81  \\
J163617.87+440808.5 &  1-93793     & 9026-12701 & 249.0745 & 44.1357 & -20.64 & 0.031 & 10.79  \\
J163716.42+442505.6 & 1-94784     & 9026-9101 & 249.3184 & 44.4182 & -20.80 & 0.031 & 10.72  \\
J161211.26+293426.8 & 1-561325    & 9028-9102 & 243.0469 & 29.5741 & -21.19 & 0.054 & 11.03  \\
J162852.06+424843.2 & 1-135285    & 9029-12704 & 247.2170 & 42.8120 & -20.28 & 0.032 & 10.53  \\
J162954.43+413616.3 & 1-569178    & 9029-9101 & 247.4768 & 41.6045 & -20.27 & 0.031 & 10.67  \\
J161116.61+450716.7 & 1-209772    & 9031-12703 & 242.8192 & 45.1213 & -20.18 & 0.055 & 10.52  \\
J154137.28+441619.9 & 1-247804    & 9035-1901 & 235.4053 & 44.2722 & -17.39 & 0.037 & 9.44  \\
J155627.25+420604.1 & 1-248003   & 9036-3701 & 239.1136 & 42.1011 & -20.78 & 0.039 & 10.84  \\
... & ... & ... & ... & ... & ... & ... \\

\\
\hline
\hline
\end{tabular}
\vspace{4pt}
\caption{Main properties of the 236 type-2 AGN. ($^a$)The magnitude corresponds to the g band provided by SDSS (DR14). ($^b$) The redshift and \oiii\ luminosity was obtained by Pipe3D \citep{Sanchez2016a,Sanchez2021}. ($^c$) Stellar masses from NSA catalog derived from Sersic fluxes. ($^d$) The \habc\ luminosity was obtained from our emission lines fitting (Cortes-Suarez et al. in prep.). 
Note. — The entire type-2 AGN Catalog will be available online.}
\label{tab:type2}
\end{table*}
   
\section{Changing-look AGN candidates}
\label{appendix:B}

\begin{figure}
  \includegraphics[width=\columnwidth]{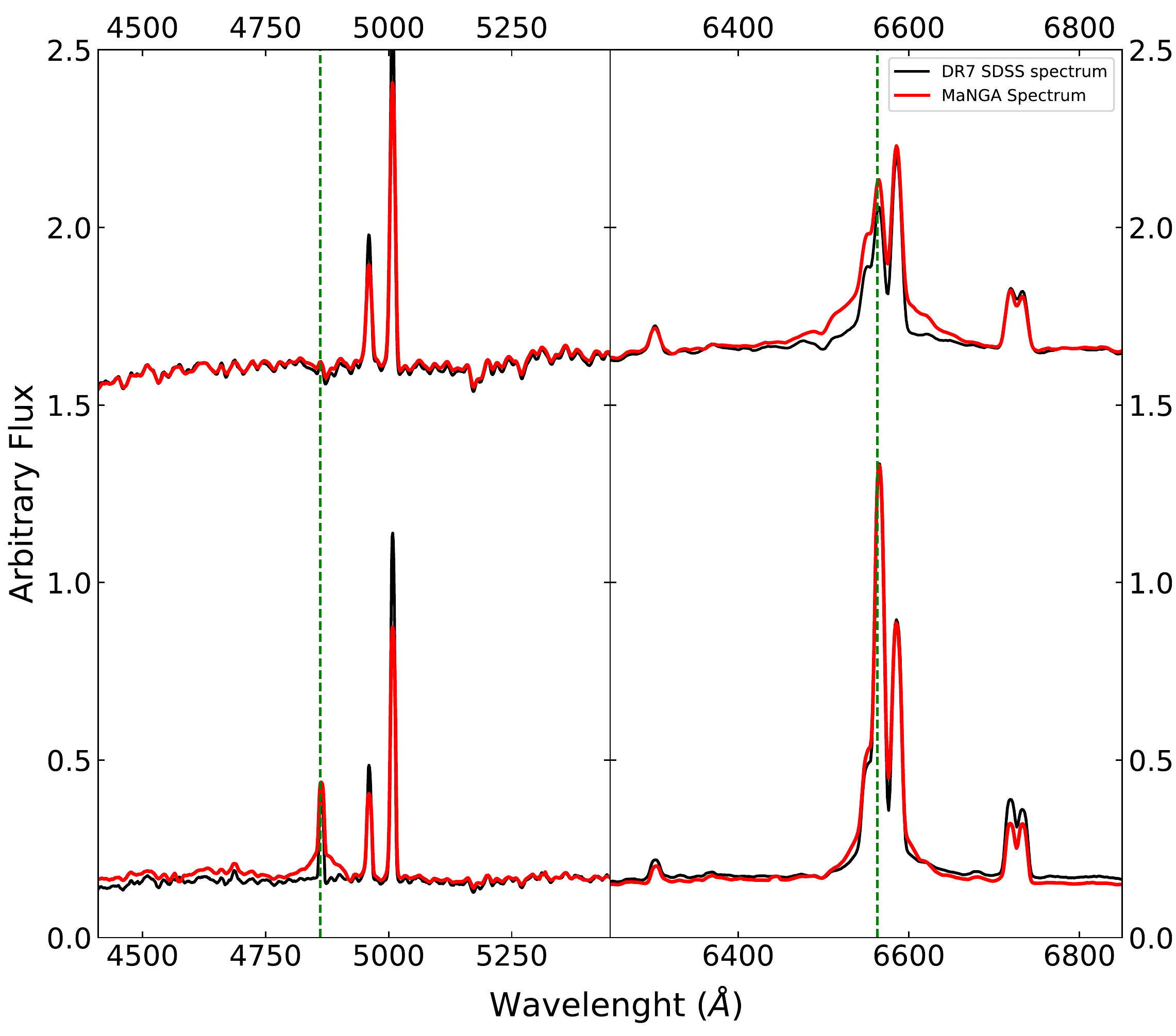}
  \caption{Two examples of possible changing-look AGN. Red spectra comes from DR7 SDSS spectroscopic data and Black spectra from MaNGA data. Left panel correspond to the \hb\ region and right panel to the \ha\ region.}
\label{fig:changelook}
\end{figure}

In Figure \ref{fig:changelook} we report two examples of broad \ha\ and \hb\ type-1 AGN that were not detected by \citetalias{Oh2015} and \citetalias{Stern2012} but whose broad \ha\ components have significantly changed or are now visible in the MaNGA DR15 data.

\begin{figure}
  \includegraphics[width=\columnwidth]{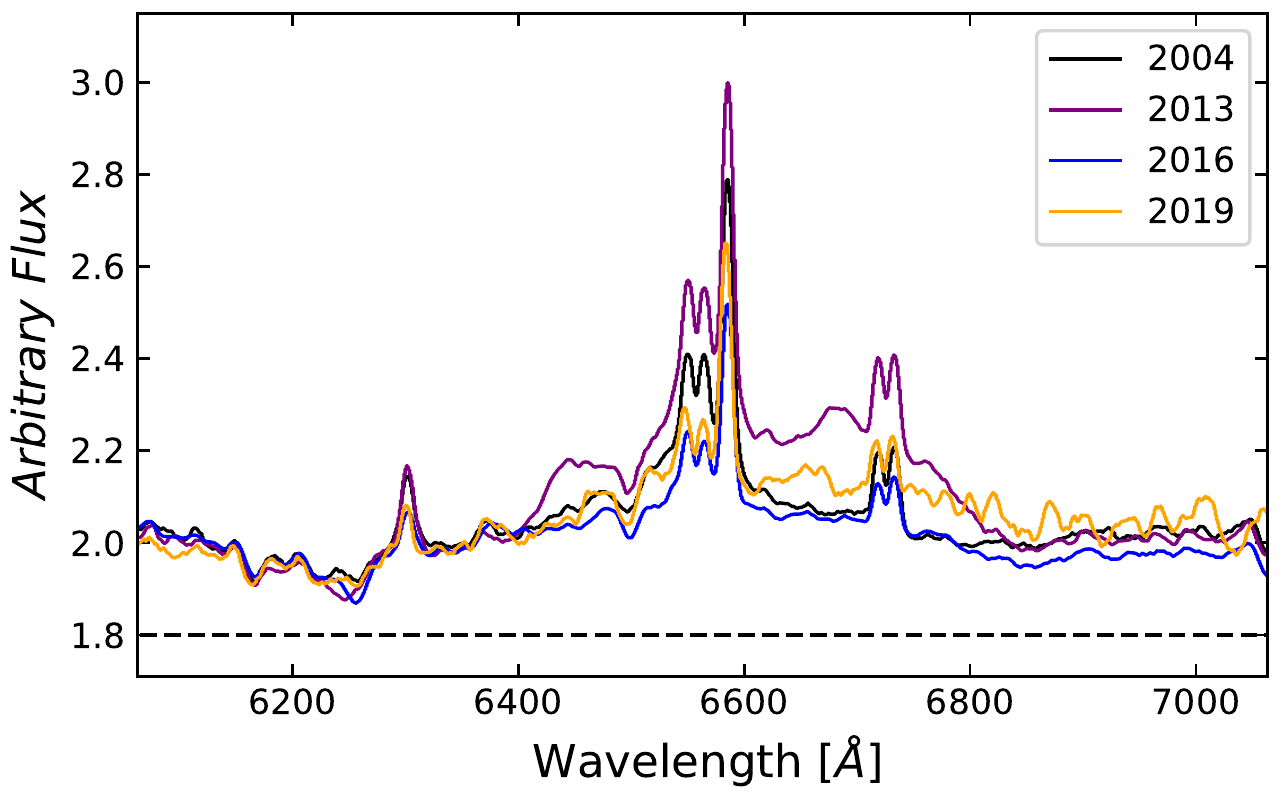}
  \caption{Comparison of four different epoch spectra for one of our type-1 AGN, 1-604860. Black spectrum is from SDSS DR7 observed in 2004, Purple from SDSS DR12 observed in 2013, Blue from MaNGA observed in 2016 and Yellow from OAN-SPM observed in 2019. Fluxes were normalizaed at 6350\AA\ for comparisons.}
\label{fig:8714}
\end{figure}

Since there is a time base of 8-16 years between observations, we speculate for a change in the physical conditions associated to the emission of these lines or even a turning off of the nuclear activity in these objects. An example is shown in Figure \ref{fig:8714} where we compare four different single-epoch spectra for the MaNGA galaxy 1-604860. The SDSS DR7 spectrum (black) was observed in 2004, the SDSS DR12 (purple) in 2013, the MaNGA spectrum (blue) in 2016, and an additional spectrum obtained at the Observatorio Astron\'omico Nacional 
San Pedro Mártir (yellow; OAN-SPM, México) in 2019. A direct inspection shows that the narrow lines fluxes change while the broad components show different level of prominence and shape along this time base. In the DR12 observation, the broad \ha\ line shows two broad components, while in the DR15 MaNGA observations these components are barely present. In the OAN-SPM observation, the broad component located beetween 6600-6800\AA\ seems to be emerging again. 

These changes are thought to be likely associated to either rapid outflow
or inflow with light from the inner disk and BLR that is being obscured by dusty clouds \citep[][]{Guo2016},
or possible changes in accretion rate of the central black hole, changes in accretion disk structure, or tidal disruptions \citep[][]{Merloni2015,Kokubo2015,MacLeod2016,Ross2018}.

\bsp	
\label{lastpage}
\end{document}